\ifdefined\tab
\documentclass[20pt,landscape]{extreport} 
\else
\documentclass[11pt]{report} 
\fi

\usepackage{cancel}
\usepackage[margin=1cm,footskip=0.7cm]{geometry}
\usepackage{indentfirst}
\usepackage[colorlinks=true,urlcolor=blue,linkcolor=blue]{hyperref}
\usepackage{lipsum}
\ifdefined\Vpubli
\usepackage[latin1]{inputenc} 
\else
\usepackage[utf8]{inputenc}
\fi
\usepackage[T1]{fontenc}

\usepackage{amsmath}

\usepackage{amssymb}
\usepackage{stmaryrd}
\usepackage[squaren,Gray]{SIunits}
\usepackage{epsfig}
\usepackage{times}
\usepackage{ifthen}
\usepackage{comment} 
\usepackage{mdwlist}
\usepackage{upquote,textcomp}
\usepackage[T1]{fontenc}

\usepackage{lmodern}
\usepackage{framed}
\usepackage{array}
\usepackage{fancyvrb}
\usepackage{tikz}

\usepackage{graphics}
\usepackage{xr}

\usepackage[squaren,Gray]{SIunits}
\usepackage{amstext}

\usepackage{hyperref}
\hypersetup{
    colorlinks=true,
    linkcolor=blue,
    filecolor=magenta,      
    urlcolor=cyan,
    pdftitle={SYG~: sujets de recherche},
    pdfpagemode=FullScreen,
    }

\usepackage{xcolor}

\newcommand{\dau}{https://www-l2ti.univ-paris13.fr/~dauphin/}
\newcommand{\tq}{\textquotesingle}

\ifx\Vpubli\undefined

\fi
\newcommand{\bb}[1]{{\mathbb{#1}}}

\def\supp{{\rm supp}\ }

\DeclareMathOperator{\signe}{signe}

\DeclareMathOperator*{\argmin}{arg\,min}

\DeclareMathAlphabet{\pazocal}{OMS}{zplm}{m}{n}

\newcommand{\un}[1]{\underline #1}
\newcommand{\itb}{\item[$\bullet$]}
\newcommand{\1}{{\mathbf 1}}

\newcommand{\Rr}{{\mathbb R}}

\newcommand{\beq}{\begin{equation}}
\newcommand{\eeq}{\end{equation}}
\newcommand{\beqs}{\begin{displaymath}}
\newcommand{\eeqs}{\end{displaymath}}
\newcommand{\beqns}{\begin{eqnarray*}}
\newcommand{\eeqns}{\end{eqnarray*}}
\newcommand{\beqn}{\begin{eqnarray}}
\newcommand{\eeqn}{\end{eqnarray}}

\DefineVerbatimEnvironment{octave}{Verbatim}{fontfamily=courier,frame=leftline,framerule=1mm}
\DefineVerbatimEnvironment{matlab}{Verbatim}{frame=leftline,framerule=1mm}

\newcommand{\ww}[1]{{\widehat{#1}}}

\newcommand{\mb}[1]{{\mbox{#1}}}
\newcommand{\sm}[1]{{\mbox{\footnotesize #1}}}

\newcommand{\br}[1]{{\mathrm{\mathbf{#1}}}}
\newcommand{\p}[1]{{{\pazocal{#1}}}}

\newcommand{\deb}{{\setcounter{enumi}{\value{suite}}}}
\newcommand{\finenum}{{\setcounter{suite}{\value{enumi}}}}
\newcommand{\sol}{{{\bf \underline {Solution}~:}}}
\newcommand{\imp}{{{\bf \underline {Implémentation}~:}}}
\newcommand{\ind}{{{\bf \underline {Indication}~:}}}

\ifdefined\Vex
\newcommand{\inputex}[1]{\input{./ex/#1}} 
\else
\newcommand{\inputex}[1]{}                 
\fi

\ifdefined\Vcours
\newcommand{\inputcours}[1]{\input{./cours/#1}}
\else
\newcommand{\inputcours}[1]{}
\fi

\ifdefined\Vqcm
\newcommand{\inputqcm}[1]{\input{./qcm/#1}}  %
\else
\newcommand{\inputqcm}[1]{}  
\fi

\newboolean{correction}
\ifdefined\corrected
\setboolean{correction}{true}
\else
\setboolean{correction}{false}
\fi

\ifthenelse{\boolean{correction}}
           {\includecomment{corr}}
           {\excludecomment{corr}}

\newcommand{\msm}{$\mbox{m}.\mbox{s}^{-1}$}
\newcommand{\de}{{({\mbox{\small de}})}}
\newcommand{\moy}{{({ \mbox{\small moy})}}}
\newcommand{\der}{{({ \mbox{\small fin}})}}
\newcommand{\ce}{{({ \mbox{\small ce}})}}
\newcommand{\PMCT}{{\mbox{PMCT}}}

\newcommand{\bz}{{\mathbf z}}
\newcommand{\deux}{{\text{\tiny{|2}}}}

\DeclareMathOperator{\OA}{OA}

\title{
Introduction à la reconnaissance de la parole  \\
}
\author{{\small Gabriel Dauphin}}

\begin{document}



\maketitle
\tableofcontents

\chapter*{Introduction}

Ce cours présente un ensemble de techniques qui sont utilisées en travaux pratiques 
pour distinguer les sons correspondant aux mots {\tt un}, {\tt deux} et {\tt trois}. 
Ces techniques utilisent la notion de description temps-fréquence qui est implicite dans la 
décomposition d'un son en trames, explicite dans la notion de {\tt spectrogramme} et qui est 
utilisée dans les différents descripteurs présentés dans le chapitre~\ref{ch:desc}.
Le chapitre~\ref{ch:traitement_son} a pour objectif de décrire la façon des sons sont représentées, cela amène 
à décrire brièvement les organes humains de la perception et le prétraitement qui est appliqué 
aux données associées aux sons. 
Le chapitre~\ref{ch:classification_par_descripteurs} a pour objectif de présenter la classification, à savoir son objectif, la façon dont on l'évalue et 
la manière dont elle utilise des valeurs calculées pour chaque son  que l'on appelle ici {\em descripteur}. 
Le chapitre~\ref{ch:desc} présente différentes techniques permettant de calculer les valeurs de descripteurs à partir des données 
associés aux sons. Cela amène à décrire brièvement les organes humains de la génération d'un son car cela explique 
l'intérêt porté à l'utilisation d'une description fréquentielle de portions de signaux temporels. 

En tant que document de cours, il a pour objectif de préciser les notions à connaître. Les notions 
à connaître le plus précisément sont la compréhension et la connaissance des {\em formules encadrées}, la compréhension 
des {\em graphiques et schémas} présentés et la connaissance et la signification des {\em termes techniques}.

\chapter{Représentation d'un son}
\label{ch:traitement_son}

Pour décrire et discuter de la façon dont les sons sont représentées, la 
section~\ref{sec:questce_quun_son} discute de ce qu'est physiquement un son et donne 
une brève description des organes humains utilisées pour la perception d'un son. 
La section~\ref{sec:son_temps} rappelle comment le traitement numérique du signal permet 
de représenter un son.
La section~\ref{sec:N:son_trames} montre une façon un peu différente de représenter le son au moyens 
de {\em trames}, c'est-à-dire de petites portions du signal sonore. 
La section~\ref{sec:detection_silence} montre une façon simplifiée de détecter les phases de silence et de paroles, 
il se trouve que ceci était un des objectifs de ce cours. Cette détection est utilisée pour réaliser un 
prétraitement permettant d'améliorer la performance de la classification. 
La section~\ref{sec:implementation} donne quelques indications sur la façon dont les techniques proposées sont implémentées
dans les travaux pratiques qui illustre ce cours.

\section{Qu'est-ce qu'un son~?}
\label{sec:questce_quun_son}

La section~\ref{ssec:model_physique} présente une description du son en tant qu'onde sonore se propageant 
dans l'air. 
La section~\ref{ssec:oreille} présente une très brève description du fonctionnement de l'oreille 
en tant qu'organe de perception des sons. 
La section~\ref{ssec:audibilite} présente des courbes d'audibilité, c'est-à-dire les seuils hauts et bas 
de perception du son.

\subsection{Modélisation physique}
\label{ssec:model_physique}
Un son est généralement décrit comme plus ou moins aigu ou plus ou moins fort. Nous allons voir
que la première notion correspond à une fréquence et la deuxième à un niveau sonore. 

Le son est une variation de la pression de l'air.~\footnote{Un son peut aussi se propager à travers d'autres matériaux et dans ce cas 
c'est le matériau traversé qui se met à vibrer, en revanche le son ne peut pas se propager à travers le vide.}
Cette variation se propage sous la forme d'une onde. Une modélisation sphérique de cette onde est décrite par 
\beqn\label{eq:propag_onde}
P(r,t)=P_0+\frac{1}{r}\left [g_1(t-\frac{r}{c})+g_2(t+\frac{r}{c})\right ]
\eeqn
\begin{itemize}
\item $P_0$ est la pression atmosphérique moyenne. L'unité dans laquelle elle s'exprime est le Pa (Pascal). La pression de l'air est environ $10^5$Pa.
\item $r$ est la distance en mètres (m) entre le point où on mesure la pression et la  source de ce signal sonore. 
\item $t$ est l'instant en secondes~(s) pour lequel on mesure la pression $P(r,t)$. 
\item $c$ est la vitesse ($\mbox{m}.\mbox{s}^{-1}$) de propagation du son à travers l'air. C'est environ de $340$\msm. 
\item $g_1$ et $g_2$ peuvent a priori être des fonctions quelconques. Pour modéliser la propagation d'un signal sonore, on 
considère $g_2=0$ et pour un son stationnaire par exemple dans une caisse de résonnance, $g_1$ et $g_2$ sont similaires. 
\end{itemize}

Pour modéliser une note de musique qui dure un temps infini, on utilise 
\beqn
\boxed{
g_1(\tau)=\cos(2\pi f_0\tau)
}
\eeqn
Cette note est caractérisée par sa fréquence $f_0$ en Hz. 
En effet, avec cette modélisation on observe que $P(r,t+\frac{1}{f_0})=P(r,t)$, 
c'est-à-dire qu'en chaque point la pression évolue dans le temps avec une périodicité de $\frac{1}{f_0}$. 
{\small 
\beqns
P(r,t+\frac{1}{f_0})=P_0+\frac{1}{r}\cos(2\pi f_0(t+\frac{1}{f_0}-\frac{r}{c}))
=P_0+\frac{1}{r}\cos(2\pi f_0(t-\frac{r}{c}))=P(r,t)
\eeqns
}
Et $f_0$ est reliée à une longueur d'onde notée $\lambda_0$ mesurée en mètres.
\beqn\label{eq:long_onde}
\boxed{
\lambda =\frac{c}{f_0}
}
\eeqn
Ceci traduit une évolution quasi-périodique du signal dans l'espace $P(r+\lambda,t)\approx P(r,t)$. 
{\small 
\beqns
P(r+\lambda,t)=P_0+\frac{1}{r+\lambda}\cos(2\pi f_0(t-\frac{r}{c}-\frac{\lambda}{c}))
=P_0+\frac{1}{r+\lambda}\cos(2\pi f_0(t-\frac{r}{c}))\approx P(r,t)
\eeqns
}
\`A l'oreille, ce son apparaît comme une note de musique. Il s'agit d'un La lorsque $f_0=440$Hz. 
Cette fréquence est donc liée au caractère plus ou moins aigu du son. 

L'intensité sonore d'un son correspond au produit de la surpression par la vitesse. Dans le cadre de ce cours, 
on considère qu'il s'agit du carré de la partie variable de la pression. 
\beqns
I=\left (P(r,t)-P_0\right )^2
\eeqns
Cette intensité sonore est en Watt par mètres carrés ($\mbox{W}.\mbox{m}^{-2}$).
Dans le cadre de ce cours, on attache de l'importance au fait que cette intensité sonore est proportionnelle à $\frac{1}{r^2}$. 
\beqn
\boxed{
I=\frac{I_e}{r^2}
}
\eeqn
$I_e$ étant une quantité inconnue à estimer. 
Aussi lorsqu'on s'éloigne de la source en doublant la distance, l'intensité sonore est divisée par $4$. 

On appelle niveau sonore, le fait d'utiliser une échelle en décibel pour mesurer l'intensité sonore. 
\beqn\label{eq:N:niveau_sonore}
\boxed{
L=10\log_{10}\left (\frac{I}{I_0}\right )
}
\eeqn
où $I_0$ est une intensité sonore de référence. On peut remarquer que dans l'équation~(\ref{eq:N:niveau_sonore}), 
l'échelle en décibel n'est pas calculé avec un facteur multiplicatif de $20$ mais de $10$, parce que l'intensité sonore est 
déjà un terme carré. Et en effet lorsqu'on s'éloigne d'une source sonore en multipliant par deux la distance, 
le niveau sonore est diminué de $10\log_{10}(4)=6$dB.~\footnote{C'est en effet la même diminution que pour 
le rapport signal sur bruit, si on divisait par deux le bruit sans modifier le signal source.}

\subsection{Fonctionnement schématique de l'oreille}
\label{ssec:oreille}

\begin{figure}
\begin{center}
\begin{tabular}{c}
\includegraphics[width=0.5\linewidth]{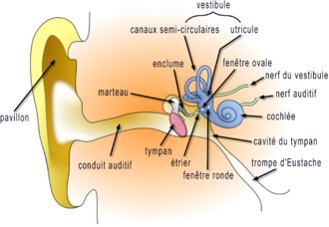}
\end{tabular}
\end{center}
\caption{Schéma détaillé de l'oreille.}
\label{fig:N:oreille}
\end{figure}

La figure~\ref{fig:N:oreille} (extraite d'une publication) 
représente un schéma détaillé de l'oreille. Schématiquement, le son s'introduit dans le conduit auditif, 
fait vibrer le tympan qui actionne le marteau, l'enclume et l'étrier qui à son tour fait vibrer le liquide présent dans la cochlée. 
Cette cochlée est tapissée de cellules ciliées qui transmettent les vibrations vers des nerfs auditifs. Les cellules au début 
de la cochlée sont sensibles aux sons aigus tandis que celles en fin de cochlée sont sensibles aux sons graves. 

\subsection{Courbes d'audibilité humaine}
\label{ssec:audibilite}

\begin{figure}
\begin{center}
\begin{tabular}{c}
\includegraphics[width=0.5\linewidth]{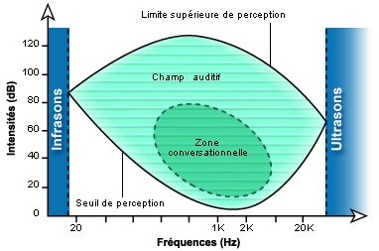}
\end{tabular}
\end{center}
\caption{Courbe d'audibilité humaine.}
\label{fig:N:courbe_audibilite}
\end{figure}
	
Dans la figure~\ref{fig:N:courbe_audibilite} (extraite d'une publication), l'échelle des ordonnées est le niveau sonore d'un son écouté. L'échelle 
des abscisses est une échelle de fréquences (Hz) mais avec une distribution logarithmique. \`A gauche ce sont les infrasons qui
correspondent aux sons qu'on ne peut entendre parce que leur fréquence est trop basse. \`A droite ce sont les ultrasons 
qu'on ne peut pas non plus entendre parce que leur fréquence est trop haute. Entre ces deux fréquences, il y a deux courbes, 
si le niveau sonore d'un son est supérieur à la limite donnée par la courbe inférieure alors on peut l'entendre et s'il 
est supérieur à la limite donnée par la courbe supérieure alors ce son peut abîmer l'oreille. 
Ces courbes sont obtenues en expérimentant sur un grand nombre de patients, elles indiquent un sensibilité moyenne, mais 
il y a en réalité des différences importantes d'un individu à un autre. 
Ces courbes nous amènent à affirmer qu'une assez grande partie de l'information sonore est obtenue en ne considérant 
que les fréquences inférieures à $4$kHz et qu'a priori la totalité de l'information sonore est obtenue en 
ne considérant que les fréquences inférieures à $20$kHz.

\section{Représentation temporelle d'un son}
\label{sec:son_temps}

\begin{figure}
\begin{center}
\begin{tabular}{c}
\includegraphics[width=0.5\linewidth]{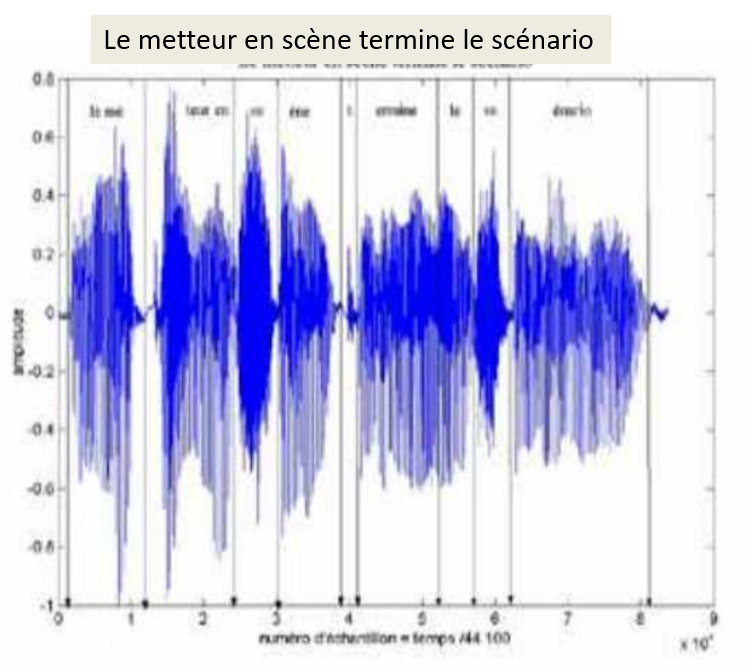}
\end{tabular}
\end{center}
\caption{Représentation temporelle du signal sonore.}
\label{fig:N:son_temps}
\end{figure}

En tant qu'onde sonore, un son est un signal temps continu à valeurs continus. Mais lorsqu'il est 
converti en signal électrique avec un capteur acoustique appelé microphone puis converti en signal numérique avec 
échantillonnage et quantification, le signal devient temps discret. 
Du fait du critère de Shannon-Nyquist, la fréquence d'échantillonnage $f_e$ est entre $40$kHz et $50$kHz (en général 
$44.1$kHz mais, 
elle peut parfois être de $8$kHz. Le signal est souvent quantifié sur $8$ bits, mais cela peut être $10$ ou $12$ bits. 
Le signal échantillonné est noté $x_n$, chaque indique $n$ correspondant à un instant $t_n=nT_e$ où $T_e=\frac{1}{f_e}$.
La figure~\ref{fig:N:son_temps} (extraite d'une publication), indique l'évolution au cours du temps de l'intensité sonore. 
En ordonnée, il n'y a pas d'unité, c'est a priori une fraction de la valeur maximale. Mais le carré du signal est proportionnel à l'intensité sonore. 
Le signal représenté est celui obtenu en enregistrant la phrase {\em Le metteur en scène termine le scénario.}

On appelle {\em durée} du signal, le temps qui s'écoule entre le début 
du signal sonore et la fin. Cette durée est égale à $NT_e$ où ici $N$ est 
le nombre d'échantillons~\footnote{\'Etant donné 
la valeur de $f_e$, il est tout à fait satisfaisant d'approcher cette durée 
à un multiple de $T_e$.} représenté sur la figure~\ref{fig:N:son_temps}. 
\'Etant donné la valeur élevée de $f_e$, on a en fait un très grand nombre d'échantillons, 
bien plus qu'il n'est possible de distinguer à l'oeil en observant la figure. L'allure générale permet aussi de distinguer les
 moments où il y a du silence, des moments où 
il y a un signal sonore important~: le silence correspond à des valeurs très faibles en valeur absolues et le signal sonore important correspond à des 
valeurs importantes du signal. 
Comme chaque intervalle de temps dans le graphe correspond aussi à un intervalle de 
temps durant lequel un son est prononcé, il est possible de faire correspondre chaque syllabe à un morceau du signal. On peut ainsi visualiser pour certaines sons des particularités, le premier {\em t} et les deux {\em s} correspondent à des parties très foncées du graphe. 

Bref, cette représentation temporelle du signal suggère l'étude séparée de chaque morceau de signal correspondant à un son ou à une syllabe. Mais comme on ne connaît 
pas a priori quand commence et quand finit le son ou la syllabe, la 
section~\ref{sec:N:son_trames} montre que l'on découpe le signal en une succession
de trames. 

Les notations utilisées pour décrire l'ensemble du signal sont les suivantes. 
\begin{itemize}
\item $T$ durée du signal en secondes.
\item $N$ nombre total d'échantillons.
\item $t_n=nT_e$ échelle de temps du signal pour $n\in\{0\ldots N-1\}$. 
\end{itemize}
On a la relation entre durée du signal et période d'échantillonnage. 
\beqn
\boxed{
T=NT_e
}
\eeqn
On a aussi l'échantillonnage du signal.
\beqn
\boxed{
x_n=x(t_n)
}
\eeqn

Pour illustrer des notions traitement du signal, on s'interroge ici sur la façon 
de rendre un son plus grave au moyen d'un traitement numérique du vecteur $x_n$. 
\begin{itemize}
\item En diminuant $f_e$, le son est plus grave. 
Mais sa durée est aussi diminuée d'autant puisqu'elle est égale à $\frac{N}{f_e}$. 
\item Si on effectue un sur-échantillonnage tout en conservant la fréquence 
d'échantillonnage alors le son devient plus grave mais sa durée est 
augmentée d'autant. 
\item Si on effectue une modulation, c'est-à-dire qu'on multiplie le graphe
par une sinusoïde d'une certaine fréquence alors on déplace effectivement 
le son à la fois en le rendant plus aigu et plus grave. Si ce son a un spectre
très étroit alors il est possible de supprimer la composante aigüe ou la composante grave. Et la durée est alors inchangée.  
\end{itemize}

\section{Représentation du signal sonore en une succession de trames}
\label{sec:N:son_trames}

\subsection{Découpage en trames sans chevauchement}

\begin{figure}
\begin{center}
\begin{tabular}{c}
\includegraphics[width=0.8\linewidth]{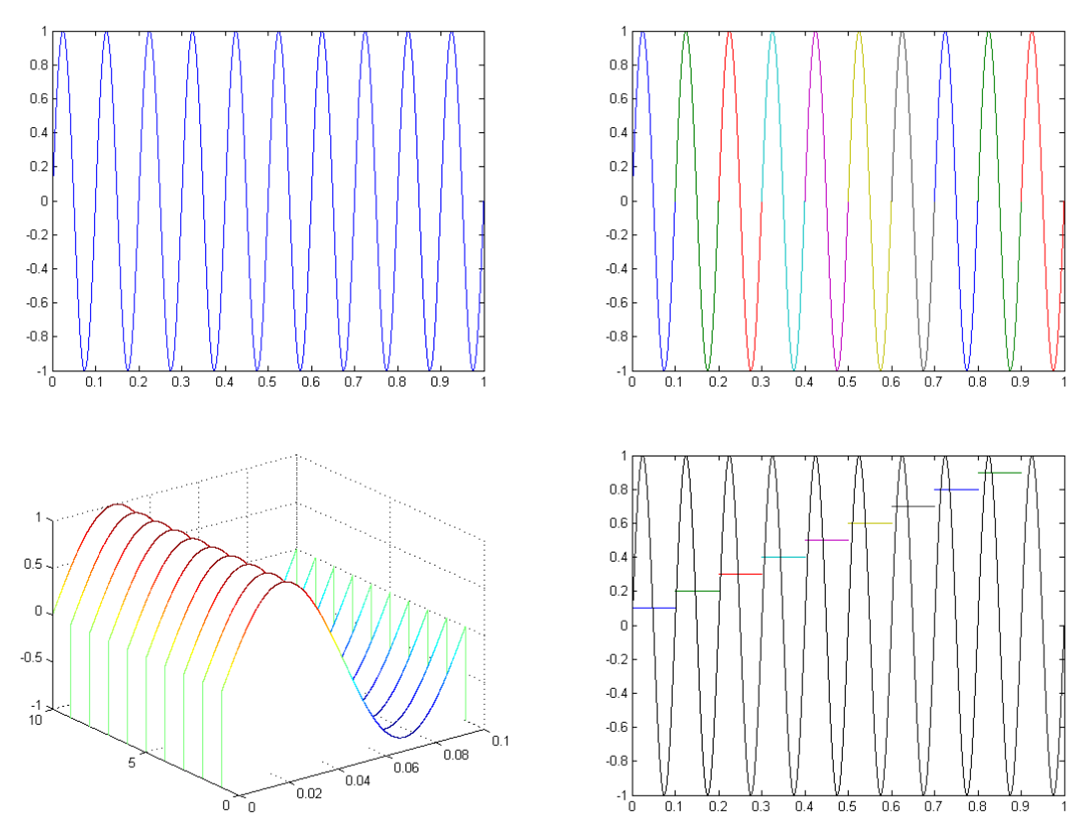}
\end{tabular}
\end{center}
\caption{Découpage d'une sinusoïde en trames qui ne se chevauchent pas.}
\label{fig:N:son_trames_simples}
\end{figure}
	
La figure~\ref{fig:N:son_trames_simples} donne quatre représentation 
d'un même signal défini par 
\beqns
x_n=\sin(20\pi nT_e)
\eeqns
C'est le signal représenté en haut à gauche de cette figure où sur une seconde, il
y $10$ périodes de la sinusoïde qui sont répétées et où donc, la fréquence de la 
sinusoïde est de $10$Hz et la période $T=0.1$s. L'échelle des ordonnées est 
une fraction de l'amplitude maximale. L'échelle des abscisses est la seconde. 

Dans cet exemple, on découpe le signal d'une seconde en $10$ trames chacune 
dure ici $0.1$s. La figure en haut à droite représente chaque trame avec une couleur 
différente~: bleu pour la première trame, vert pour la deuxième, rouge pour la troisième, cyan pour la quatrième, violet pour la cinquième, etc...

Le graphe en bas à gauche est une autre représentation de ces trames. L'axe pointant vers la droite est le temps écoulé depuis le début de chaque trame. 
L'axe pointant vers la gauche et commençant à zéro au milieu indique le numéro 
de la trame considérée. L'axe vers le haut indique la valeur du signal pour la trame considérée à l'instant de la trame considérée. La valeur zéro est ici à mi-hauteur. Ici la couleur ne distingue plus les trames entre elles, elle dépend de la valeur prise par le signal. Dans cet exemple le signal est identique sur chaque trame, c'est pour cela qu'on retrouve exactement le même signal sur chaque courbe. 

La courbe en noir du graphe en bas à droite est la même que le graphe en haut à gauche, il s'agit du signal $x_n$ représenté en fonction du temps. Sur ce graphe, il y a en plus des traits horizontaux de couleurs et de hauteurs différentes
qui permettent d'identifier chaque trame. Ce sont ces traits qui permettent de lire 
les instants associés au début et à la fin de chaque trame en trouvant l'abscisse 
du début et de la fin du trait horizontal. 

\subsection{Notations pour définir les trames}

On utilise ici les notations suivantes. Ces notations ne sont pas 
universelles, il est donc nécessaire de les redéfinir quand on les utilise. 
\begin{itemize}
\item $K$ désigne le nombre de trames. 
\item $T_K$ est la durée d'une trame. 
\item $N_K$ est le nombre d'échantillons dans une trame, il est identique pour toutes les trames d'un même son.
\item $k\in \{0\ldots K-1\}$ est un indice qui identifie la trame. 
\item $t_k^\de$ instant associé au premier échantillon de la trame $k$.
\item $t_k^\der$ instant associé au dernier échantillon de la trame $k$. 
\item $t_k^\ce$ instant associé au centre de la trame $k$. 
\item $t_{k,n}$ échelle de temps associée à la trame $k$ donnant 
les instants associées aux échantillons $n=0$ jusqu'à $n=N_K-1$. 
\end{itemize}

Du fait qu'il n'y a pas de chevauchement, on a
\beqn\label{eq:N:T_K}
\boxed{
T_K=\frac{T}{K}\mbox{ et }N_K=\frac{N}{K}
}
\eeqn

D'une façon générale, le début, le centre et la fin d'une trame 
$k$
sont définies par
\beqn\label{eq:N:t_k_de}
\boxed{
\mbox{Pour }k\in\{0\ldots K-1\},\quad\left \{
\begin{array}{l}
t_k^\de=k\frac{T}{K}\\[0.2cm]
 t_k^\der=t_k^\de+T_K-T_e\\[0.2cm]
t_k^\ce=\frac{1}{2}\left (t_k^\de+t_k^\der\right )
\end{array}
\right. 
}\eeqn

Avec ces définitions et pour une trame $k$, on peut définir une échelle de temps et déterminer les valeurs du signal
\beqn\label{eq:N:ech_temps_trame}
\boxed{
\begin{array}{c}
\mbox{Pour }k\in\{0\ldots\}\mbox{ et }n\in \{0\ldots N_K-1\}\\[0.5cm]
t_{k,n}=t_k^\de+nT_e \mbox{ et }x_{k,n}=x(t_{k,n})=x\left[\frac{t_{k,n}}{T_e}\right ]
\end{array}
}
\eeqn
Dans la dernière expression en bas à droite de l'équation~(\ref{eq:N:ech_temps_trame}), l'expression entre crochet est un indice 
\beqns
n'=\frac{t_{k,n}}{T_e}=k\frac{N}{K}+n
\eeqns
et cet indice est utilisé pour retrouver dans le signal $x_n$ la valeur recherchée 
$x_{k,n}=x_{n'}$.

\subsection{Découpage en trames avec chevauchement}

\begin{figure}
\begin{center}
\begin{tabular}{c}
\includegraphics[width=0.8\linewidth]{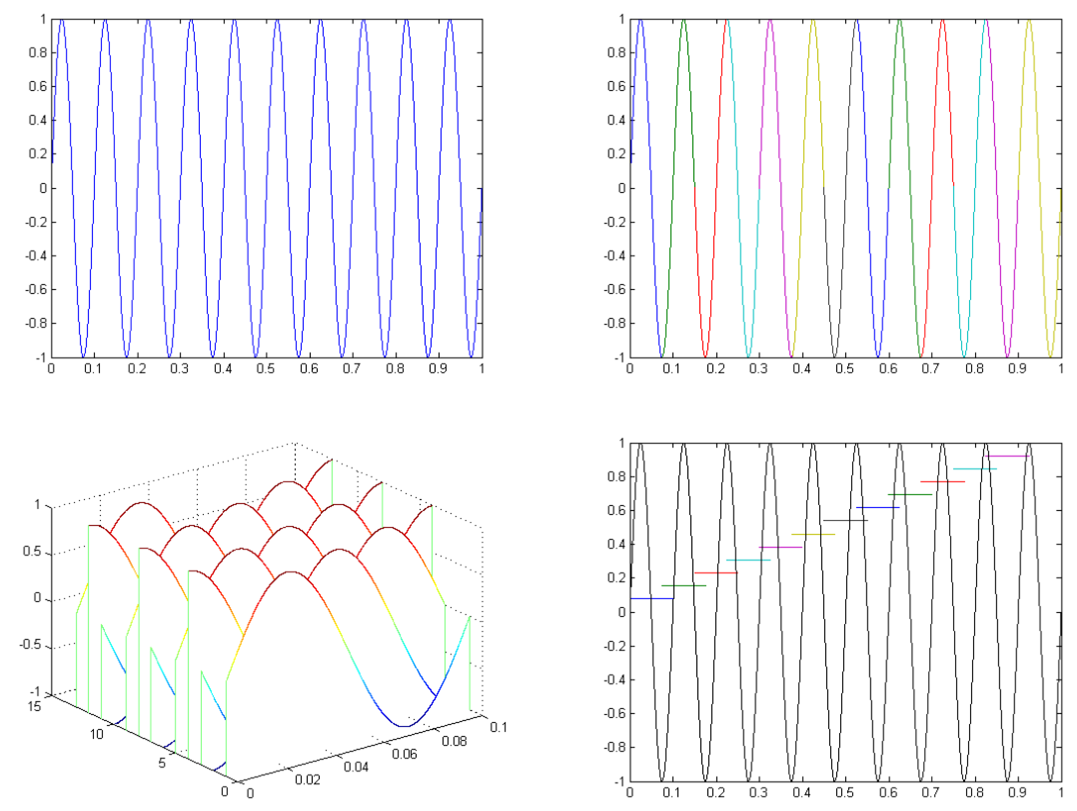}
\end{tabular}
\end{center}
\caption{Découpage d'une sinusoïde en trames qui se chevauchent.}
\label{fig:N:son_trames_chevauchement}
\end{figure}

\`Etant donné un nombre total d'échantillons $N$, un nombre de trames $K$ et 
du choix de chevauchement utilisé, on a les deux propriétés suivantes. 
\begin{itemize}
\item Les instants de débuts des trames ne sont pas modifiées.
\beqn
\boxed{
t_k^\de=k\frac{T}{K}
}
\eeqn
\item La longueur des trames est augmenté d'un facteur $\frac{1}{1-\alpha}$. 
\beqn\label{eq:N:T_K_alpha}
\boxed{
T_K=\frac{T}{K}\frac{1}{1-\alpha}\mbox{ et }N_K=\frac{N}{K}\frac{1}{1-\alpha}
}
\eeqn
\item Hormis la première trame, toutes les trames partagent avec la précédente
$\alpha N_K$ échantillons. Et hormis la dernière trame, toutes les trames partagent avec la suivante $\alpha N_K$ échantillons. 
\end{itemize}

La dernière affirmation se calcule ainsi pour $k\in\{0\ldots K-1\}$
\beqns
t_k^\der-t_{k+1}^\de+T_e=\left (k\frac{T}{K}+\frac{T}{K}\frac{1}{1-\alpha}-T_e\right ) -(k+1)\frac{T}{K}+T_e=\frac{T}{K}\left (\frac{1}{1-\alpha}-1\right )
=\alpha T_K
\eeqns

Ainsi les équations (\ref{eq:N:t_k_de}) et (\ref{eq:N:ech_temps_trame})
ne sont pas modifiées quand on met un chevauchement. En revanche l'équation~(\ref{eq:N:T_K})
est remplacée par l'équation~(\ref{eq:N:T_K_alpha}).

Lors de l'implémentation, ces formules ont été un peu modifiées pour tenir 
compte de ce que le découpage en trames ne tombe pas exactement sur un échantillon 
et qu'il faut donc commencer à partir de l'échantillon le plus proche. 

\section{Signaux sonores et silences}
\label{sec:detection_silence}
\subsection{Puissance moyenne à court terme}
\label{ssec:PMCT}

\begin{figure}
\begin{center}
\begin{tabular}{c}
\includegraphics[width=0.8\linewidth]{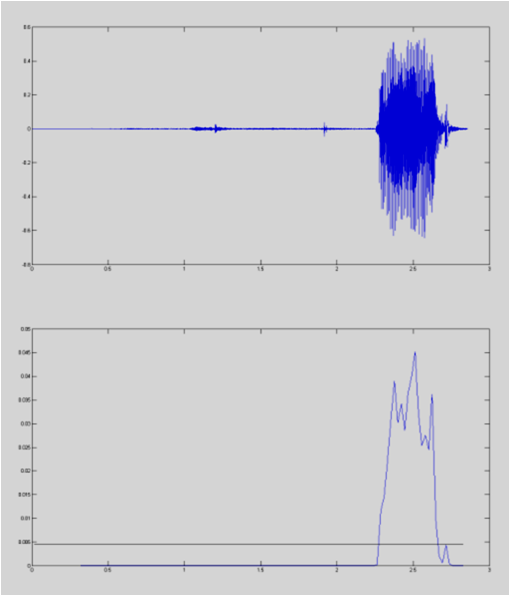}
\end{tabular}
\end{center}
\caption{Signal temps discret représentant un signal sonore commençant et terminant par un silence.}
\label{fig:N:son_puissance}
\end{figure}

En traitement du signal, on définit une puissance moyenne qui évolue dans le temps
qui est appelée puissance moyenne à court terme. 
Cela se fait en introduisant une fenêtre notée $w_T$ qui est un signal temps 
continu de durée $T$ centré en zéro. 
La notion de fenêtre a déjà été présentée lors du cours sur la 
synthèse des filtres numérique à réponse impulsionnelle finie avec notamment la fenêtre rectangulaire et triangulaire. 
\`A temps continu, la fenêtre rectangulaire de largeur $T$ est ainsi définie.
\beqns
w_T(\tau)=\1_{\left [-\frac{T}{2},\frac{T}{2}\right]}(\tau)
\eeqns
La fenêtre triangulaire de largeur $T$ est ainsi définie.
\beqns
w'_T(\tau)=\left (1-\frac{2t}{T}\right)\1_{\left [-\frac{T}{2},\frac{T}{2}\right]}(\tau)
\eeqns
Ici la fenêtre est utilisée, non pour modifier une réponse impulsionnelle mais un signal, ici $x^2(t)$. L'action de la fenêtre 
porte sur l'intervalle de temps centré en $t$. Aussi elle est d'abord centrée en $t$ puis normalisée.
\beqns
w_T(\tau)\mapsto w_T(t-\tau)\mapsto \frac{1}{\int_\Rr w_T(t-\tau)\,d\tau}\,w_T(t-\tau)
\eeqns
Ainsi modifiée la fenêtre est ensuite multipliée par $x^2(\tau)$
\beqns
x^2(\tau)\mapsto x^2(\tau)\frac{1}{\int_\Rr w_T(t-\tau)\,d\tau}\,w_T(t-\tau)
\eeqns

Cette puissance moyenne à court terme, notée $\sm{PMCT}$~\footnote{{\tt PMCT}
est défini dans \cite{Camastra07} p.~41-42.}~\footnote{Une interprétation en terme d'énergie
existe dans page~3, section~2.1 dans~\cite{Peeters04}.}
 est 
\beqn\label{eq:PMCT_TC}
P(t)=\frac{{\int_\Rr {x^2(\tau)w_T(t-\tau)\,d\tau }}}{{\int_\Rr w_T(\tau)\,d\tau}}
\eeqn
Avec une fenêtre rectangulaire, $P(t)$ devient 
\beqn\label{eq:PMCT_TC_rect}
\boxed{
P(t)=\frac{1}{T}{\int_{t-\frac{T}{2}}^{t-\frac{T}{2}} x^2(\tau)\,d\tau}
}
\eeqn

Ici le découpage en trame du signal se prête justement à un calcul très similaire.
\begin{itemize}
\item $T$ est remplacée par la longueur de la trame $T_K$ qui compte $N_K$ échantillons. 
\item $t$ est remplacée par $t_k^\ce$, le milieu de la trame $k$ qui se trouve à peu près à l'échantillon $\frac{N_K}{2}$ de 
la trame $k$. 
\end{itemize} 
 Cela permet ainsi d'approcher la  discrétisation de l'équation~(\ref{eq:PMCT_TC}).
\beqn
\PMCT_k=\frac{{\sum_{n=0}^{N_K-1} x_k^2[n]w_{N_K}[n-\frac{N_K}{2}]}}{{\sum_{n=0}^{N_K-1} w_{N_K}[n-\frac{N_K}{2}]}}
\eeqn
où $w_{N_K}[n]$ est une fenêtre de support $\{-\frac{N_K}{2}\ldots \frac{N_K}{2}\}$ {\bf mais} avec des indices allant de $0$ à $N_K-1$.  

Avec une fenêtre rectangulaire, 
\beqn
\boxed{
\PMCT_k=\frac{1}{N_K}
{\sum_{n=0}^{N_K-1} x_k^2[n]}
}
\eeqn

\subsection{Détection du début et de la fin d'un signal sonore.}
\label{ssec:det_deb_fin_son}

La figure~\ref{fig:N:son_puissance} montre un exemple de signal sonore commençant et terminant par un silence. 
On souhaite dans cet exemple retrouver l'instant à partir duquel le signal sonore commence et l'instant auquel ce
 signal sonore s'arrête.

On se donne un seuil, il est ici fixé à $10\%$ de la valeur maximale.
\beqn\label{eq:seuil}
\boxed{
s_e=\frac{1}{10}\max_k({\PMCT_k})
}
\eeqn
Ce seuil est utilisé pour déterminer deux numéros de trames~: $k^\de$ la première trame pour laquelle la puissance $\PMCT_k$ est supérieure 
$s_e$ et $k^\der$ la dernière  pour laquelle ceci est vrai. 
\beqn
\boxed{
k^{\de}=\underset{k}{\min}\,\{k|\,\PMCT_k>s_e\}
\mbox{ et }
k^{\der}=\underset{k}{\max}\,\{k|\,\PMCT_k>s_e\}
}
\eeqn
Les instants $t^\de$ et $t^\der$ associés au début et à la fin du signal sonore sont les centres des trames $k^\de$ et $k^\der$. 
\beqn
\boxed{
\left\{\begin{array}{l}
t^{\de}=t_{k}^{\ce}\mbox{ avec }k=k^{\de}\\[0.3cm]
t^{\der}=t_k^{\ce}\mbox{ avec }k=k^{\der}
\end{array}
\right . 
}
\eeqn
où $t_k^{\ce}$ désigne l'instant associé au centre de la trame $k$. 

\subsection{Normalisation d'un signal sonore}

La normalisation du signal a pour but de rendre comparable 
un son qui serait prononcé de façon forte et un son prononcé à voix plus basse. 
Cette normalisation consiste à imposer que la puissance moyenne soit fixe et en l'occurrence égale à $1$, mais 
cette étape se fait après avoir retiré les silences du signal.~\footnote{Si on effectuait cette normalisation du 
signal avant de retirer les silences, un son avec beaucoup de silences 
que l'on aurait normalisé  avec ses silences, aurait une 
puissance moyenne plus forte, une fois retirés les silences.}

Pour comparer les signaux, il est souhaitable que les signaux soient comparable en terme de puissance moyenne. Cette moyenne étant calculée sans considérer les silences. 
Pour obtenir que les signaux soient comparés, on effectue une {\em normalisation}, c'est-à-dire qu'on chacun des signaux par une valeur appropriée. 
Cette valeur est déterminée avec les étapes suivantes. 
\begin{itemize}
\item On utilise la technique de détection du début et de fin du signal sonore présentée en section~\ref{ssec:det_deb_fin_son}. 
\item On calcule la puissance moyenne en ne considérant que les trames qui se situent entre la première et la dernière des trames associées au signal sonore. 
On note $\bb{K}$ l'ensemble des trames où le son est plus fort que le seuil fixé par l'équation~(\ref{eq:seuil})
et $|\bb{K}|$ le nombre de ces trames. 
\beqns
\boxed{
P_\moy=\frac{1}{|\bb{K}|}{\sum\limits_{k\in \bb{K}} \sm{PMCT}_k}
}
\eeqns
\item La normalisation consiste à multiplier le signal par $\frac{1}{\sqrt{P_\moy}}$. 
\beqn
\boxed{
x_n\mapsto \frac{1}{\sqrt{P_\moy}} x_n
}
\eeqn
\end{itemize}

\section{Indications sur l'implémentation proposée}
\label{sec:implementation}

\begin{figure}
\begin{center}
\begin{tabular}{c}
\includegraphics[width=0.6\linewidth]{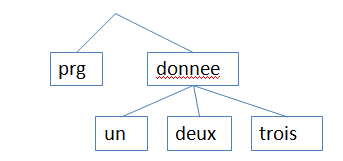}
\end{tabular}
\end{center}
\caption{Organisation des données dans l'implémentation}
\label{fig:N:projet_son1}
\end{figure}

Dans la simulation proposée, le répertoire principal est subdivisé en {\tt prg} contenant les programmes Matlab (fichiers {\tt .m} ou {\tt .p}) et {\tt donnee} qui contient 
les fichiers audio répartis en trois répertoires  {\tt un},  {\tt deux } et {\tt trois}. 
Cette organisation est schématisée sur la figure~\ref{fig:N:projet_son1}.

\chapter{Obtenir une classification grâce à des descripteurs}
\label{ch:classification_par_descripteurs}

L'objectif au cours de la séance de TP correspondante est de classer les sons à partir de descripteurs. Un descripteur peut 
soit être une valeur pour l'ensemble du signal, soit un ensemble de valeurs, une pour chaque trame.
On suppose dans cette partie qu'on dispose d'un certain nombre de descripteurs qui nous donnent des valeurs, une valeur par trame. 
La classification des sons est alors exposée en trois parties. 
La section~\ref{sec:distance_entre_sons} permet à partir de descripteurs d'évaluer numériquement 
une similarité entre deux sons. En fait on parle plutôt de distance entre sons, la distance correspondant ici à un défaut de similarité. 
une distance entre deux sons. 
La section~\ref{sec:knn} permet à partir de distances entre sons de prédire l'identité d'un son quelconque, 
c'est-à-dire s'il ressemble plus à un son de type {\tt un}, {\tt deux} ou {\tt trois}. 
Enfin la section~\ref{sec:evaluation_classification} montre comment on évalue généralement un classifieur.


\section{Calcul d'une distance entre deux sons}
\label{sec:distance_entre_sons}

\begin{figure}
\begin{center}
\begin{tabular}{c}
\begin{tikzpicture}
[node distance = 10mm,thick, main/.style = {draw, circle},
  rect/.style={draw,rectangle,fill=green!50},rect_y/.style={draw,rectangle},
  minimum size=0mm]
	\node (H1) {};
	\node (L1)[below of=H1] {};
	\node (L11) [below of=L1] {};
		\node (L12) [below of=L11] {};
		\node (L13) [below of=L12] {};
		\node (L14) [below of=L13] {};
	  \node (sons) [right of=L1] {sons}; 
		\node[rect_y] (xn) [below of=sons] {$x_a[n]$};   
		\node (xmn) [below of=xn] {};  
		\node[rect_y] (xpn) [below of=xmn] {$x_b[n]$};  
		\node (L2) [right of=sons] {};
	  \node (trames) [right of=L2] {trames}; 
		\node[rect_y] (xkn) [below of=trames] {$x_a[k,n]$};   
		\node (xmkn) [below of=xkn] {};  
		\node[rect_y] (xpkn) [below of=xmkn] {$x_b[k,n]$};  
		\node (L3) [right of=trames] {};
		\node (sans_silence) [right of=L3] {$\begin{array}{c}\mbox{sans le}\\ \mbox{silence}\end{array}$}; 
		\node[rect_y] (xkn1) [below of=sans_silence] {$x_a[k,n]$};   
		\node (xmkn1) [below of=xkn1] {};  
		\node[rect_y] (xpkn1) [below of=xmkn1] {$x_b[k,n]$};  
		\node (L4) [right of=sans_silence] {};
		\node (normalisation) [right of=L4] {$\begin{array}{c}\mbox{normali-}\\ \mbox{sation}\end{array}$}; 
		\node[rect_y] (xkn2) [below of=normalisation] {$x_a[k,n]$};   
		\node (xmkn2) [below of=xkn2] {};  
		\node[rect_y] (xpkn2) [below of=xmkn2] {$x_b[k,n]$};  
		\node (L5) [right of=normalisation] {};
		\node (H5) [above of=L5] {};
		\node (L51) [below of=L5] {};
		\node (L52) [below of=L51] {};
		\node (L53) [below of=L52] {};
		\node (L54) [below of=L53] {};
		\node (descripteur) [right of=L5] {$\begin{array}{c}\mbox{calcul des}\\ \mbox{descripteurs}\end{array}$}; 
		\node[rect_y] (zkj) [below of=descripteur] {$z_{a}^{(j)}[k]$};   
		\node (zmkj) [below of=zkj] {};  
		\node[rect_y] (zpkj) [below of=zmkj] {${z}_{b}^{(j)}[k]$};  		
		\node (L6) [right of=descripteur] {};
		\node (distance) [right of=L6] {$\begin{array}{c}\mbox{calcul de la}\\ \mbox{distance}\end{array}$}; 
		\node (L67l) [below of=distance] {}; 
		\node[rect_y] (d) [below of=L67l] {$d(x_a,x_b)$};   
		
		\begin{scope}[-latex]
\draw[thick,->,blue] (xn) to   (xkn);
\draw[thick,->,blue] (xkn) to   (xkn1);
\draw[thick,->,blue] (xkn1) to   (xkn2);
\draw[thick,->,green] (xkn2) to   (zkj);
\draw[thick,->,blue] (xpn) to   (xpkn);
\draw[thick,->,blue] (xpkn) to   (xpkn1);
\draw[thick,->,blue] (xpkn1) to   (xpkn2);
\draw[thick,->,green] (xpkn2) to   (zpkj);
\draw[thick,->,red] (zpkj) to   (d);
\draw[thick,->,red] (zkj) to   (d);
\draw[gray,dashed,-] (H1) to   (H5);
\draw[gray,dashed,-] (H5) to   (L54);
\draw[gray,dashed,-] (L14) to   (L54);
\draw[gray,dashed,-] (H1) to   (L14);
\end{scope}
\end{tikzpicture}
\end{tabular}
\end{center}
\caption{Schéma décrivant la mise en oeuvre du calcul de la distance entre deux sons noté $x$ et $x'$. 
La partie encadrée en pointillé est la préparation des signaux
transformant $x_a[n]$ et $x_b[n]$ en $x_a[k,n]$ et $x_b[k,n]$, $n$ étant dans $x_a[n]$ et $x_b[n]$ l'indice de l'échantillon pour  l'ensemble du son 
et pour $x_a[k,n]$ et $x_b[k,n]$, l'indice dans la trame $k$. La lettre $j$ désigne un type de descripteur et celui-ci est calculé pour les deux sons et pour chaque 
$d(x_a,x_b)$ désigne la valeur de la distance entre les deux sons.  }
\label{fig:dist_sons}
\end{figure}
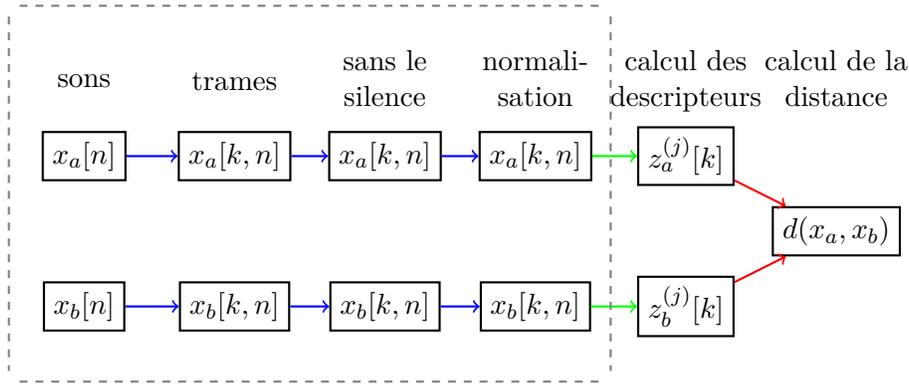

La figure~\ref{fig:dist_sons} schématise la façon de calculer une distance (i.e. une non-similarité) entre deux sons. 
La partie encadrée en pointillé figure la préparation des signaux avec les tâches suivantes. 

\begin{minipage}{0.95\linewidth}
\begin{itemize}
\item Chaque signal sonore est décrit avec $x_a[n]$ et $x_b[n]$ par un signal temps discret échantillonné à la même fréquence d'échantillonnage, mais 
pas forcément de même durée. 
\item Chaque signal est découpé en trames, de longueur et de chevauchement identique, le nombre de trames est a priori différent d'un signal à l'autre. 
\item Les trames silencieuses en début et fin de son sont retirées. 
\item Les valeurs des signaux au sein de chaque trames sont normalisées globalement au niveau du son, c'est-à-dire que si une trame est 
en moyenne deux fois plus forte qu'une autre trame du même son, elle reste deux fois plus forte après la normalisation. 
Par contre quand on compare deux sons, le fait de multiplier par deux toutes les valeurs d'un des deux sons, ne modifie pas la distance entre les 
deux sons. 
\end{itemize}
\end{minipage}

On rajoute les notations suivantes~: 

\begin{minipage}{0.95\linewidth}
\begin{itemize}
\item $x_a$ et~$x_b$ désignent les sons~$a$ et~$b$. 
\item $x_a[n]$ et~$x_b[n]$ désignent les~$N_a$ et~$N_b$ valeurs des signaux associés aux sons~$a$ et~$b$. 
\item $x_a[kn]$ et~$x_b[kn]$ désignent les~$N_k$ valeurs des sons $a$ et $b$ pour la trame~$k$, c'est l'équivalent pour le son~$a$ de ce qui était noté précédemment~$x_k[n]$. 
\item $J$ désigne le nombre total de descripteur. 
\item $j\in\{1\ldots J\}$ désigne un descripteur spécifique. $\PMCT$ est un exemple de descripteur, les autres seront décrits dans le 
chapitre~\ref{ch:desc}. 
\item ${z}_a^{(j)}[k]$ et~${z}_b^{(j)}[k]$  désigne la valeur du descripteur~$j$ pour la trame~$k$ et les sons~$a$ et~$b$. 
\end{itemize}
\end{minipage}

Les calculs figurés par les flèches vertes joignant $x_a[kn]$ et $x_b[kn]$ à  ${z}_a^{(j)}[k]$ et ${z}^{(j)}_b[k]$ correspondent aux 
calculs des valeurs des descripteurs, 
Les flèches rouges joignant  ${z}_{a}^{(j)}[k]$ et ${z}_{b}^{(j)}[k]$ à $d(x_a,x_b)$ figurent les calculs permettant d'évaluer la distance à partir 
des valeurs des descripteurs, trois techniques sont proposées dans cette section. 

\subsection{Les descripteurs par trames sont transformés en descripteurs par son}
\label{ssec:mu_sigma}
On rajoute ici un autre sens au mot descripteur. Précédemment, il s'agissait d'un descripteur par trame, ici il s'agit d'un descripteur 
pour l'ensemble d'un son. Ce descripteur est identifié par $j$ correspondant aux valeurs du descripteur par trame qui sont 
résumés dans une seule valeur. Ce descripteur est aussi identifié par la façon de faire ce résumé~: $\mu$ pour la moyenne, $\sigma$ 
pour l'écart-type, $\gamma_1$ et $\gamma_2$ pour les coefficients d'asymétrie et d'aplatissement.

Ces quatre descripteurs de sons sont calculés à partir des valeurs d'un descripteur obtenu pour chaque trame d'un son. 

\begin{minipage}{0.95\linewidth}
\begin{itemize}
\item La moyenne d'un descripteur $j$ est la moyenne des valeurs de $z_k^{(j)}$ pour chaque trame $k$. 
\beqn\label{eq:mu}
\boxed{
\mu^{(j)}=\frac{1}{K}{\sum_{k=0}^{K-1} z_k^{(j)}}
}
\eeqn
\item 
L'écart-type d'un descripteur $j$ est la racine carré de la moyenne des variances.
\beqn\label{eq:sigma}
\boxed{
\sigma^{(j)}=\sqrt{\frac{1}{K}{\sum_{k=1}^K \left (z_k^{(j)} -\mu^{(j)}\right)^2}}
}
\eeqn
\item Le coefficient d'asymétrie d'un descripteur $j$ est 
\beqn
\gamma_1^{(j)}=\frac{1}{(\sigma^{(j)})^3}{\frac{1}{K}{\sum_{k=1}^K \left (z_k^{(j)} -\mu^{(j)}\right)^3}}
\eeqn
\item Le coefficient d'aplatissement d'un descripteur $j$ est 
\beqn
\gamma_2^{(j)}=\frac{1}{(\sigma^{(j)})^4}{\frac{1}{K}{\sum_{k=1}^K \left (z_k^{(j)} -\mu^{(j)}\right)^4}}
\eeqn
\end{itemize}
 \end{minipage}

Ces quatre descripteur de sons permettent de définir différentes distances entre sons suivant ce qui est pris en compte. 
Par exemple en prenant en compte la moyenne et l'écart-type des descripteur $j\in\{1\ldots J\}$, on a la distance suivante~: 
\beqn
\boxed{
d(x_a,x_b)=\sqrt{{\sum_{j=1}^J \left( \mu_a^{(j)}-\mu_b^{(j)}\right )^2+\left( \sigma_a^{(j)}-\sigma_b^{(j)}\right )^2}}
}
\eeqn

\subsection{La distance entre deux sons est obtenue en comparant trame par trame, les valeurs des descripteurs.}
\label{ssec:dist_trames}

Les sons $a$ et $b$ n'ont pas a priori la même durée et donc pas le même nombre de trames. On note $\min(K_a,K_b)$ le nombre de trames du son 
le plus court. On définit la distance comme la moyenne des distances entre les descripteurs trame par trame. Cette distance est 
en fait la racine carré de la somme des carrés des différences. Chaque son est décrit par $J{\times}K_a$ et $J{\times}K_b$ valeurs, 
on arrive à la définition suivante. 
\beqn
\boxed{
d(x_a,x_b)=\sqrt{\frac{1}{\min(K_a,K_b)}
{\sum_{k=0}^{\min(K_a,K_b)-1}{\sum_{j=1}^J\left (z_a^{(j)}[k]-z_b^{(j)}[k]\right )^2}}}
}
\eeqn

Un important défaut de cette technique est qu'elle amène à considérer un son comme différent, lorsqu'il est prononcé avec une vitesse différente ou simplement une vitesse qui varie différemment. 

\subsection{La distance entre deux sons est obtenue en cherchant la déformation temporelle minimisant la différence entre les deux sons.}
\label{ssec:dist_def_temp}

\subsubsection{Notion de déformation temporelle dynamique }

\begin{figure}
\begin{center}
\begin{tabular}{cc}
\includegraphics[width=0.5\linewidth]{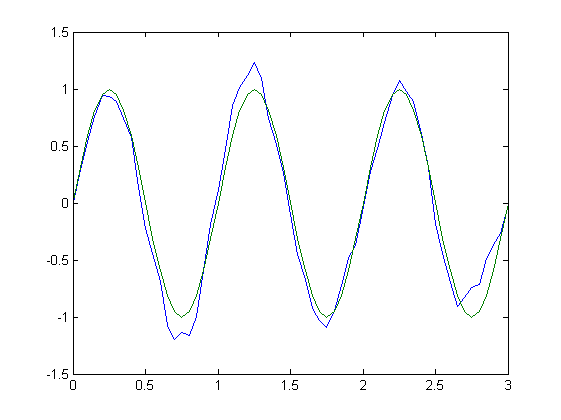}
\includegraphics[width=0.5\linewidth]{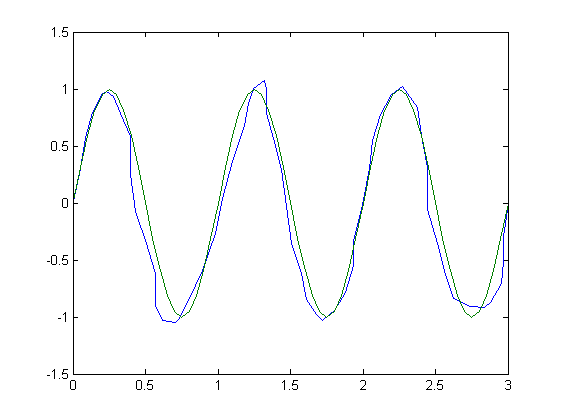}\\
\includegraphics[width=0.5\linewidth]{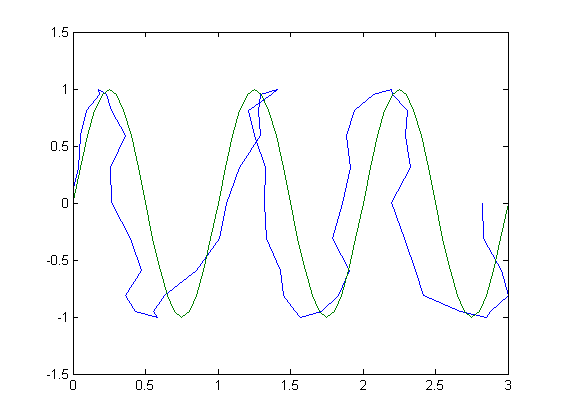}
\includegraphics[width=0.5\linewidth]{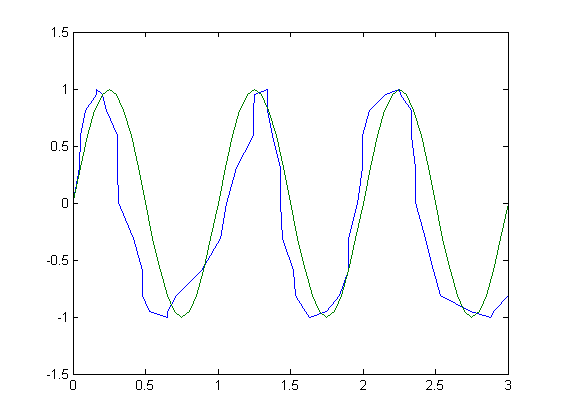}
\end{tabular}
\end{center}
\caption{Quatre déformations possibles d'un même signal sinusoïdal représenté sur les quatre figures en vert. 
Le temps est représenté sur l'axe horizontal et les valeurs des signaux sont représentées sur l'axe vertical. En haut à gauche, le signal bleu 
est perturbé par un bruit additif, en haut à droite, il est perturbé par une bruit additif de moindre ampleur ajouté à la fois sur les valeurs du signal 
et celles du temps~; en bas à gauche, il est perturbé par un bruit additif ajouté seulement sur les valeurs de temps~; en bas à droite, le bruit 
additif ajouté seulement sur les valeurs de temps est corrigé par le fait qu'un évènement postérieur à un autre évènement doit rester postérieur après 
la perturbation.}
\label{fig:def_temp_1}
\end{figure}

La figure~\ref{fig:def_temp_1} représente un même signal sinusoïdal colorié en vert $x_v(t)$ et 
différentes déformations de ce signal sinusoïdal, $x_b(t)$, colorées en 
bleu. 

Il est d'usage lorsqu'on évalue la distance entre deux courbes définies sur un même intervalle de temps $[0,T]$, de calculer la 
moyenne quadratique de la différence entre $x_v(t)$ et $x_b(t)$. 
\beqn
\boxed{
d_1(x_v,x_b)=\sqrt{\frac{1}{T}{\int_0^T \left(x_v(t)-x_b(t)\right )^2\,dt}}
}
\eeqn
Cette évaluation est souvent choisie pour sa simplicité, elle a une interprétation pertinente lorsque le signal déformé $x_b(t)$ resulte 
d'un bruit blanc gaussien centré additif, car dans ce cas $d_1(x_v,x_b)$ est une estimation de l'écart-type de cette gaussienne. Le graphe en haut à gauche 
de la figure~\ref{fig:def_temp_1} montre en bleu un signal déformé par un bruit blanc additif, on y observe que parfois la courbe bleue est au dessus 
de la courbe verte et parfois elle est en dessous. 

Il existe dans la littérature scientifique une technique permettant d'évaluer la distance avec les différences en hauteur entre les signaux mais
avec les distances euclidienne entre les courbes, c'est-à-dire pour chaque point de la courbe $x_b(t)$, on utilise la distance entre ce point et 
le point le plus proche de $x_v(t)$. Cette évaluation est pertinente lorsqu'on considère que la déformation de $x_b(t)$ résulte d'une part 
d'un variation de  la valeur du signal $x_b(t^b_n)-x_v(t^v_n)$ mais aussi une variation de la base de temps $t^b_n-t^v_n$. Le graphe en haut à droite est obtenu en ajoutant deux bruits blancs 
gaussiens centrés additifs indépendants sur les valeurs du signal et sur les instants associés à ces valeurs. 

Le graphe en bas à gauche est obtenu en ne modifiant que les bases de temps $t^b_n$ par rapport à $t^v_n$, il s'agit d'un bruit blanc gaussien centré 
additif. C'est-à-dire qu'on a l'égalité entre les valeurs des signaux $x_b$ et $x_v$ mais les valeurs ont lieu à des instants différents. 
\beqns
x_b(t^b_n)=x_v(t^v_n)
\eeqns
Sur ce graphe, la courbe bleue présente des boucles suggérant que le temps s'inverse parfois. Ceci n'est pas une modélisation réaliste. 

Le graphe en bas à droite est obtenu aussi en ne modifiant que les bases de temps $t^b_n$ par rapport à $t^v_n$, mais pour plus de réalisme, l'ordre des 
évènements est respecté, il n'y a donc plus de boucles.
\beqns
t^v_n\leq t^v_{n'}\quad\Rightarrow\quad t^b_n\leq t^b_{n'}
\eeqns
Concrètement, les variations de $t^b_n$ sont limitées par la contrainte que $t^b_{n}\leq t^b_{n+1}$ doit toujours être vrai. 
C'est ce type de  déformation du signal que l'on cherche à prendre en compte. Quand on prononce un son plus lentement pour plus rapidement ou 
avec une vitesse qui varie au cours de la prononciation du son, il s'agit du même son. La distance devrait théoriquement ne pas 
être sensible à ce type de déformation. 
On appelle cette déformation, une déformation temporelle dynamique.~\footnote{Cette notion est présentée en détail dans~\cite{Cassisi12}.}

\subsubsection{Table d'indexation}

\begin{figure}
\begin{center}
\begin{tabular}{c}
\includegraphics[width=0.5\linewidth]{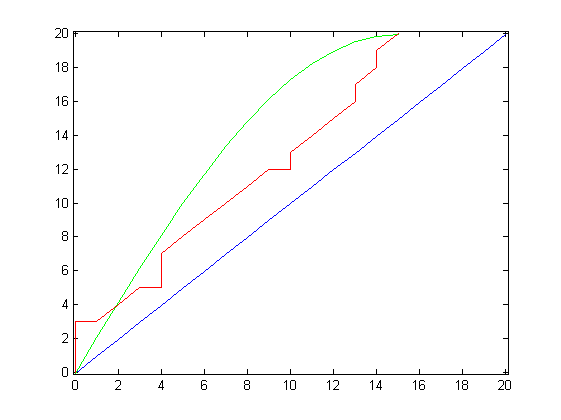}
\end{tabular}
\end{center}
\caption{En vert et bleu, deux signaux sont définis sur des intervalles de temps différents $(t_n,x_n)$, $(t'_n,y_n)$.
En rouge, le signal $(t_n,x_n)$ est transformé par une table d'indexation en 
$(t'_{\Phi_{y}[n]},x_{\Phi_x[n]})$ de façon à approcher le signal $(t'_n,y_n)$.}
\label{fig7}
\end{figure}

La figure~\ref{fig7} illustre ce que l'on cherche à faire avec deux tables d'indexation. 
Le trait bleu associé à $x_n$ est défini sur l'intervalle de temps $[0,20]$, tandis que la courbe verte associée à $y_n$ 
est définie sur l'intervalle de 
temps $[0,15]$. Le signal $x_{\Phi_x[n]}$ est défini pour les entiers $\{0\ldots 21\}$ et est à valeurs parmi les 
valeurs de $\{x_0\ldots x_{20}\}$, ce sont des points qui sont sur la courbe rouge. La courbe rouge est en fait 
obtenue avec reliant les points $(t_{\Phi_y[n]},x_{\Phi_x[n]})$.
\beqns
\begin{array}{llcl}
\Phi_x[0]=0 & \Phi_y[0]=0 & (t_{\Phi_y[0]},x_{\Phi_x[0]})=(t_0,x_0)=(0,0) & \mbox{position initiale en bas à gauche}\\
& & \vdots& \\
\Phi_x[3]=3 & \Phi_y[1]=0 & (t_{\Phi_y[3]},x_{\Phi_x[3]})=(t_0,x_3)=(0,3) & \mbox{déplacement vertical {\bf V}}\\
\Phi_x[4]=3 & \Phi_y[4]=1 & (t_{\Phi_y[4]},x_{\Phi_x[4]})=(t_1,x_3)=(1,3) & \mbox{déplacement horizontal {\bf H}}\\
\Phi_x[5]=4 & \Phi_y[5]=2 & (t_{\Phi_y[5]},x_{\Phi_x[5]})=(t_2,x_4)=(2,4) & \mbox{déplacement diagonal {\bf D}}\\
&&\vdots &
\end{array}
\eeqns 
Cet exemple illustre le fait que deux tables d'indexation permettent de modéliser un déplacement horizontal et vertical. 

En pratique, on n'utilise pas la base de temps, on modifie juste les indices avec les deux tables 
d'indexation de façon à minimiser le carré des différences entre $x_{\Phi_x[n]}$ et $y_{\Phi_y[n]}$. 
\beqn
y_{\Phi_y[n]}\approx x_{\Phi_x[n]}
\eeqn

On introduit de nouvelles notations. 
\begin{itemize}
\item $N_\Phi$: Les tables d'indexation sont définies sur une plage d'indices communs 
\beqns
n\in\{0\ldots N_\Phi-1\}
\eeqns
\item $N_x$ et $N_y$ sont le nombre d'indices pour $x_n$ et $y_n$. 
Chaque table d'indexation est à valeurs dans les indices sur lesquels $x_n$ et $y_n$ sont définis
\beqns
\Phi_x[n]\in \{0\ldots N_x-1\} \mbox{ et }\Phi_y[n]\in \{0\ldots N_y-1\}
\eeqns
\end{itemize}

En pratique, on impose que le premier indice de $\Phi_x$ corresponde au point de départ de $x_n$, le dernier de $\Phi_x$
avec le point d'arrivée de $x_n$ et de même pour $\Phi_y$ avec $y_n$. 
\beqn\label{eq:phi_contr1}
\left \{
\begin{array}{ll}
\Phi_x[0]=0,\,&\Phi_x[N_{\Phi}-1]=N_x-1\\[0.2cm]
\Phi_{y}[0]=0,\,&\Phi_{y}[N_{\Phi}-1]=N_y-1\\[0.2cm]
\end{array}
\right . 
\eeqn

On introduit la contrainte modélisant le non-retournement du temps 
\beqn\label{eq:phi_contr2}
\Phi_x[n]\leq \Phi_x[n+1] \mbox{ et }\Phi_{y}[n]\leq \Phi_{y}[n+1]
\eeqn

On rajoute une autre contrainte qui indique que la différence de temporalité entre les deux signaux ne doit pas 
être trop grande. 
Cette contrainte a aussi l'avantage de diminuer la complexité numérique de l'algorithme utilisé pour calculer cette nouvelle distance. 
\beqn\label{eq:phi_contr3}
\left |\Phi_x[n]-\Phi_{y}[n]\right |\leq \delta
\eeqn
$\delta$ étant un paramètre à définir pour calculer cette distance. 

\subsubsection{Distance définie au moyen d'un problème d'optimisation}

Pour chaque choix de $\Phi_x$ et $\Phi_y$, on a une valeur de distance qui est obtenue classiquement avec la moyenne quadratique. 
\beqn
\label{eq:tw_d1}
\boxed{
d_{\Phi_x,\Phi_{y}}(x,y)=\sqrt{\frac{1}{N_x+N_y}{\sum_{n=0}^{N_{\Phi}-1} \left(x_{\Phi_x[n]}-y_{\Phi_{y}[n]} \right )^2}}
}
\eeqn
La distance prenant  en compte de la déformation temporelle dynamique est alors défini en choisissant 
$\Phi_x$ et $\Phi_y$ de façon à respecter les contraintes définies par les équations~(\ref{eq:phi_contr1}),~(\ref{eq:phi_contr2}) et~(\ref{eq:phi_contr3}). 
et à minimiser $d_{\Phi_x,\Phi_{y}}(x,y)$.
\beqn\label{eq:prob_min}
\boxed{
d_\delta(x,y)=\underset{\Phi_x,\Phi_y}{\min}\,d_{\Phi_x,\Phi_{y}}(x,y)
}
\eeqn

\subsubsection{Algorithme résolvant ce problème d'optimisation}

\begin{figure}
\begin{center}
\begin{tabular}{c}
\includegraphics[width=0.5\linewidth]{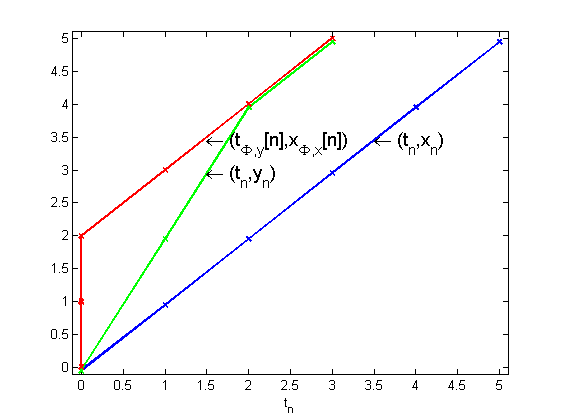}
\includegraphics[width=0.5\linewidth]{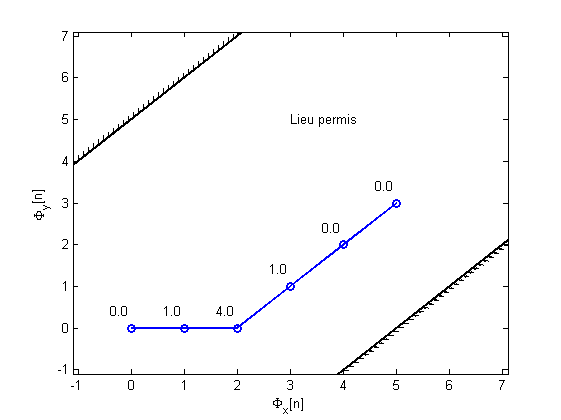}
\end{tabular}
\end{center}
\caption{\`A gauche~: représentation temporelle des signaux $x_n$, $y_n$ et $x_n$ transformés par déformation temporelle. 
\`A droite~: représentation des tables d'indexation permettant de générer la déformation temporelle du signal $x_n$. 
}
\label{fig8}
\end{figure}

La gauche de la figure~\ref{fig8} est similaire à la figure~\ref{fig7} avec en bleu et vert deux signaux $(t_n,x_n)$ et $(t_n,y_n)$ et 
en rouge, la déformation du premier signal pour se rapprocher du second~: $(t_{\Phi_y[n]},x_{\Phi_x[n]})$. 
La droite de cette figure montre les valeurs prises par~$\Phi_x[n]$ et~$\Phi_y[n]$ en représentant ces deux tables 
comme une succession de points formant un chemin reliant au départ~$(0,0)$ au point d'arrivée~$(N_x,N_y)=(5,3)$.
Ces contraintes correspondent à  l'équation~(\ref{eq:phi_contr1}). 
On satisfait la contrainte de l'équation~(\ref{eq:phi_contr2}) en imposant que le chemin soit composé seulement de 
déplacement verticaux, horizontaux ou diagonaux et en l'occurrence seulement d'un pas à la fois. 
La contrainte indiquée par l'équation~(\ref{eq:phi_contr3}) est indiquée par les traits noirs hachurés à l'extérieur de la zone autorisée sur la 
droite de la figure~\ref{fig8}.

Les valeurs utilisées pour les signaux représentés sur la figure~\ref{fig8} sont 
\beqns
\begin{array}{llllll}
x_0=0 & x_1=1 & x_2=2 & x_3=3 & x_4=4 & x_5=5\\[0.2cm]
y_0=0 & y_1=2 & y_2=4 & y_3=5\\[0.2cm]
\end{array}
\eeqns
Les tables d'indexations créées sont~: 
\beqns
\begin{array}{llllll}
\Phi_x[0]=0 & \Phi_x[1]=1 & \Phi_x[2]=2 & \Phi_x[3]=3 & \Phi_x[4]=4 & \Phi_x[5]=5 \\[0.2cm]
\Phi_y[0]=0 & \Phi_y[1]=0 & \Phi_y[2]=0 & \Phi_y[3]=1 & \Phi_y[4]=2 & \Phi_y[5]=3 \\[0.2cm]
\end{array}
\eeqns

On appelle arête ({\em edge}) les traits associés au déplacement sur le graphe de droite. On appelle noeud ({\em node}, {\em vertice}) les 
points indiqués par des ronds bleus sur le graphe de droite. 
La localisation du point $n$ est déterminé par $\Phi_x[n]$ et $\Phi_y[n]$, il se trouve que ces valeurs déterminent aussi 
\beqns
(x_{\Phi_x[n]}-y_{\Phi_y[n]})^2
\eeqns
Les nombres en bleus situés à côté de chaque noeud donnent la valeur de cette quantité. 
Ce sont les valeurs de $x_n$ et $y_n$ qui donnent à chaque point du graphe de droite une valeur pour $(x_{\Phi_x[n]}-y_{\Phi_y[n]})^2$
dépendant uniquement de la position du point. 
La distance entre les deux signaux telle que définie dans l'équation est à la somme des quantités inscrites à côté de chaque noeud. 
L'équation~(\ref{eq:prob_min}) indiquant que la distance est obtenue à partir du chemin minimisant $d_{\Phi_x,\Phi_y}(x,y)$
 définie par l'équation~(\ref{eq:tw_d1}). Il se trouve que le graphe à droite de la figure~\ref{fig8} donne une interprétation de cette minimisation. 
$d_{\Phi_x,\Phi_y}(x,y)$ est la somme des valeurs associées à chaque point et qui ne dépend que de la position des points. 
La minimisation consiste donc à trouver le chemin minisant la somme des valeurs de chaque noeud en utilisant que des déplacements verticaux, horizontaux ou diagonaux et menant d'un point de départ à un point d'arrivée tout en restant dans le domaine délimité par les zones hachurées. 

L'algorithme permettant cette minimisation s'appelle l'algorithme de Dijkstra. 
Il consiste dans un premier temps à découvrir tous les noeuds possibles.
\begin{itemize}
\item Le premier noeud regardé est le point de départ et on lui affecte la valeur associée $(x_0-y_0)^2$. 
\item On répète la tâche suivante
\begin{enumerate}
\item\label{it:vois} Aller au voisin et lui donner la valeur de la distance entre le point de départ et ce voisin. 
\item On indique aussi à ce voisin le noeud précédent où l'on était. 
\item Si sur ce voisin il y a déjà une valeur, alors cette valeur doit être diminuée si la distance trouvée est moindre et 
on lui change l'origine que si l'on change sa valeur. 
\item Revenir au noeud précédent. 
\item Aller à l'étape~\ref{it:vois} s'il y a un voisin disponible et non-exploré dans cette tâche. 
\item Revenir au premier voisin et s'autoriser à aller à une profondeur supplémentaire. 
\end{enumerate}
\end{itemize}
Lorsque ces tâches ne peuvent plus être répétés parce qu'on a été au maximum de la profondeur possible, alors 
on part du point d'arrivée et on remonte au point de départ en suivant les indications laissées à chaque noeud.

\begin{figure}
\begin{center}
\begin{tabular}{c}
\includegraphics[width=\linewidth]{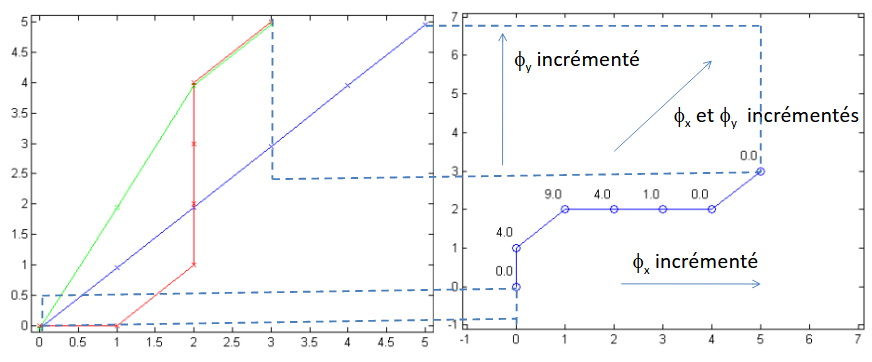}
\end{tabular}
\end{center}
\caption{\`A gauche~: représentation temporelle des signaux $x_n$, $y_n$ et $x_n$ transformés par déformation temporelle. 
\`A droite~: représentation des tables d'indexation permettant de générer la déformation temporelle du signal $x_n$. 
}
\label{fig8bis}
\end{figure}

La figure~\ref{fig8bis} est assez similaire à la figure~\ref{fig8} mais avec d'autres valeurs pour les signaux $x_n$, $y_n$, $\Phi_x[n]$ et $\Phi_y[n]$.

\section{Prédiction par le plus proche voisin}
\label{sec:knn}
\subsection{Principe}

\begin{figure}
\begin{center}
\begin{tabular}{c}
\includegraphics[width=\linewidth]{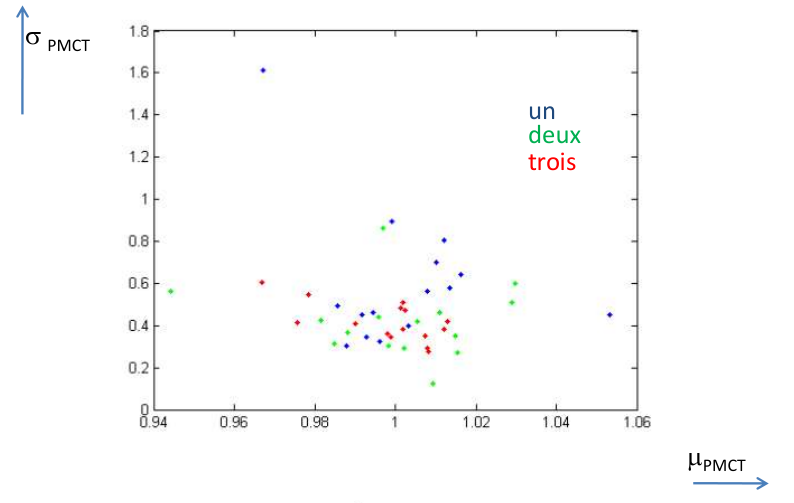}
\end{tabular}
\end{center}
\caption{Visualisation de sons à travers les valeurs des indicateurs $\mu_\sm{PMCT}$ et $\sigma_\sm{PMCT}$}
\label{fig_knn}
\end{figure}

On appelle {\bf classifieur} le programme ou le procédé par lequel il est estimé que le son appartient à telle classe. 

On appelle {\bf prédiction} le fait pour un classifieur d'estimer qu'un son appartient à une classe en utilisant uniquement les données 
numériques disponible pour ce son. 

La prédiction par le plus proche voisin est un cas particulier d'un classifieur appelé $k$-NN pour {\em Nearest Neighbor} ce qui signifie 
justement les $k$ plus proche voisins. Ce classifieur est performant mais très coûteux en terme de complexité numérique dès lors que la 
base de donnée est assez importante. Il est aussi particulièrement simple à implémenter, ce qui est la raison principal de son choix ici. 
Il est utilisé ici dans le cas où $k=1$. 

Ce classifieur consiste pour chaque son inconnu appelé requête, de parcourir tous les sons de la base de données et pour chaque son de 
calculer sa distance avec la requête. Le classifieur détermine la classe comme étant celle du son ayant la plus petite distance avec la requête. 
Cette classe est appelée la prédiction faite par le classifieur pour cette requête. 
Il est intéressant de remarquer que ce choix de faire une classification en fonction d'un seul élément (ou de $k$ éléments dans la version plus 
générale de cet algorithme) permet de s'adapter à une situation ou une classe regrouperait des éléments très hétérogènes.

La figure~\ref{fig_knn} montre comment une vingtaine de sons peuvent être visualisés sur un graphe. Chaque point correspond à un son, sa couleur 
(bleu pour {\tt un}, vert pour {\tt deux}, rouge pour {\tt trois}) indique le mot que ce son vocalise. Cette information est ce que l'on cherche à prédire avec 
la prédiction par le plus proche voisin. La localisation de chaque point dépend des valeurs de deux indicateurs, $\mu_{\sm{PMCT}}$ et $\sigma_{\sm{PMCT}}$. 
Ces indicateurs ont été présentés dans les sections~\ref{ssec:PMCT} et~\ref{ssec:mu_sigma}, c'est un programme informatique qui à partir des
données enregistrées permet de calculer ces deux valeurs et ainsi de positionner ces points sur ce graphique.
Pour illustrer le fonctionnement de cette technique de prédiction, voici son fonctionnement. 
\begin{enumerate}
\item  Sélectionnez dix points que vous numérotez de $1$ à $10$. 
\item Considérer un autre 
point qui ne figure pas parmi ces points. Ce point est appelé requête. 
\item Mesurez avec une règle la distance entre ce point requête et chacun de ces 10 points numérotés. 
\item Regardez la couleur du point qui a la distance avec le point requête la plus faible (et qui est donc sur le graphique 
le plus proche du point requête parmi les 10 points numérotés).
\item La prédiction consiste à attribuer au point requête la couleur du point le plus proche et par suite la classe, {\tt un}, {\tt deux} ou 
{\tt trois} suivant la 
couleur observée. 
\end{enumerate}

\subsection{Formalisation de la technique du plus proche voisin}

 Pour formaliser cette technique du plus proche voisin, on adopte les notations suivantes. 
\begin{itemize}
\item $x$ désigne en général un son défini pour plus de généralité par ses valeurs  $x_n$. 
\item $\chi$ désigne un ensemble de sons. 
\item $y$ désigne la classe à laquelle le son $x$ appartient, ({\tt un}, {\tt deux} ou {\tt trois}), est indiqué par son appartenance respective à $\chi_1, \chi_2,\chi_3$. 
\item $\br{d}(x_a,x_b)$ est une valeur positive correspondant à 
la distance entre les sons $x_a$ et $x_b$. En fait pour plus de généralité, ce calcul repose sur la réalisation des 
étapes suvantes. 
\begin{itemize}
\item Découpage de $x_a$ et $x_b$ en trames décrit à la section~\ref{sec:N:son_trames}.
\item Calcul d'un indicateur par trame pour $x_a$ et $x_b$, illustré par $\sm{PMCT}$ décrite à la section~\ref{ssec:PMCT}.
\item Mise en  oeuvre d'une des trois distance présentées dans les sections~\ref{ssec:mu_sigma},~\ref{ssec:dist_trames},~\ref{ssec:dist_def_temp}. 
\end{itemize}
\item $\br{d}(x,\chi)$ désigne la distance entre le son $x$ et un ensemble de sons regroupés dans $\chi$. 
Du fait du choix de la technique du plus proche voisin, cette distance est ici définie par 
\beqn
\boxed{
\br{d}(x,\chi)=\min\limits_{x'\in \chi} \br{d}(x,x')
}
\eeqn
où $\min\limits_{x'\in \chi} \br{d}(x,x')$ désigne la valeur la plus faible parmi les distances mesurées entre un son donné $x$ et 
tous les sons $x'$ contenus dans $\chi$. 
\item $\ww{x}$ désigne la classe prédite. Cette classe est prédite de cette façon. 
\beqn
\boxed{
\ww{x}=\argmin_{c\in\{1,2,3\}}\br{d}(x,\chi_c)
}
\eeqn
où $\chi_1,\chi_2,\chi_3$ regroupent des sons dont on sait qu'ils sont respectivement associés à {\tt un}, {\tt deux} et {\tt trois}. 
et $\argmin\limits_{c\in\{1,2,3\}}\br{d}(x,\chi_c)$ est ainsi défini. 
\beqns
\argmin\limits_{c\in\{1,2,3\}}\br{d}(x,\chi_c)=\left \{
\begin{array}{lclcl}
1 & \mbox{si} & \br{d}(x,\chi_1)\leq \br{d}(x,\chi_2) &\mbox{et}& \br{d}(x,\chi_1)\leq \br{d}(x,\chi_3)\\
2 & \mbox{si} & \br{d}(x,\chi_2)\leq \br{d}(x,\chi_1) &\mbox{et}& \br{d}(x,\chi_2)\leq \br{d}(x,\chi_3)\\
3 & \mbox{si} & \br{d}(x,\chi_3)\leq \br{d}(x,\chi_1) &\mbox{et}& \br{d}(x,\chi_3)\leq \br{d}(x,\chi_2)\\
\end{array}
\right . 
\eeqns

\end{itemize}

\subsection{Illustration avec l'implémentation proposée en TP}

\begin{figure}
\begin{center}
\begin{tabular}{c}
\includegraphics[width=\linewidth]{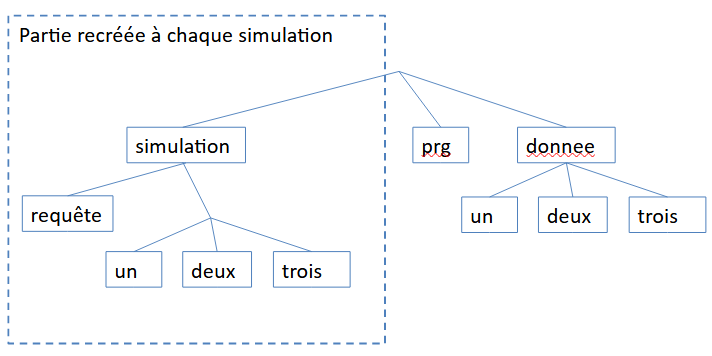}
\end{tabular}
\end{center}
\caption{Schéma correspondant aux différents dossiers existant et créés pour réaliser la prédiction en utilisant la technique du plus proche voisin. 
}
\label{fig9bis}
\end{figure}

La figure~\ref{fig9bis} illustre la façon  dont l'algorithme est ici implémenté. Chaque bloc représente un répertoire 
contenant des fichiers. 
Les sons que l'on utilise pour tester et évaluer la performance d'un classifieur sont en réalité issus de la base de données. 
Il faudrait donc s'assurer qu'ils ne sont pas utilisés aussi pour l'apprentissage. 
La partie qui n'est pas encadré en pointillé n'est pas modifiée lors de la mise en oeuvre de la classification. Cette
mise en oeuvre consiste à tirer aléatoirement une partie des sons des répertoires {\tt un}, {\tt deux} et {\tt trois}, 
de les placer dans le
répertoire {\tt requête}, les autres sons sont placés dans trois nouveaux répertoires contenus dans le répertoire 
{\tt simulation} et appelé {\tt un}, {\tt deux} et {\tt trois}. 
L'information selon laquelle les sons placés dans {\tt requête} n'est pas disponible sur les fichiers sons, mais elle 
est stockée pour être utilisée ultérieurement lors de l'évaluation de l'algorithme. 

\section{\'Evaluation de la performance d'un classifieur}
\label{sec:evaluation_classification}

En pratique quand on utilise un procédé de reconnaissance vocal, on a besoin de savoir dans quelle mesure on peut lui faire confiance. 
Cette mesure est ce qu'on appelle l'{\bf évaluation de la performance}. Pour se faire, on utilise des sons dont on connaît la classe à laquelle 
ils appartiennent et on compare cette classe avec ce que le classifieur prédit. Il existe un très grand nombre de mesures de performance, on en présente 
trois types~: la sensibilité globale, la précision et le rappel. 

\subsection{Matrice de confusion}

Le rôle de la matrice de confusion est de résumer sous la forme d'une matrice les données utiles au calcul de l'évaluation de la performance. 
Ces données utiles consistent en le fait de connaître le nombre d'expérience où le son requête appartenait à telle classe et où la classe prédite est 
celle-là. Dans le cadre de cette étude, il s'agit d'une matrice $3{\times}3$ puisqu'il y a trois sons possibles. 

\begin{itemize}
\item Chaque {\bf ligne} est associée aux expérimentations sur les sons qui en réalité appartiennent à telle classe. 
\item Chaque {\bf colonne} est associée aux expérimentations sur les sons qui ont été prédit comme étant de telle classe. 
\item La somme de toutes les cases est égale au nombre total d'expérimentations. 
\item \`A la case $i,j$, il y a le nombre d'expérimentations pour lesquelles le son appartient en réalité à la classe~$i$ et 
le classifieur l'a prédit comme étant de la classe~$j$. 
\item $m$ est un indice faisant référence à un son $x$. 
\item $M$ est le nombre total de sons utilisés pour les différentes requêtes. 
\item $\1(\ldots)$ est la fonction caractéristique, elle prend la valeur $1$ si les trois points de suspension sont remplacés par 
une affirmation exacte et par $0$ si l'affirmation est fausse. 
\item $\1(y_m=c)$ vaut $1$ si  le son numéro $m$ est de la classe $c$ et $\1(y_m=c)$ vaut $0$ si ce n'est pas le cas. 
\item $\1(\ww{x}_m=c)$ vaut $1$ si le son numéro $m$ est prédit par le classifieur comme étant de classe $c$ et $\1(\ww{x}_m=c)$ vaut $0$ 
si ce n'est pas le cas. 
\item $\1(y_m=i)\1(\ww{x}_m=j)$ vaut $1$ is le son numéro $m$ est d'une part de la classe $i$ et d'autre part a été prédit par le classifieur 
comme étant de la classe $j$. 
\end{itemize}

Les composantes de la matrices de confusion sont~: 
\beqn\label{eq:matrice_confusion}
\boxed{
C_{ij}={\sum_{m=1}^M\1(y_m=i)\1(\ww{x}_m=j)}
}
\eeqn

\subsection{Sensibilité globale}

La sensibilité globale, notée $\OA$ pour {\em Overall Assessment}, est 
le rapport entre le nombre de classifications correctes sur le nombre totales de requêtes, noté $M$. 
Les classifications correctes sont celles pour lesquelles $y_m=\ww{x}_m$. 

\beqn
\label{eq:OA}
\boxed{
\OA=\frac{C_{11}+C_{22}+C_{33}}{3}
}
\eeqn

\subsection{Précision et rappel}

La précision et le rappel sont des notions très utilisées pour les classifieurs binaires, c'est-à-dire pour lesquels, il n'y a que deux classes. 
Mais ces notions sont parfois étendues au cas plus général où on considère un nombre plus important de classes, c'est ce que nous utilisons ici. 

Pour chaque classe~$c$, on calcule deux indicateurs. La {\bf précision} $\p{P}_c$ est le rapport entre le nombre de requête correctement attribuées à la classe~$c$
et le nombre de requêtes que le classifieur évalué a prédit comme étant de la classe~$c$. 
Le {\bf rappel} $\p{R}_c$  est le rapport entre le nombre de requête correctement attribué à la classe~$c$ 
et le nombre de requêtes qui dans la réalité correspondent à la classe~$c$. 
\beqn
\boxed{
\left\{
\begin{array}{l}
\p{P}_1=\frac{C_{11}}{C_{11}+C_{21}+C_{31}}\\[0.3cm]
\p{P}_2=\frac{C_{22}}{C_{12}+C_{22}+C_{32}}\\[0.3cm]
\p{P}_3=\frac{C_{33}}{C_{13}+C_{23}+C_{33}}\\[0.3cm]
\end{array}
\right.
\quad
\mbox{et} \quad 
\left\{
\begin{array}{l}
\p{R}_1=\frac{C_{11}}{C_{11}+C_{12}+C_{13}}\\[0.3cm]
\p{R}_2=\frac{C_{22}}{C_{21}+C_{22}+C_{23}}\\[0.3cm]
\p{R}_3=\frac{C_{33}}{C_{31}+C_{32}+C_{33}}\\[0.3cm]
\end{array}
\right.
}
\eeqn
Ainsi la précision indique la probabilité  qu'un échantillon, dont le classeur évalué indique qu'il s'agit de la classe~$i$, 
soit effectivement de la classe~$i$. 
Par contre,  le rappel indique la probabilité pour qu'un échantillon dont on sait qu'il est de la classe~$i$, soit détecté 
comme tel par le classifieur évalué.

\subsection{Validation croisée}

\begin{figure}
\begin{center}
\begin{tabular}{c}
\includegraphics[width=0.8\linewidth]{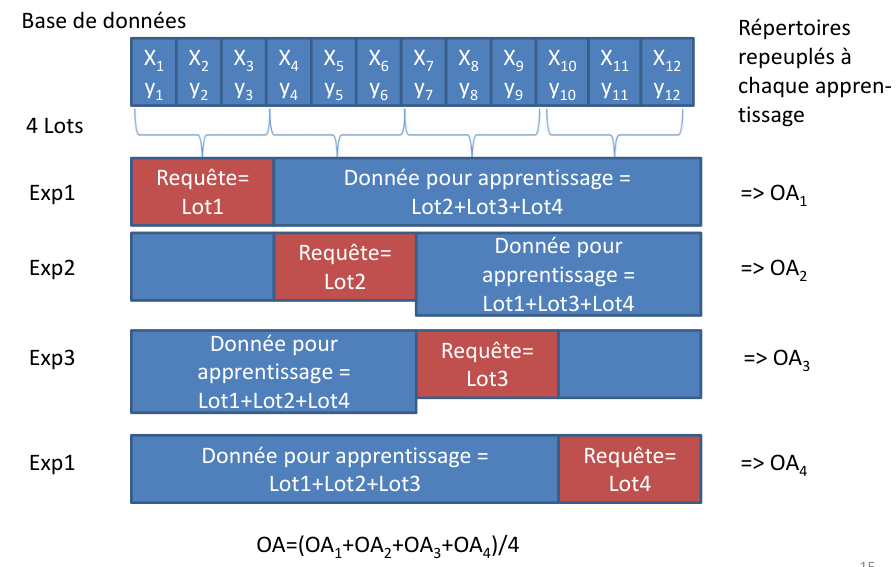}
\end{tabular}
\end{center}
\caption{Schéma décrivant ce qu'est une validation croisée.}
\label{fig:O:valid_croisee}
\end{figure}

La {\bf validation croisée} est une technique souvent utilisée pour la mise en oeuvre d'un algorithme de classification, 
elle n'est pas destinée à améliorer l'efficacité d'un classifieur, elle est destinée à évaluer plus précisément sa 
performance vis-à-vis d'une base de données en particulier  lorsqu'il est compliqué d'augmenter le nombre de sons dont on connaît la
classe à laquelle chacun appartient. 
La figure~\ref{fig:O:valid_croisee} illustre le fonctionnement de la validation croisée avec un schéma. 
En haut de cette figure se trouve 12 sons figurés par $x_1\ldots x_{12}$ et $12$ valeurs notées $y_1\ldots y_{12}$ indiquant les véritables classes à laquelle chaque son appartient.
Ici il s'agit d'une validation croisée en quatre blocs, ce qui amène à réaliser quatre apprentissages de classifieurs. 
{\tt Exp1} désigne en haut la première expérience avec le premier apprentissage. Les trois premiers sons de la base 
d'apprentissage $(x_1,y_1),(x_2,y_2),(x_3,y_3)$ servent de base pour tester le classifieur et sont affichés sur fond rouge. Le classifieur testé est entraîné 
en utilisant tous les autres sons $(x_4,y_4),\ldots,(x_{12},y_{12})$ qui sont eux affichés sur fond bleu. Le test du classifieur permet de calculer une matrice 
de confusion spécifique à {\tt Exp1} et celle-ci permet de calculer la sensibilité globale $\sm{OA}_1$.
Les trois autres expériences sont fait de la même façon, mais en utilisant successivement comme sons tests les 
sons $(x_4,y_4),(x_5,y_5),(x_6,y_6)$, puis $(x_4,y_4),(x_5,y_5),(x_6,y_6)$ et enfin $(x_4,y_4),(x_5,y_5),(x_{12},y_{12})$.

\chapter{Descripteurs de trames}
\label{ch:desc}

La partie précédente ayant montrée comment un descripteur par trame permet de définir une technique de classification, cette 
partie traite de la détermination d'un descripteur par trame. 
L'inspiration qui a guidé l'élaboration des descripteurs provient de l'étude des caractéristiques des ondes sonores émises par un individu,
c'est ce qui constitue la section~\ref{sec:model_parole} et qui montre l'intérêt de trouver la fréquence fondamentale et la plage de fréquences présentes
dans le son émis. 
La section~\ref{sec:estimation_spectrale} présente différentes heuristiques permettant de décrire ce contenu spectral en indiquant la fréquence fondamentale, la 
fréquence moyenne et la largeur de spectre. 
La section~\ref{sec:banc_filtres} présente une technique plus systématique pour décrire le contenu spectral, c'est en utilisant un banc de filtres, chaque 
filtre constitue une heuristique donnant une information partielle sur le contenu spectral du son.

\section{Modélisation de la production de la parole}
\label{sec:model_parole}

La présentation de la modélisation de la façon dont un humain produit un son est 
faite en trois parties. La section~\ref{ssec:onde_stationnaire} décrit le phénomène physique 
permettant de transformer une source sonore en un son dont le caractère plus ou moins aigü du son 
peut être déterminé assez précisément au moyen d'une note. 
La section~\ref{ssec:appareil_phonatoire} décrit brièvement les organes humains intervenant pour 
la production d'un son. 
Enfin la section~\ref{ssec:spectrogramme} présente la notion d'analyse temps-fréquence illustré 
avec la présentation d'un spectrogramme.

\subsection{Onde stationnaire}
\label{ssec:onde_stationnaire}

On dit qu'une onde est stationnaire si en chaque point, on observe que  la surpression évolue dans le temps comme une sinusoïde. 
Par exemple l'équation suivante décrit une évolution sinusoïdale dont l'amplitude crête à crête dépend de l'indice $x$ et vaut 
$\cos(2\pi \frac{x}{\lambda_0})$. 
\beqns
P(x,t)=\Delta P_0\cos(2\pi f_0t)\cos(2\pi \frac{x}{\lambda_0})
\eeqns
où $\lambda$ et $f_0$ sont 
Cette onde peut approximativement correspondre à un cas particulier de l'équation~(\ref{eq:propag_onde}) lorsque le coefficient $\frac{1}{r}$ est 
supposé approximativement constant et~$r$ est remplacé par~$x$. 
La trigonométrie nous indique
que le produit de deux cosinus est une différence entre deux sinus,
 $\cos(a)\cos(b)=\frac{1}{2}\cos(a-b)+\frac{1}{2}\cos(a+b)$. 
Cela permet de voir qu'une onde stationnaire est la somme d'une onde se propageant dans le sens des $x$ croissants et 
une autre dans le sens des $x$ décroissants.
\beqn\label{eq:onde_stationnaire}
P(x,t)=\frac{\Delta P_0}{2}\cos\left (2\pi f_0(t-\frac{x}{c})\right )
+\frac{\Delta P_0}{2}\cos\left (2\pi f_0(t+\frac{x}{c})\right )
\eeqn
où $c=\lambda_0 f_0$ est la vitesse de propagation du son dans l'air. 
Cette modélisation correspond à $g_1=g_2$ dans l'équation~(\ref{eq:propag_onde}).

\begin{figure}[htbp]
\begin{center}
\begin{tabular}{c}
\includegraphics[width=0.5\linewidth]{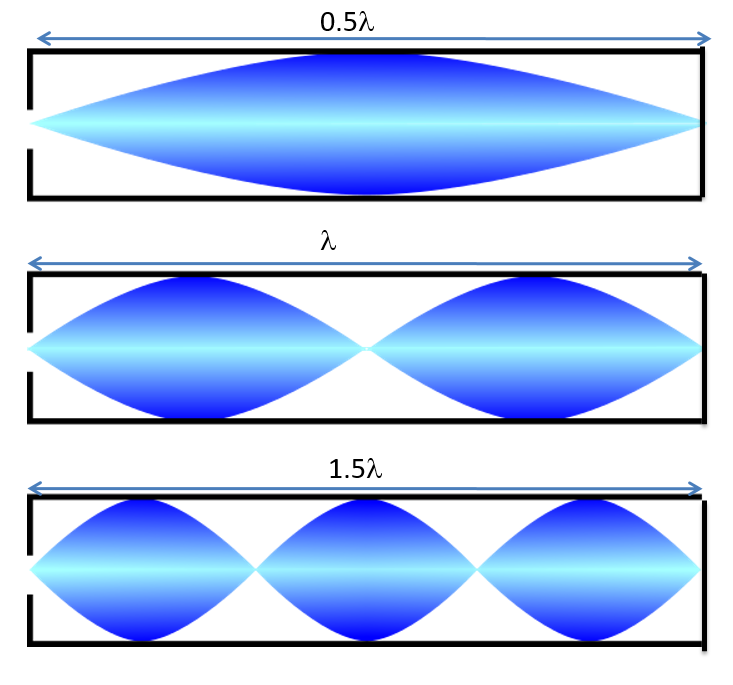}
\end{tabular}
\end{center}
\caption{Pour une même taille donnée, un même obstacle peut être le siège de trois ondes stationnaires. 
La vitesse moyenne des particules d'air doit être nulle aux deux extrémités de la caisse de résonnance. }
\label{fig:onde_stationnaire}
\end{figure}

La figure~\ref{fig:onde_stationnaire} illustre de manière simplifiée la façon dont la forme géométrique d'une caisse de 
résonance détermine une fréquence de résonance. 
Le haut de la figure schématise al caisse de résonance qui est un pavé. La coupe est un rectangle. La zone colorée 
en bleu schématise une onde stationnaire qui consiste en une variation spatiale et temporelle avec une relation 
entre la longueur d'onde et la fréquence de la vibration indiquée dans l'équation~\ref{eq:long_onde} ($\lambda=c/f$ où 
$c$ est ici la vitesse de propagation du son dans l'air, approximativement 1km toutes les $3$~secondes). Les trois dessins 
illustrent qu'avec une même forme géométrique (un pavé de longueur $d$), $3$ longueurs d'onde sont montrés. 
\begin{center}
\begin{tabular}{|c|c|c|}
\hline 
$\lambda=2d$ & $f=\frac{c}{2d}$ & $d=\frac{\lambda}{2}$\\[0.1cm]
\hline
$\lambda=d$ & $f=2\frac{c}{2d}$ & $d={\lambda}$\\[0.1cm]
\hline
$\lambda=\frac{2}{3}d$ & $f=3\frac{c}{2d}$ & $d=\frac{3\lambda}{2}$\\[0.1cm]
\hline
\end{tabular} 
\end{center}

Cette schématisation illustre un phénomène plus général, à savoir que l'émission d'un son à une certaine fréquence de base
produit généralement des harmoniques qui sont des vibrations à des fréquences qui sont multiples de cette fréquence de base. 
La relation entre longueur du rectangle et longueur d'onde est ici déterminé par le fait que la vibration 
du son est nul sur la paroi (i.e. la paroi est fixe). Dans une onde stationnaire modélisée par l'équation~(\ref{eq:onde_stationnaire}),
c'est à cet endroit que la pression est soit maximale soit minimale.
L'annexe~\ref{app:onde_stationnaire_1D} formalise cette description de manière plus précise.  

Une onde stationnaire peut aussi exister avec un rectangle ouvert sur un côté. Dans ce cas c'est la pression qui est fixée et la vibration 
qui est maximale à l'endroit où le rectangle est ouvert. Les relations entre longueurs d'onde, fréquences et longueur du rectangle sont alors modifiées. 
\begin{center}
\begin{tabular}{|c|c|c|}
\hline 
$\lambda=4d$ & $f=\frac{c}{4d}$ & $d=\frac{\lambda}{4}$\\[0.1cm]
\hline
$\lambda=\frac{4}{3}d$ & $f=3\frac{c}{4d}$ & $d=\frac{3\lambda}{4}$\\[0.1cm]
\hline
$\lambda=\frac{4}{5}d$ & $f=5\frac{c}{4d}$ & $d=\frac{5\lambda}{4}$\\[0.1cm]
\hline
\end{tabular} 
\end{center}

\subsection{Modélisation de l'appareil phonatoire}
\label{ssec:appareil_phonatoire}

\begin{figure}[htbp]
\begin{center}
\begin{tabular}{c}
\includegraphics[width=0.5\linewidth]{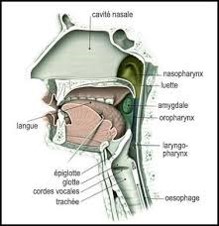}
\end{tabular}
\end{center}
\caption{Diagramme représentant les organes de la parole }
\label{fig:parole_formation}
\end{figure}

On appelle appareil phonatoire l'ensemble des organes humains permettant de produire un son. 
L'objet de cette partie est une brève description de son fonctionnement. 

La figure~\ref{fig:parole_formation} représente une coupe de la tête d'une personne, (coupe dite sagittale), avec la gorge en bas à droite, la bouche au milieu à gauche, les dents le long d'une ligne horizontale placée légèrement au dessus de la bouche. 
Cette figure permet de modéliser la façon dont un son est formé. 
\begin{itemize}
\item Le son est produit par la vibration d'un organe musculaire appelé corde vocale ou plis vocaux. Cet organe musculaire 
permet un contrôle de la fréquence du son émis à ce niveau.  
\item L'intensité du son est modulé par l'intensité du flux d'air sortant du poumon situé plus bas. 
\item Une première caisse de résonance est localisée dans la bouche et sa forme géométrique est modulée notamment par 
la position de la langue et l'ouverture de la bouche. Cette forme géométrique entraîne la formation de vibrations. 
\item Une deuxième caisse de résonance est localisée dans un espace libre au dessus de la bouche au niveau du nez, espace 
appelé cavité nasale. Cette deuxième caisse de résonance entraîne la formation d'un son avec une certaine fréquence. 
\end{itemize}

Ainsi l'anatomie du corps humain permet de modifier l'intensité d'un son et la différence entre deux fréquences 
de résonance.


\subsection{Spectrogramme et notion de description temps-fréquence}
\label{ssec:spectrogramme}

\begin{figure}[htbp]
\begin{center}
\begin{tabular}{c}
\includegraphics[width=0.5\linewidth]{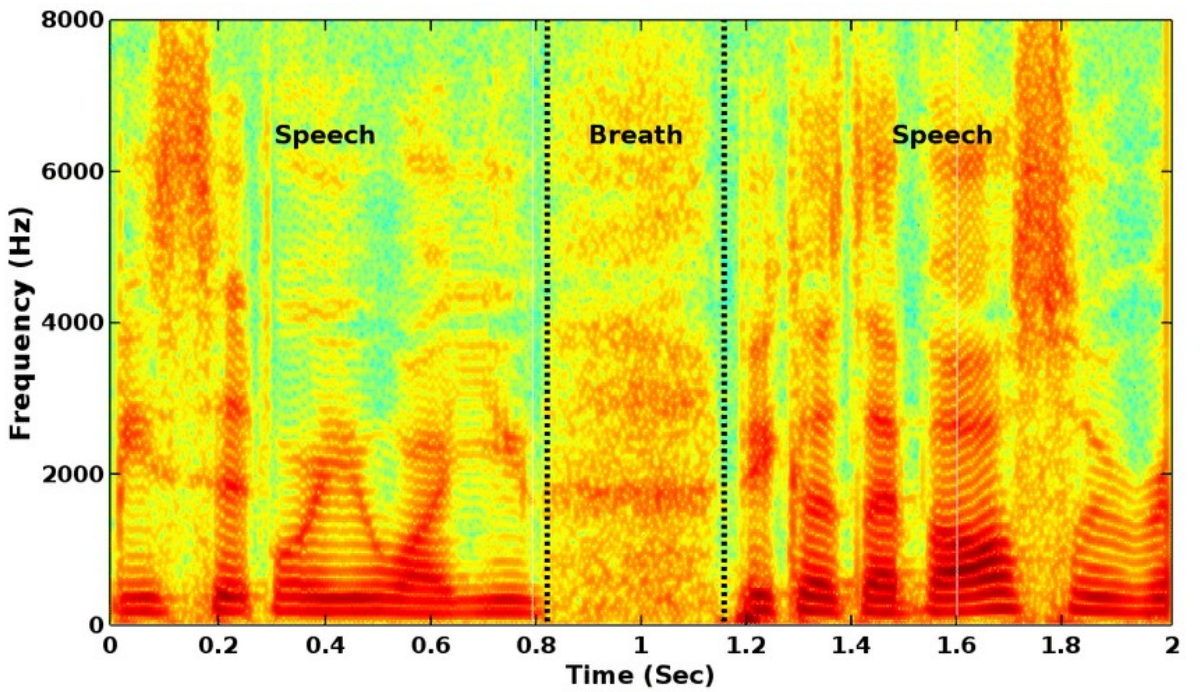}
\end{tabular}
\end{center}
\caption{spectrogramme d'un signal sonore.}
\label{fig:spectrogramme}
\end{figure}

La figure~\ref{fig:spectrogramme} est un exemple de spectrogramme, c'est une visualisation d'un son, ici 
constitué d'une partie parlée à gauche, un silence au milieu, et de nouveau une partie parlée à droite. 
L'axe horizontal est le temps en seconde. L'axe vertical est la fréquence en Hertz. 
Il s'agit d'une représentation temps-fréquence, ce qui peut paraître contradictoire dans le cadre de ce cours. 
En effet une description temporelle d'un signal ne dépend pas de la fréquence et une description fréquentielle d'un signal 
au moyen de son spectre ne dépend pas du temps. En fait la position latérale d'un point détermine une portion du signal sur un intervalle
de temps. Tous les points d'une même verticale concernent la même portion du signal. 
Cette portion de signal est transformée en un spectre qui permet d'indiquer comment l'énergie de cette portion du signal est décomposée en 
une somme  de signaux sinusoïdaux. Chaque point de cette verticale représente une de ces sinusoïdes, c'est la hauteur de ce point qui en déterminant 
une fréquence détermine aussi la sinusoïde. Et un point 
 est d'autant plus sombre ou rouge que l'amplitude du signal sinusoïdal est plus importante. Sur ce spectrogramme, on observe une alternance de 
bandes sombres/rouges horizontales qui sont à la fois régulièrement espacées et avec des intensités qui diminues lorsque la fréquence augmente. 
Ce phénomène est lié au caisses de résonance qui produisent des sons avec une fréquence de base et des harmoniques qui sont des multiples 
de la fréquence de base.

Historiquement la recherche en reconnaissance de la parole s'est d'abord donnée comme guide le fait 
de modéliser les organes qui produisent ou reconnaissent la parole. En pratique la complexité numérique 
et la difficulté à modéliser limitaient la précision des modèles obtenues et par suite la performance. 
Il semble aujourd'hui admis que les techniques de machine learning permettent plus facilement d'obtenir 
des techniques efficaces. Il reste cependant de toutes ces études l'importance apportée à l'utilisation 
d'un certain type de descripteurs et par exemple l'importance d'estimer la fréquence fondamentale. 
L'utilité de s'intéresser à la fréquence fondamentale pour reconnaître une voyelle est 
expliquée en détail dans~\cite{Meunier07}.

\section{Estimation spectrale}
\label{sec:estimation_spectrale}

L'objet de cette partie est de définir 
une application transformant les valeurs sur une trame~$k$ en une valeur donnée, appelée valeur du 
descripteur. 
La section~\ref{ssec:estimation_frequence_fondamentale} présente des techniques permettant 
d'estimer la fréquence fondamentale. 
La section~\ref{sec:descripteur_spectre} presente d'autres techniques permettant d'estimer 
d'autre caractéristiques du spectre, par exemple sa largeur. 

\subsection{Estimation de la fréquence fondamentale}
\label{ssec:estimation_frequence_fondamentale}

Dans cette partie, les deux indicateurs présentés ont pour objectif d'estimer la fréquence fondamentale.~\footnote{Pour un son musical assez simple, cette 
fréquence correspond au {\em pitch} défini section~C.3, p.~215-217 de~\cite{Rocchesso03}.} 

Une autre technique existe pour mesurer la fréquence fondamentale en s'appuyant sur l'autocorrélation.~\footnote{Cette 
technique est brièvement présentée p.~45-46 dans~\cite{Camastra07}.}

\subsubsection{Zero Crossing Rate}

\begin{figure}[htbp]
\begin{center}
\begin{tabular}{c}
\includegraphics[width=0.5\linewidth]{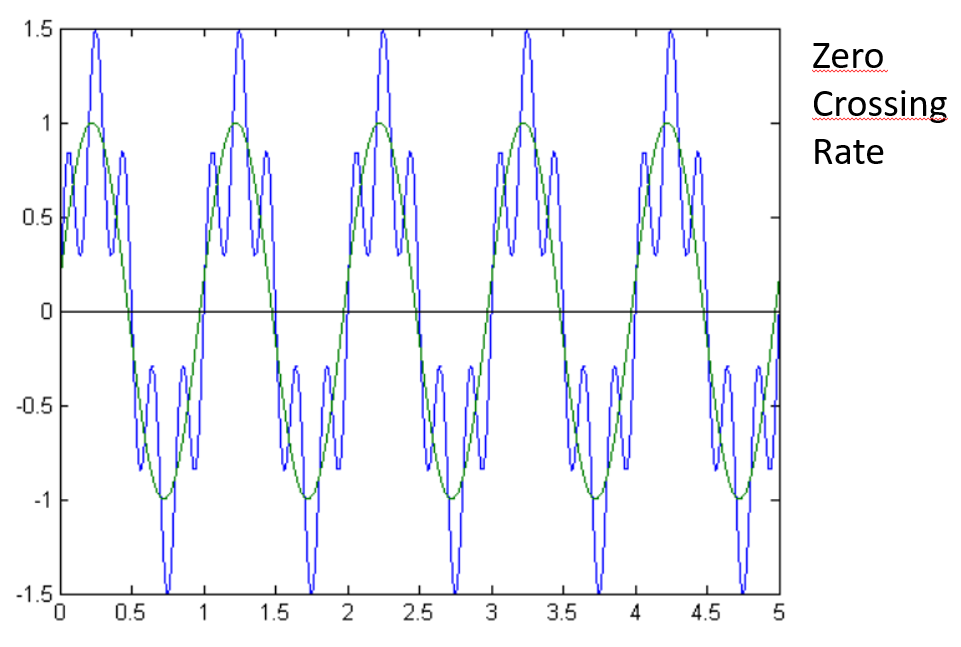}
\end{tabular}
\end{center}
\caption{Graphe illustrant le fonctionnement de l'indicateur $\mb{ZCR}$. }
\label{fig:zero_crossing_rate}
\end{figure}

La figure~\ref{fig:zero_crossing_rate} montre l'évolution de deux signaux en fonction du temps. 
Le premier en trait plus foncé/bleu est la superposition de 2 sinusoïdes. Le deuxième 
 en trait plus clair/vert est une sinusoïde.  Le descripteur~\footnote{Cette technique 
appelée {\em zero crossing rate} est évoquée dans la section~2.4 p.~12 de~\cite{Peeters04}, 
et définie dans~\cite{Camastra07} p.~43-44.} 
$\mb{ZCR}$ détermine le nombre de fois où la courbe traverse l'axe horizontal par une unité de temps. 
Ici il y a deux traversées toutes les secondes. 

Le nombre de passage par zéro n'est en fait pas une notion qui se prête à une formalisation 
utilisant le temps continu. 
Le nombre moyen de passages par zéro pour la trame $k$ est donnée par
\beqn
\label{eq:ZCR}
\boxed{
Z_k=\frac{1}{N_K}{\sum_{n=0}^{N_K-1} \1\left (\signe(x_k[n+1])\neq \signe(x_k[n])\right)}
}
\eeqn
Le fait de normaliser par la longueur de la trame ($N_K$) fait que $Z_k$ n'est pas augmenté lorsque 
les trames sont plus longues. 

Ce nombre moyen de passage par zéro est utilisé pour estimer la fréquence fondamentale, c'est la fréquence 
de la sinusoïde qui aurait ce nombre moyen de passages par zéros. 
\beqn
\label{eq:f_ZCR}
\boxed{
f^{(\sm{ZCR})}_k=\frac{Z_k f_e}{2}
}
\eeqn
En effet, pour une sinusoïde de fréquence $f^{(\sm{ZCR})}_k$, sa période est $T=\frac{1}{f^{(\sm{ZCR})}_k}$ 
et le nombre moyen de passages par zéros est $2$ divisé par le nombre de points sur une période $Tf_e$, 
aussi 
\beqns
f^{(\sm{ZCR})}_k=\frac{1}{T}=\left(\frac{2}{T f_e}\right ) \frac{f_e}{2}=\frac{Z_k f_e}{2}
\eeqns

\subsubsection{Fréquence moyenne du spectre}

\begin{figure}[htbp]
\begin{center}
\begin{tabular}{c}
\includegraphics[width=0.5\linewidth]{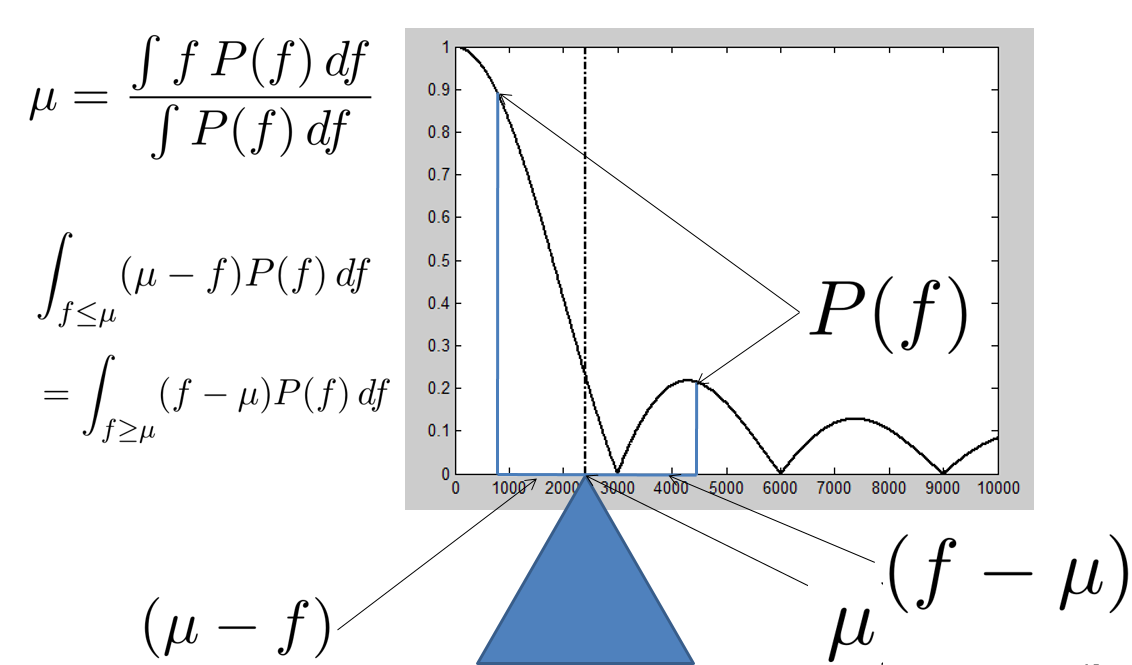}
\end{tabular}
\end{center}
\caption{Illustration du calcul de la fréquence moyenne et 
de sa signification comme séparant en deux parties 
égales la surface sous la courbe.}
\label{fig:illustration_frequence_moyenne}
\end{figure}

La notion de temps-fréquence évoquée lors de la description du spectrogramme dans la section~\ref{ssec:spectrogramme}
permet de définir une densité spectrale d'énergie qui dépend du temps. Ici cela se fait en 
appliquant une fenêtre rectangulaire avec l'intervalle $\left [t-\frac{T}{2},t-\frac{T}{2}\right]$
sur le signal $x(t)$. 
\beqn
\label{eq:DSP_t}
\boxed{
\left|\ww{X}_t(f)\right|^2=\left |\mb{TF}\left [x(t)\1_{\left [t-\frac{T}{2},t-\frac{T}{2}\right]}(t)\right](f)
\right|^2
}
\eeqn
Ensuite cette densité spectrale dépendant du temps et de la fréquence est une utilisée 
comme pondération dans $\mu(t)$ qui est appelée la fréquence moyenne (son unité est bien le Hertz). 
\beqn\label{eq:mu_t}
\boxed{
\mu(t)=\frac{{\int\limits_0^{+\infty} f\left |\ww{X}(f)\right |^2\,df}}
{{\int\limits_0^{+\infty} \left |\ww{X}(f)\right |^2\,df}}
}
\eeqn
On peut voir ce calcul de moyenne comme assez similaire au calcul de l'espérance 
d'une variable aléatoire à partir de la densité de probabilité d'une variable aléatoire~: 
$E[T]={\int_\bb{R} tf_T(t)\,dt}$ où ici $f$ jouerait le rôle de $t$ et 
$\frac{\left |\ww{X}(f)\right |^2}{{\int\limits_0^{+\infty} \left |\ww{X}(f)\right |^2\,df}}$
jouerait le rôle de $f_T(t)$.

La figure~\ref{fig:illustration_frequence_moyenne} donne une justification 
de la définition de $\mu(t)$ dans l'équation~(\ref{eq:mu_t}).
La courbe en noir notée $P(f)$,
correspondant en réalité à $|\ww{X}_t(f)|^2$, 
est la densité spectrale d'énergie d'une portion du signal 
$x(t)\1_{[t-\frac{T}{2},t+\frac{T}{2}]}(t)$. 
Le trait semi-pointillé vertical est positionné 

Cette notion est adaptée aux signaux à temps discret en 
considérant la portion du signal~\footnote{Cette application 
de la transformée de Fourier discrète sur 
le signal découpé peut être vu comme 
l'utilisation d'une transformée 
de Fourier à court terme, p.~99 de~\cite{Rocchesso03}.} 
 sur une trame $x_k[n]$ et en lui calculant la transformée de Fourier discrète considérant 
que la longueur de la trame est suffisante pour que cette transformée de Fourier discrète
soit une bonne approximation de la transformée de Fourier de cette portion de signal. 
On utilise ici un indice $l$ pour les fréquences et on ne considère que la première 
moitié des $N_K$ indices parce que la deuxième moitié des indices correspond aux fréquences 
négatives qui n'est pas prise en compte dans l'équation~(\ref{eq:mu_t}).
Cette première moitié des indices est notée 
\beqn\label{eq:N_K_2}
\boxed{
N_K^{\deux}=\left \lfloor \frac{N_K}{2} \right\rfloor
}
\eeqn
où $\lfloor\ldots \rfloor$ désigne l'arrondi par en dessous. 
Les $N_K^{\deux}$ premières fréquences sont 
$\left [0\ldots \frac{N_K^{\deux}f_e}{N_K}\right]$. 
Ces différentes valeurs sont écrites au moyen d'un nouvel indice $l$ dont dépend la fréquence. 
\beqn\label{eq:f_l}
\boxed{
f_l=\frac{lf_e}{N_K}\mbox{ pour }l\in \{0\ldots N_K^{\deux}
}
\eeqn
La pondération utilisée est la densité spectrale de puissance de la trame $k$.
\beqn
\boxed{
\left |\ww{X}_k[l]\right |^2= \left |
\sm{TFD}\left [x_k[n]\right][l]\right |^2
}
\eeqn
Un estimateur de la fréquence moyenne du spectre $\mu_k$~\footnote{Cet estimateur 
est défini dans la section~6.1.1 p.~13 de~\cite{Peeters04}.}
 est  
\beqn\label{eq:mu_sp}
\boxed{
\mu_k=\frac{{\sum_{l=0}^{N_K^{\deux}}\,\,f_l\,\,\left |\widehat{X}_k[l]\right |^2}}
{{\sum_{l'=0}^{N_K^{\deux}}\left |\widehat{X}_k[l']\right |^2}}
}
\eeqn 
Un indice $l'$ est utilisé parce qu'il n'y a en fait aucun lien entre le compteur utilisé dans la sommation au niveau du numérateur et
le compteur utilisé dans la sommation au niveau du dénominateur. 

\subsection{Autre descripteurs du spectre}
\label{sec:descripteur_spectre}

\subsubsection{Largeur du spectre}

\begin{figure}[htbp]
\begin{center}
\begin{tabular}{c}
\includegraphics[width=0.5\linewidth]{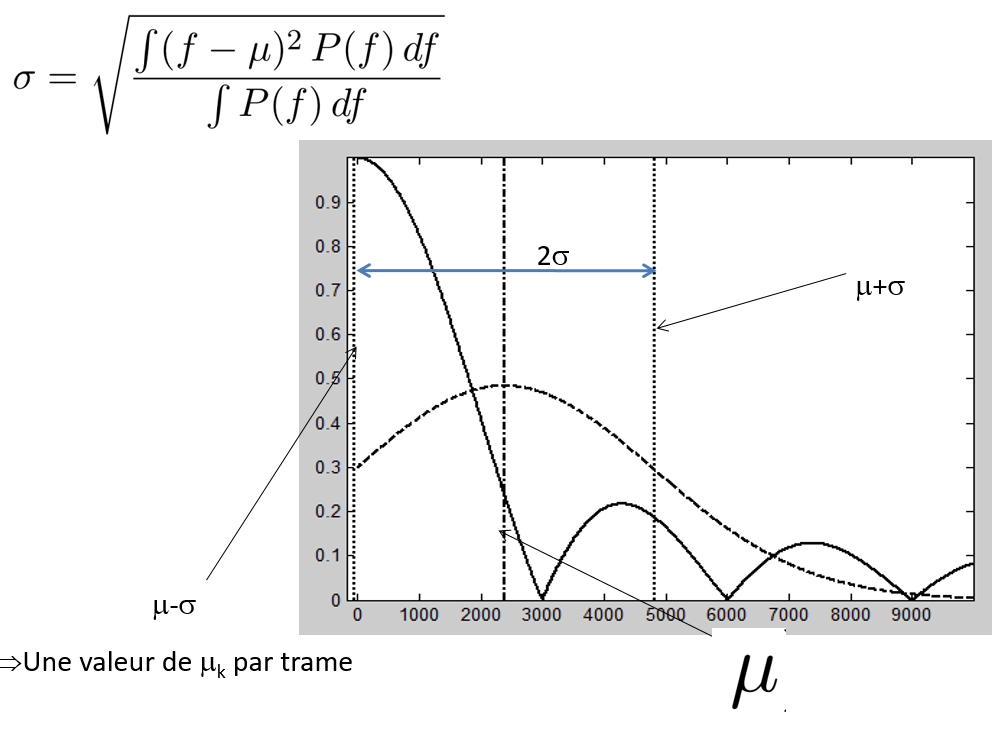}
\end{tabular}
\end{center}
\caption{Illustration graphique sur un exemple du sens donné à la notion d'estimation de la largeur d'un spectre.}
\label{fig:illustration_largeur_spectre}
\end{figure}

La largeur du spectre est estimée connaissant $\mu(t)$ avec 
\beqn
\label{eq:largeur_spectre}
\boxed{
\sigma(t)=\frac{\sqrt{{\int\limits_{\bb{R}_+}\left(f-\mu(t)\right)^2\left |\ww{X}_t(f)\right|^2\,df}}}{\sqrt{{\int\limits_{\bb{R}_+}\left |\ww{X}_t(f)\right|^2\,df}}}
}
\eeqn

La figure~\ref{fig:illustration_largeur_spectre} représente l'exemple d'un spectre d'une porte. 
La fréquence est en abscisse. Ce spectre a pour équation un sinus cardinal avec un demi-lobe principal de $3000$Hz. 
Cette figure illustre deux descripteurs $\mu$ et $\sigma$. $\mu$ est la fréquence moyenne et $\sigma$ est l'écart-type. 
Les deux sont calculés comme si les spectres étaient nuls pour les fréquences négatives (s'ils étaient calculés sur les 
vrais spectres, comme leur module est symétrique, la moyenne serait nulle. 
$\mu$ est indiqué par un trait vertical en pointillé. $\sigma$ est représenté par deux traits verticaux autour de celui associé à $\mu$. 
Ces deux traits verticaux indiquent que si le spectre était une gaussienne (courbe en traits pointillés), $66\%$ de sa surface serait incluse dans ces deux traits
verticaux en pointillés. Ce graphe n'est en fait pas tout à fait exact, d'une part la proportion de $66\%$ et l'axe de symétrie au centre de la gaussienne n'est 
ici pas valable puisqu'on ne considère que les fréquences positives, d'autre part la moyenne et le calcul de la largeur se fait en considérant 
la densité spectrale et non le module de la transformée de Fourier et cette densité spectrale n'est donc pas exactement un sinus cardinal.

De même pour le calcul de la moyenne temporelle, l'équation~(\ref{eq:largeur_spectre}) décrivant l'estimation de la largeur du spectre peut 
se voir comme l'estimation de l'écart-type d'une variable aléatoire positive de densité de probabilité $f_T(t)$
\beqns
\sqrt{\mb{var}(T)}=\sqrt{E[(T-E[T])^2]}=\sqrt{{\int\limits_{\bb{R}_+}(t-E[T])^2f_T(t)\,dt}}
\eeqns
où ce qui remplace ici la variable $t$ est $f$ et la densité de probabilité $f_T(t)$ est 
\beqns
\frac{1}{{\int\limits_{\bb{R}_+}\left |\ww{X}_t(f)\right|^2\,df}}\, \left |\ww{X}_t(f)\right|^2
\eeqns

En utilisant les mêmes notations que pour (\ref{eq:mu_sp}), une estimation de la largeur du spectre~\footnote{Cet estimateur est défini dans la section~6.1.2 p.~13 de~\cite{Peeters04}.} est
\beqn
\label{eq:sigma_sp}
\boxed{
\sigma_k=\sqrt{\frac{{\sum_{l=0}^{N_K^\deux}\,\,\left (f_l-\mu_k\right )^2\,\,
\left |\ww{X}_k[l]\right |^2}}{{\sum_{l'=0}^{N_K^\deux}\left |\widehat{X}_k[l']\right |^2}}}
}
\eeqn


\subsubsection{Coefficient d'asymétrie du spectre}

Un estimateur du coefficient d'asymétrie ({\em skewness})~\footnote{Cet estimateur est défini dans la section~6.1.3 p.~13 de~\cite{Peeters04}.} est donné 
par 
\beqns
\gamma_1(t)=\frac{1}{\sigma^3(t)}\frac{{\int\limits_{\bb{R}_+}\left(f-\mu(t)\right)^3\left |\ww{X}_t(f)\right|^2\,df}}{{\int\limits_{\bb{R}_+}\left |\ww{X}_t(f)\right|^2\,df}}
\eeqns
qui devient avec les notations utilisées dans (\ref{eq:sigma_sp}).
\beqn
\gamma_1[k]=\frac{1}{\sigma_k^3}{\frac{{\sum_{l=0}^{N_K^\deux}\,\,\left (f_l-\mu_k\right )^3\,\,
\left |\ww{X}_k[l]\right |^2}}{{\sum_{l'=0}^{N_K^\deux}\left |\widehat{X}_k[l']\right |^2}}}
\eeqn
De même, un estimateur du coefficient d'aplatissement ({\em kurtosis})~\footnote{Cet estimateur est défini dans la section~6.1.4 p.~14 de~\cite{Peeters04}.} est donné par
\beqns
\gamma_1(t)=\frac{1}{\sigma^4(t)}\frac{{\int\limits_{\bb{R}_+}\left(f-\mu(t)\right)^4\left |\ww{X}_t(f)\right|^2\,df}}{{\int\limits_{\bb{R}_+}\left |\ww{X}_t(f)\right|^2\,df}}
\eeqns
qui devient avec les notations utilisées dans (\ref{eq:sigma_sp}).
\beqn
\gamma_2[k]=\frac{1}{\sigma_k^4}{\frac{{\sum_{l=0}^{N_K^\deux}\,\,\left (f_l-\mu_k\right )^4\,
\,\left |\widehat{X}_k[l]\right |^2}}{{\sum_{l'=0}^{N_K^\deux}\left |\widehat{X}_k[l']\right |^2}}}
\eeqn

\section{Utilisation de bancs de filtres}
\label{sec:banc_filtres}

\begin{figure}[htbp]
\begin{center}
\begin{tabular}{c}
\includegraphics[width=0.5\linewidth]{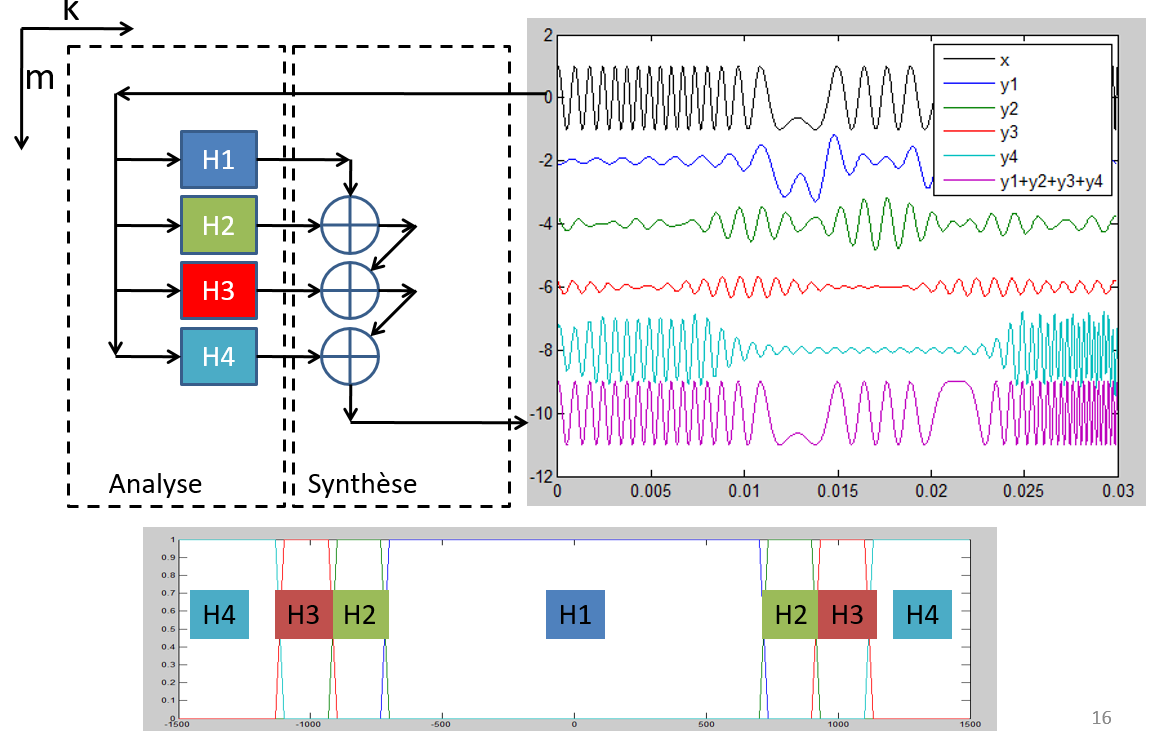}
\end{tabular}
\end{center}
\caption{Schéma d'un banc de filtres }
\label{fig:banc_de_filtre}
\end{figure}

On introduit de nouvelles notations pour décrire un banc de filtres. 
\begin{itemize}
\item $l$ est un indice qui compte les raies du spectre des trames. 
\item $f_l$ est la valeur de la fréquence associée à la raie~$l$. 
\item $m$ est un indice qui numérote les filtres utilisés dans le banc de filtres. 
\item $\ww{H}_m(f)$ est la réponse fréquentielle du filtre numéro $m$. 
\end{itemize}

La figure~\ref{fig:banc_de_filtre} schématise un banc de filtres. Le bas de la figure montre les réponses fréquentielles de $4$~filtres notés~$\ww{H}_1(f)$, 
$\ww{H}_2(f)$, $\ww{H}_3(f)$ et $\ww{H}_4(f)$. Le filtre~$\ww{H}_1$ est un passe-bas. Les filtres $\ww{H}_2(f)$ et~$\ww{H}_3(f)$ sont des filtres passe-bande.
Le filtre $\ww{H}_4(f)$ est un filtre passe-haut. Une particularité de ces filtres est que la somme de leur réponses fréquentielles est un filtre passe-tout 
généralement un simple filtre à retard dont le module de la réponse fréquentielle vaut~$1$. 

Dans la partie supérieure de la figure~\ref{fig:banc_de_filtre}, on retrouve à gauche les~$4$ filtres $\ww{H}_1(f)$, 
$\ww{H}_2(f)$, $\ww{H}_3(f)$ et $\ww{H}_4(f)$. Ces filtres reçoivent tous en entrée le signal $x(t)$, celui-ci est visualisé en haut à droite en noir. 
Le signal en bleu représenté juste en dessous est $y_1(t)$, c'est la sortie du filtre de réponse fréquentielle $\ww{H}_1(f)$. On observe que ce signal 
présente des variations de moindre amplitudes, c'est la conséquence de ce que $\ww{H}_1(f)$ est un passe-bas. 
Les signaux représentés en dessous en vert, rouge et cyan, sont notés $y_2(t), y_3(t), y_4(t)$. Ce sont respectivement les sorties des filtres
$\ww{H}_2(f)$, $\ww{H}_3(f)$ et $\ww{H}_4(f)$. On y retrouve des variations parfois significativement plus importantes. 
La décomposition du signal $x(t)$ en $4$ signaux $y_1(t), y_2(t), y_3(t), y_4(t)$ est appelé {\em analyse}, terme indiqué en bas à gauche de la figure~\ref{fig:banc_de_filtre}. Ces signaux, $y_1(t), y_2(t), y_3(t), y_4(t)$ sont ensuite additionnés pour former le signal visualisé en bas à droite en violet et noté 
$y_1(t)+y_2(t)+y_3(t)+y_4(t)$. Tel qu'on peut l'observer, ce signal est identique à $x(t)$. En réalité, ce dernier signal est généralement retardé par rapport 
à $x(t)$. Il est en effet impossible de réaliser une décomposition en banc de filtres avec des filtres causaux sans introduire un retard. 
La formation d'une reconstruction retardée de $x(t)$ à partir des signaux $y_1(t), y_2(t), y_3(t), y_4(t)$ s'appelle {\em synthèse}, ce qui est indiqué en 
bas du deuxième bloc. 

En regardant plus attentivement le bas de la figure~\ref{fig:banc_de_filtre}, on peut observer que les supports~\footnote{Par support, on entend ici l'intervalle sur lequel une fonction ici la réponse fréquentielle est non nulle.} des réponses 
fréquentielles des différents filtres ne sont pas disjoints, on dit que ces filtres se chevauchent. En fait l'allure même du signal $y_1(t)$ avec ces variations atténuées mais existantes laisse penser qu'il y a probablement un chevauchement assez important. Et en effet la notion de banc de filtre 
n'est pas du tout contradictoire avec un chevauchement même assez important. 
Pour la suite de ce document et pour plus de simplicité, on va considérer des bancs de filtres sans chevauchement où les réponses
fréquentielles sont en fait des portes définies par $\1_{[f_1,f_2]}(f)$. 

On dit qu'un banc de filtres suit une échelle linéaire lorsque les intervalles associés aux différentes portes caractérisant les réponses fréquentielles 
des filtres $\ww{H}_m(f)$ sont de longueur identiques. 
\beqn\boxed{
[f_{\min},f_{\max}]=\bigcup_{m}[f_m,f_{m+1}]\quad\mbox{où}\quad 
\left \{ \begin{array}{l}
f_{\min}=f_1,\,\,f_{\max}=f_{M+1}\\[0.3cm]
f_m=f_{\min}+(m-1)\frac{f_{\max}-f_{\min}}{M}\\[0.3cm]
\ww{H}_m(f)=\1_{[f_m,f_{m+1}]}(f)
\end{array}
\right. 
}
\eeqn

\begin{figure}
\begin{center}
\begin{tabular}{cc}
\includegraphics[width=0.5\linewidth]{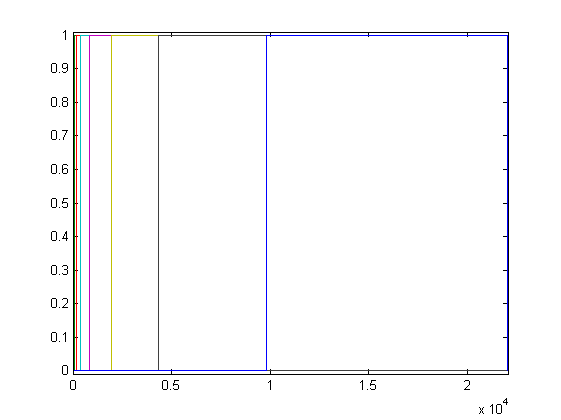}&
 \includegraphics[width=0.5\linewidth]{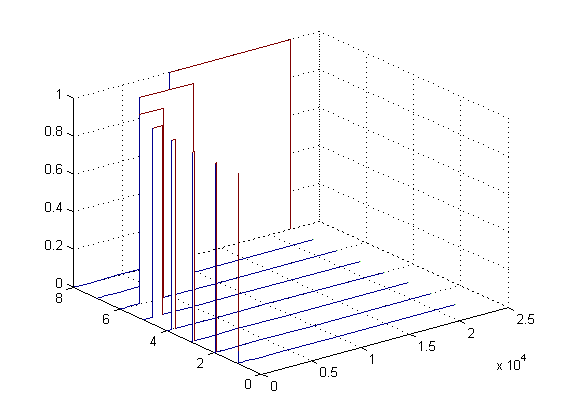}
\end{tabular}
\end{center}
\caption{\`A gauche~: les courbes $\widehat{H}_m(f)$ en fonction de $f$ pour $m\in\{1,\ldots,8\}$ sont superposées
sur un même graphe.
 \`A droite~: ces courbes ont chacune leur abscisse.}
\label{fig9}
\end{figure}

La sensibilité de l'oreille humaine vis-à-vis de la décomposition fréquentielle d'un son et la façon dont les sons sont produits 
par les organes humains amènent à choisir pour les intervalles caractérisant les filtres $\ww{H}_m(f)$ une échelle logarithmique. 
Ce terme d'échelle logarithmique signifie que les intervalles sont de longueurs {\em croissantes} et que les bornes des ces intervalles 
suivent une loi {\em géométriques}. 
\beqn\boxed{
[f_{\min},f_{\max}]=\bigcup_{m}[f_m,f_{m+1}]\quad\mbox{où}\quad 
\left \{ \begin{array}{l}
f_{\min}=f_1,\,\,f_{\max}=f_{M+1}\\[0.3cm]
f_m=f_{\min}\left (\frac{f_{\max}}{f_{\min}}\right)^{\frac{m-1}{M}}\\[0.3cm]
\ww{H}_m(f)=\1_{[f_m,f_{m+1}]}(f)
\end{array}
\right. 
}
\eeqn

En pratique pour un banc de filtre suivant une échelle linéaire, on peut choisir $f_{\min}=0$ et $f_{\max}=\frac{f_e}{2}$, 
et pour une échelle logarithmique, il faut que $f_{\min}>0$, on peut choisir $f_{\min}=\frac{1}{T_K}$ et $f_{\max}=\frac{f_e}{2}$,
$T$ étant la durée de la trame.  

La figure~\ref{fig9} montre les réponses fréquentielles de huit filtres $\ww{H}_m$ qui suivent une échelle logarithmique, à gauche 
en les superposant sur un même graphe 2D et à droite en mettant chaque courbe le long d'un axe différent. 

\begin{figure}
\begin{center}
\begin{tabular}{cc}
\includegraphics[width=0.5\linewidth]{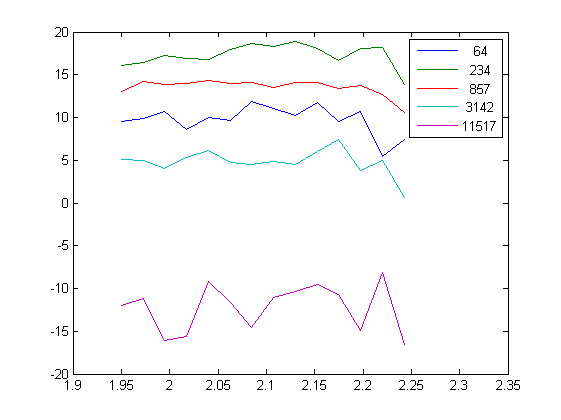} &\includegraphics[width=0.5\linewidth]{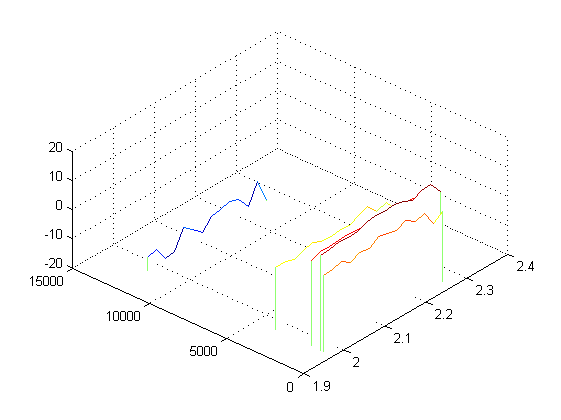}
\end{tabular}
\end{center}
\caption{\`A gauche~: superposition de l'évolution des descripteurs en fonction du temps. Chaque courbe
est associée à un filtre en particulier dont la légende indique la fréquence centrale associée à ce filtre.
\`A droite~: représentation séparée de l'évolution de chaque descripteur en fonction du temps.Dans les 
deux graphes, l'échelle en temps indique les instants associés aux milieux de chaque trame.}
\label{fig10}
\end{figure}

La figure~\ref{fig10} représente un exemple de représentation de ce descripteur temps-fréquence, 
il s'agit à gauche et à droite des courbes qui dans la figure~\ref{fig9} sont appelées $y_1(t), y_2(t), y_3(t), 
y_4(t)$. La différence est qu'ici il y en a $5$ et que les filtres utilisés ici ne se chevauchent pas. 
On observe des variations d'amplitudes plus faibles pour les quatre  premières 
par rapport à la dernière associées au fréquences plus élevées. 
Cependant il est significatif que la courbe bleue foncée, associée au filtre passe-bas et située au milieu sur le graphe de gauche et en 
haut à gauche sur le graphe de droite, a des variations qui sont d'une amplitude assez significatives, là 
on aurait pu s'attendre à des variations moins importantes. Cela traduit le fait que le signal 
sonore étudié présente des variations temporelles importantes d'une trame à une autre. 

Ces bancs de filtres servent à définir des descripteurs, un par filtre du banc de filtre.
De multiples descripteurs sont proposés dans la littérature scientifique. 
On aurait pu s'attendre à définir un descripteur comme la puissance sur une portion du signal en 
sortie du filtre $\ww{H}_m(f)$
autour de l'instant $t$.
\beqn\label{eq:desc_bf}
\boxed{
D_m(t)=\frac{1}{T}{\int_{-\frac{T}{2}}^{\frac{T}{2}} |h_m(\tau)\ast x_t(\tau)|^2\,d\tau}
\quad \mbox{ où }\quad \left \{ \begin{array}{l}
h_m(\tau)\mbox{ est la réponse impulsionnelle du filtre }\ww{H}_m(f)\\[0.3cm]
x_t(\tau)=x(\tau+t)\1_{\left [-\frac{T}{2},\frac{T}{2}\right]}(\tau)
\end{array}
\right . 
}
\eeqn

Mais comme la perception de l'intensité sonore est mieux corrélée à une intensité mesurée en décibel et donc 
suivant une échelle logarithmique vis-à-vis de la puissance, il convient d'ajouter un logarithme dans l'équation~(\ref{eq:desc_bf}).
La littérature scientifique a fait le choix d'utiliser un autre descripteur dont une version en temps continu serait 
\beqn
\sm{D\_LOG}_m(t)=\frac{1}{f_{m+1}-f_m}\,{\int_{f_m}^{f_{m+1}} 10\log_{10} \left |\ww{H}_m(f) \ww{X}_t(f)\right|^2\,df}
\eeqn
Et en pratique, le descripteur utilisé, dépend de deux paramètres, $m$ pour le filtre et par 
suite la fréquence associée à ce filtre, et $k$ pour la trame considérée. 

Le descripteur utilisé dans les travaux pratiques 
a pour nom {\tt LOG\_MAG\_FB\_LOG}~\footnote{La signification du nom se retrouve en lisant 
l'expression de droite à gauche. {\tt LOG} en dernier signifie qu'on considère une échelle logarithme, puis 
que cette échelle permet de définir un banc de filtre ({\em filter bank}), qu'on calcule ensuite le module ({\em magnitude}) 
et qu'avant de considérer la moyenne on considère le logarithme des modules.}

En pratique, le descripteur est ainsi défini.
\beqn
\boxed{
\sm{LOG\_MAG\_FB\_LOG}(m,k)=
\frac{1}{\left |\bb{L}_m\right |}
{\sum_{l\in \bb{L}_m} 10\log_{10}\left |\widehat{H}_{m,l}\widehat{X}_{k,l}\right|^2}
}
\eeqn
où $\bb{L}_m$ désigne les indices $l$ pour lesquels $\widehat{H}_{m}\left(\frac{lf_e}{N_K}\right)$ est non-nul, 
ce sont a priori les indices $l$ pour lesquels la fréquence $\frac{lf_e}{N_K}$ sont contenus dans 
$\left [f_m,f_{m+1}\right]$. 
$|\bb{L}_m|$ désigne le nombre d'indices $l$ contenus dans $\bb{L}_m$. 

Remarquons que le fait de modifier un descripteur en le multipliant par une certaine constante ou 
en lui ajoutant une constante n'a aucune incidence sur la classification qui en résulte.


D'autres suggestions de descripteurs pour la reconnaissance de la parole se trouvent dans~\cite{Rabiner93}.

\newcommand{\Vv}{\mathbf V}
\newcommand{\Dd}{\mathbf D}
\newcommand{\Hh}{\mathbf H}
\newcommand{\MZ}{\mathcal Z}
\newcommand{\MX}{\mathcal X}
\newcommand{\bx}{\mathbf x}
\newcommand{\MH}{\mathcal H}
\newcommand{\MMCT}{\mbox{MMCT}}
\renewcommand\theenumi{{\small{\thesection.\arabic{enumi}}}}
\newcounter{suite}

\chapter{Travaux pratiques}

\section*{Préambule}

L'objectif de ces travaux pratiques est d'expérimenter quelques idées simples visant à classer un signal sonore comme étant phonétiquement 
similaire à  {\em un}, {\em deux} ou {\em trois}. 
Ces expérimentations utilisent des programmes dans \href{\dau TP_son.zip}{TP\_son.zip} 
et qui sont disponibles sur 
\begin{verbatim}
https://www-l2ti.univ-paris13.fr/~dauphin/Signal_Processing.html
\end{verbatim}

\section{Séance 6\\ Expérimentations et mise en place d'un prétraitement sur un son}

\subsection{Enregistrer et produire un son}

\begin{enumerate}
\item 
Enregistrez successivement plusieurs sons correspondant à {\em un}, {\em deux} et {\em trois}, puis écoutez le résultat. 
\finenum
\end{enumerate}

\begin{itemize}
\item
Cela peut se faire en connectant un micro à la carte son et avec les étapes suivantes. 
\begin{itemize}
\item {\tt enregistrement=audiorecorder(16000,8,1);}
Les paramètres correspondent à la fréquence d'échantillonnage utilisée pour l'enregistrement, le nombre de bits utilisés et le nombre 
d'entrées son. Le résultat est une structure. 
\item {\tt record(enregistrement);} permet de démarrer l'enregistrement. 
\item {\tt stop(enregistrement);} permet d'arrêter l'enregistrement. 
\item {\tt parole= play(enregistrement); } 
\end{itemize} 
\item
Cela peut aussi se faire en enregistrant ces sons sur un téléphone portable a priori au format {\tt .m4a} puis en récupérant sur 
l'ordinateur les fichiers associés. 
Les fichiers codés peuvent être transformés en fichiers de données avec une version de Matlab supérieure à 2012, 
en utilisant l'instruction {\tt audioread} puis {\tt save}. {\tt sound} permet d'écouter le son enregistré. 
\end{itemize}

\begin{corr}\relax 
\sol
\begin{verbatim}
clear
[y,fe]=audioread('un.m4a');
save un.mat y fe;
sound(y,fe); 
\end{verbatim} 
\end{corr}

\fbox{\parbox{0.8\linewidth}{
\imp
 \begin{itemize}
\item 
Choisissez un répertoire de travail, appelé ici {\tt rep} et  localisé sur un disque dur de l'ordinateur (et non sur la portion du disque dur du serveur associé 
à votre compte informatique).

\item Téléchargez le fichier \href{\dau TP_son.zip}{TP\_son.zip} {\tt TP\_son.zip}
qui fait à peu près 29Mo et placez les fichiers extraits dans {\tt rep}. 
Puis ouvrez Matlab en choisissant comme répertoire de travail le répertoire {\tt prg} à l'intérieur de {\tt rep}, avec 
la commande {\tt cd prg}.
Vérifiez que tout est bien en place en lançant {\tt verification(\tq court\tq)}. 

\item En utilisant l'instruction Matlab {\tt copyfile}, créez deux programmes {\tt sauvegarde.m} et  {\tt charger.m}. 
Le premier est chargé d'enregistrer sur le répertoire du serveur le répertoire sur lequel vous travaillez. Le
deuxième est chargé d'installer le répertoire sur lequel vous allez travailler à partir des données sauvegardées sur 
le profil.  Le programme {\tt sauvegarde.m} est à actionner  en milieu et en fin de séance. 
Dans {\tt rep}, créez un répertoire {\tt prg} contenant les programmes, un répertoire {\tt doc} contenant
les notes, et un répertoires {\tt donnees} contenant 
les répertoires {\tt un}, {\tt deux} et {\tt trois} avec les fichiers audios associés au sons correspondant. 
\end{itemize}
Le fait que tout le monde ait la même structure aide à automatiser les traitements informatiques. 
}}

\subsection{Séquence de mots}
\label{ssec:seq_mots}

\begin{enumerate}
\deb
\item\label{it:seq_mots} Réalisez une fonction 
qui convertit un vecteur composé de chiffres à valeurs dans $\{1,2,3\}$ en un séquence de mots,
qui dictent oralement la séquence de chiffres.~\footnote{Ainsi un aveugle pourrait connaître les composantes de ce vecteur 
en écoutant l'ordinateur.}
Ajoutez des silences pour que la séquence soit plus facile à comprendre.~\footnote{
Il est possible d'écouter une séquence sur 
{\tt https://www-l2ti.univ-paris13.fr/{$\sim$}dauphin/synthese\_vocale.wav}
.}
\finenum
\end{enumerate}

\ind{\em
Vous pouvez compléter le programme suivant et utiliser {\tt lireSon.m} expliqué en annexe~\ref{sec:fct}.
}
\begin{verbatim}
function synthetiseur_vocal(v)
  horiz=@(u)u(:)';
  if ~all(size(v)==[1 numel(v)])     error('paramètres invalides'), end
  if ~all(ismember(v,[1 2 3]))       error('paramètres invalides'), end
  if isempty(v)                      error('paramètres invalides'), end
  y=[]; 
  for k=1:numel(v)
    ...
  end
  sound(y,fe);
end
\end{verbatim}

\fbox{\parbox{0.8\linewidth}{
\imp
Dans le but d'aider à la compréhension et de pouvoir continuer le TP sans être bloqué,
la fonction {\tt synthetiseur\_vocal} comme toutes les fonctions demandées sont déjà réalisées, mais leur 
programme n'est pas visible, leur nom est légèrement différent, elles ont systématiquement un préfixe~\footnote{La lettre
{\tt s} est pour dire qu'il s'agit d'une solution.} qui est {\tt s\_}. 
Ainsi pour cette question, le nom de la fonction déjà réalisée est {\tt s\_synthetiseur\_vocal}.
}}

\fbox{\parbox{0.8\linewidth}{
\imp
Dans l'utilisation de {\tt lireSon}, il est important que le nom du répertoire se termine avec un {\em anti-slash} (ou un {\em slash}). 
}}

\begin{corr}
\relax
\sol\begin{verbatim}
function synthetiseur_vocal(v)
  horiz=@(u)u(:)';
  if ~all(size(v)==[1 numel(v)])     error('paramètres invalides'), end
  if ~all(ismember(v,[1 2 3]))       error('paramètres invalides'), end
  if isempty(v)                      error('paramètres invalides'), end
  y=[]; 
  for k=1:numel(v)
    if 1==v(k)     rep='..\donnees\un\';
    elseif 2==v(k) rep='..\donnees\deux\';
    else           rep='..\donnees\trois\';
    end
    [y_,fe]=lireSon(rep,0);
    y=[y,horiz(y_)]; 
  end
  sound(y,fe);
end
\end{verbatim}
\end{corr}

\subsection{Rendre le son plus aigu ou plus grave}

\begin{enumerate}
\deb
\item Montrez expérimentalement comment en modifiant la fréquence d'échantillonnage, il est possible de rendre  un son, plus aigu ou plus grave. 
Pourquoi la durée du signal est-elle modifiée~?

\begin{corr}
\relax \sol
\begin{verbatim}
clear
[y,fe]=lireSon('..\donnees\un\',0); sound(y,fe*3); sound(y,fe*2); 
sound(y,fe/2); sound(y,fe/3); 
\end{verbatim}
La durée signal est le nombre d'échantillon divisé par la fréquence d'échantillonnage {\tt fe}. 
\end{corr}
\item Montrez comment grâce à l'interpolation, il est possible de rendre ce son plus aigu ou plus grave, mais cette fois sans 
changer la fréquence d'échantillonnage. 

\begin{corr}
\relax \sol
\begin{verbatim}
clear
[y,fe]=lireSon('..\donnees\un\',0); 
y1=y(1:3:end); sound(y1,fe); 
y2=y(1:2:end); sound(y2,fe); 
y3=zeros(1,2*length(y)); y3(1:2:end)=y; y3=filter([1 1], 1, y3); 
sound(y3,fe); 
y4=zeros(1,3*length(y)); y4(1:3:end)=y; y4=filter([1 1 1], 1, y4); 
sound(y4,fe); 
\end{verbatim}
\end{corr}
\finenum 
\end{enumerate}
 On considère maintenant un son {\tt la} obtenu avec une sinusoïde à $440$Hz sur une durée d'une demi-seconde. 
\begin{enumerate}
\deb
\item En multipliant par une autre sinusoïde à $150$Hz et en utilisant un filtre passe-haut, montrez que l'on peut 
obtenir un son à $590$Hz. 

\ind  {\em
Le son {\tt la} correspond à l'échantillonnage d'un signal $x_s(t)=\sin(2\pi f_st)$ avec $f_s=440$Hz. 
La multiplication par l'échantillonnage de $x_m(t)=\cos(2\pi f_mt)$ où $f_m=150$Hz peut s'écrire comme la somme de deux 
sinusoïdes. 
\beqns
x_s(nT_e)x_m(nT_e)=\frac{1}{2}\sin\left (2\pi \frac{f_s+f_m}{f_e} n\right )+\frac{1}{2}\sin\left (2\pi \frac{f_s-f_m}{f_e} n\right )
\eeqns
où $T_e=\frac{1}{f_e}$ est la période d'échantillonnage. On peut constater que 
la partie gauche est bien une sinusoïde à $150+440$Hz$=590$Hz. Il reste donc à supprimer la partie droite qui 
est une sinusoïde à $440-150$Hz$=290$Hz. Cela peut se faire avec un filtre passe-haut avec une fréquence de coupure 
à $440$Hz. Pour faire la synthèse de ce filtre, on peut utiliser {\tt fir1}, il convient cependant de choisir 
le nombre de coefficients pour ce filtre et ainsi de choisir la résolution fréquentielle du filtre utilisé. Vous 
pouvez utiliser ici $N=1000$.
Vous pouvez compléter le programme suivant.
}
\begin{verbatim}
duree=2; fe=44100; t=0:1/fe:(duree-1/fe); 
fs=440; xs=sin(2*pi*fs*t); sound(xs,fe);
xm=sin(2*pi*150*t); y1=xs.*xm; 
N=1000; 
A=...
B=fir1(N,...
y2=filter(B,A,y1); 
sound(y2,fe); 
\end{verbatim}

\begin{corr}
\relax \sol 
\begin{verbatim}
clear
duree=2; fe=44100; t=0:1/fe:(duree-1/fe); 
fs=440; xs=sin(2*pi*fs*t); sound(xs,fe);
xm=sin(2*pi*150*t); y1=xs.*xm; 
N=1000; 
A=1; 
B=fir1(N,fs/fe*2,'high'); y2=filter(B,1,y1); 
sound(y2,fe); 
\end{verbatim}
\end{corr}
\finenum
\end{enumerate}


\fbox{\parbox{0.8\linewidth}{
\imp
Il est souhaitable de commencer toutes nouvelles séquences d'instructions et tout script par {\tt clear}. 
Sinon il y a un risque que cette séquence d'instruction utilise des valeurs de précédentes expérimentations 
stockées dans des variables qui portent le même nom. 
}}

\subsection{Sous-échantillonner le signal}

Pour l'application considérée, la reconnaissance de mots, les informations au-delà de $4$kHz ne sont pas jugées prioritaires. 
Aussi pour simplifier, les signaux sont sous-échantillonnés à $8$kHz. 
\begin{enumerate}
\deb
\item\label{it:sous_ech}
Construisez une fonction qui réalise ce sous-échantillonnage avec la syntaxe suivante
\begin{verbatim}
function [y2,fe2]=sous_ech(y1,fe1)
\end{verbatim}
\finenum
\end{enumerate}

\ind {\em
Pour respecter le critère de Shannon-Nyquist, il est nécessaire d'utiliser un filtrage pour atténuer fortement les fréquences au-delà de 
$4kHz$. Vous pouvez utiliser {\tt fir1} de façon à calculer une réponse impulsionnelle de $1000$ termes et ainsi compléter le programme 
suivant. 
\begin{verbatim}
function [y2,fe2]=sous_ech(y1,fe1)
  fe2=8000;
  t1=0:1/fe1:(length(y1)-1)/fe1;
  N=1000;
  B=fir1(N,...
  A=...
  z1=filter(B,A,y1); 
  t2=0:1/fe2:t1(end);
  y2=zeros(size(t2));
  for t2_=1:length(t2)
    t1_=find(t1>=t2(t2_),1,'first');
    y2(t2_)=...
  end
end
\end{verbatim}
}
\begin{corr}
\relax \sol
\begin{verbatim}
function [y2,fe2]=sous_ech(y1,fe1)
  fe2=8000;
  t1=0:1/fe1:(length(y1)-1)/fe1;
  N=1000;
  B=fir1(N,4000/fe1*2,'low'); A=1;
  z1=filter(B,A,y1); 
  t2=0:1/fe2:t1(end);
  y2=zeros(size(t2));
  for t2_=1:length(t2)
    t1_=find(t1>=t2(t2_),1,'first');
    y2(t2_)=z1(t1_);
  end
end
\end{verbatim}
\end{corr}

\subsection{Réhausser les fréquences plus élevées (Facultatif)}

On considère un filtre défini par sa relation entrée-sortie 
\beqns
y_n=x_n-\eta x_{n-1}
\eeqns
\begin{enumerate}
\deb
\item\label{it:rehausser} (Facultatif)
Réalisez une fonction appelée {\tt rehausser.m} qui implémente ce filtre. La syntaxe est {\tt [y,fe]=rehausser(y,fe)}. 
La valeur de $\eta$ est déterminée  de façon que la fréquence de coupure du filtre soit de $f_c=1999.5$
et un calcul présenté en annexe~\ref{sec:calcul} indique que 
ce filtre a la fréquence de coupure $f_c$ quand d'une part $f_e>4f_c$ et d'autre part
\beqns
c=\cos\left (2\pi \frac{f_c}{f_e}\right ) \mbox{ et }
\eta=2c+1-2\sqrt{c(c+1)}
\eeqns
La syntaxe de la fonction à réaliser est la suivante.
\begin{verbatim}
function y2=rehausser(y1,fe)
\end{verbatim}
\finenum
\end{enumerate}

\begin{corr}
\relax \sol
\begin{verbatim}
function y2=rehausser(y1,fe)
  fc=1999.5; c=cos(2*pi*fc/fe);
  eta=2*c+1-2*sqrt(c*(c+1));
  if (fe==8000)   if ~(abs(eta-0.96)<=0.01) error('rehausser'), end, end
  B=[1,-eta]; A=1; 
  y2=filter(B,A,y1); 
end
\end{verbatim}
\end{corr}

\subsection{Découper le son en une succession de trames}
\label{ssec:decoupe}

On considère ici un signal {\tt x} sinusoïdal de fréquence $f_0=10$Hz et d'une durée d'une seconde. 
Un signal est ici représenté par un vecteur ligne. Un signal découpé est représenté par une matrice dont 
le numéro de ligne indique la trame considérée et la colonne indique la position d'un échantillon dans chacune de ces trames. 
Lorsque la fin du signal ne coïncide pas avec la dernière trame, ici, on 
considère que la trame non complète n'est pas du tout prise en compte. 

Si le signal était temps continu, le signal découpé serait défini par 
\beqns
x_k(t)=x\left (t-t_k^{(de)}\right )\1_{[t_k^{(de)},t_k^{(fi)}]} (t)
\eeqns

\`A temps discret, il est défini par 
\beqns
x_k[n]=x\left [n-n_k^{(de)}\right ]\1_{\left [n_k^{(de)},n_k^{(fi)}\right]}[n ]
\eeqns

Des explications plus détaillées sont disponibles en annexe~\ref{ssec:decoupe_signal}.

\begin{figure}
\begin{center}
\begin{tabular}{cc}
\includegraphics[width=0.5\linewidth]{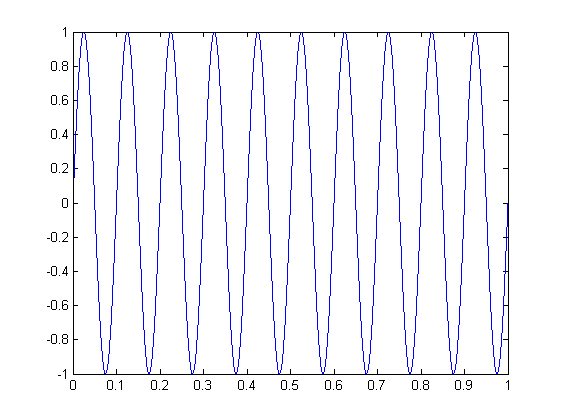} &\includegraphics[width=0.5\linewidth]{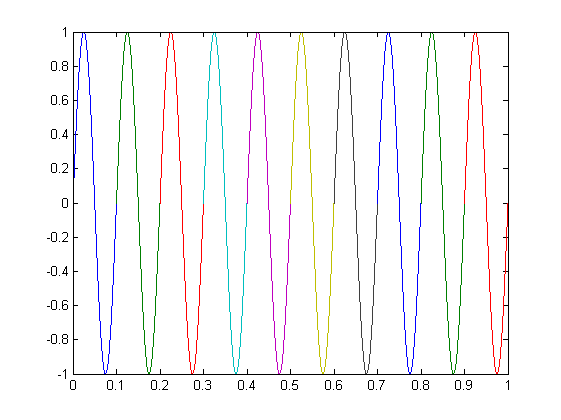}\\
\includegraphics[width=0.5\linewidth]{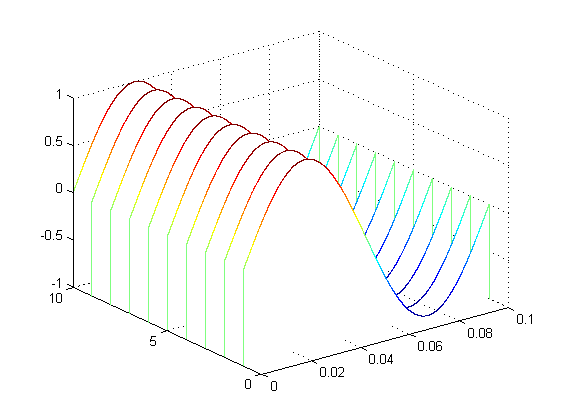} & \includegraphics[width=0.5\linewidth]{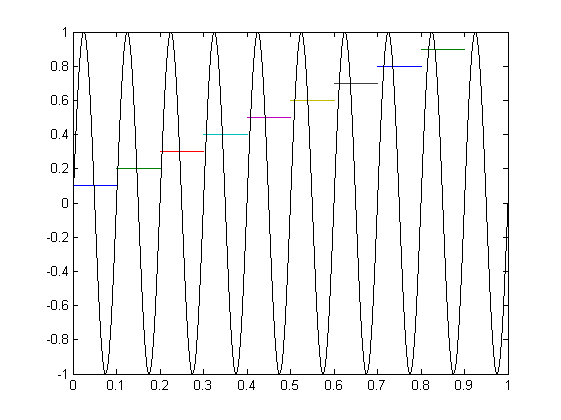}
\end{tabular}
\end{center}
\caption{En haut à gauche~: signal {\tt x} en fonction du temps. En haut à droite~: succession des trames affichées avec des couleurs différentes en fonction du temps. En bas à gauche~: succession des trames représentées cette fois en fonction d'une base de temps propre à la trame. En bas à droite~: signal {\tt x} en noir en fonction du temps et indication colorée des instants associés à chaque trame.}
\label{fig1}
\end{figure}

Quatre visualisations sont considérées ici, elles sont représentées sur la figure~\ref{fig1}~: 
\begin{itemize}
\item Graphe en haut à gauche. Signal représenté en fonction du temps en fonction du temps.
\item Graphe en haut à droite. Superposition des trames colorées sur un même graphe. 
Concrètement la courbe à gauche en bleue du graphe en haut à droite est la restriction de $x(t)$ aux cent premières 
millisecondes. La courbe en verte légèrement à droite est la restriction de $x(t)$ pour $t\in[0.1,0.2]$, etc...
Chacune de ces dix restrictions de $x(t)$ est appelée trame. 
Dans les graphes ultérieurs, il peut y avoir superposition et dans ce cas, les trames postérieures peuvent recouvrir des trames
antérieures (graphe en haut à droite). 
\item Graphe en bas à gauche. Succession des trames représentées successivement en utilisant leur propre base de temps au moyen de l'instruction Matlab {\tt waterfall}. Concrètement la courbe en bleue à gauche du graphe en haut à droite 
qui est la restriction de $x(t)$ à $t\in [0,0.1]$ se trouve 
représentée sur ce graphe en bas droite sous la forme de la courbe la plus à droite. 
Sa voisine légèrement à gauche est la restriction de $x(t)$ à $t\in [0.1,0.2]$, elle est représentée en vert sur le graphe en haut à droite.
\item Graphe à bas à droite. 
Les traits horizontaux colorés
indiquent les plages de temps associées à chaque trame. Ainsi le premier trait bleu indique que la première trame correspond 
à l'intervalle de temps $[0,0.1]$. 
Le signal en noir est identique au signal en bleu représenté sur le graphe en haut à gauche. 
\end{itemize}
Un autre ensemble de visualisations est disponible sur la figure~\ref{fig5} (p.~\pageref{fig5}).

\begin{figure}
\begin{center}
\begin{tabular}{c}
\includegraphics[width=\linewidth]{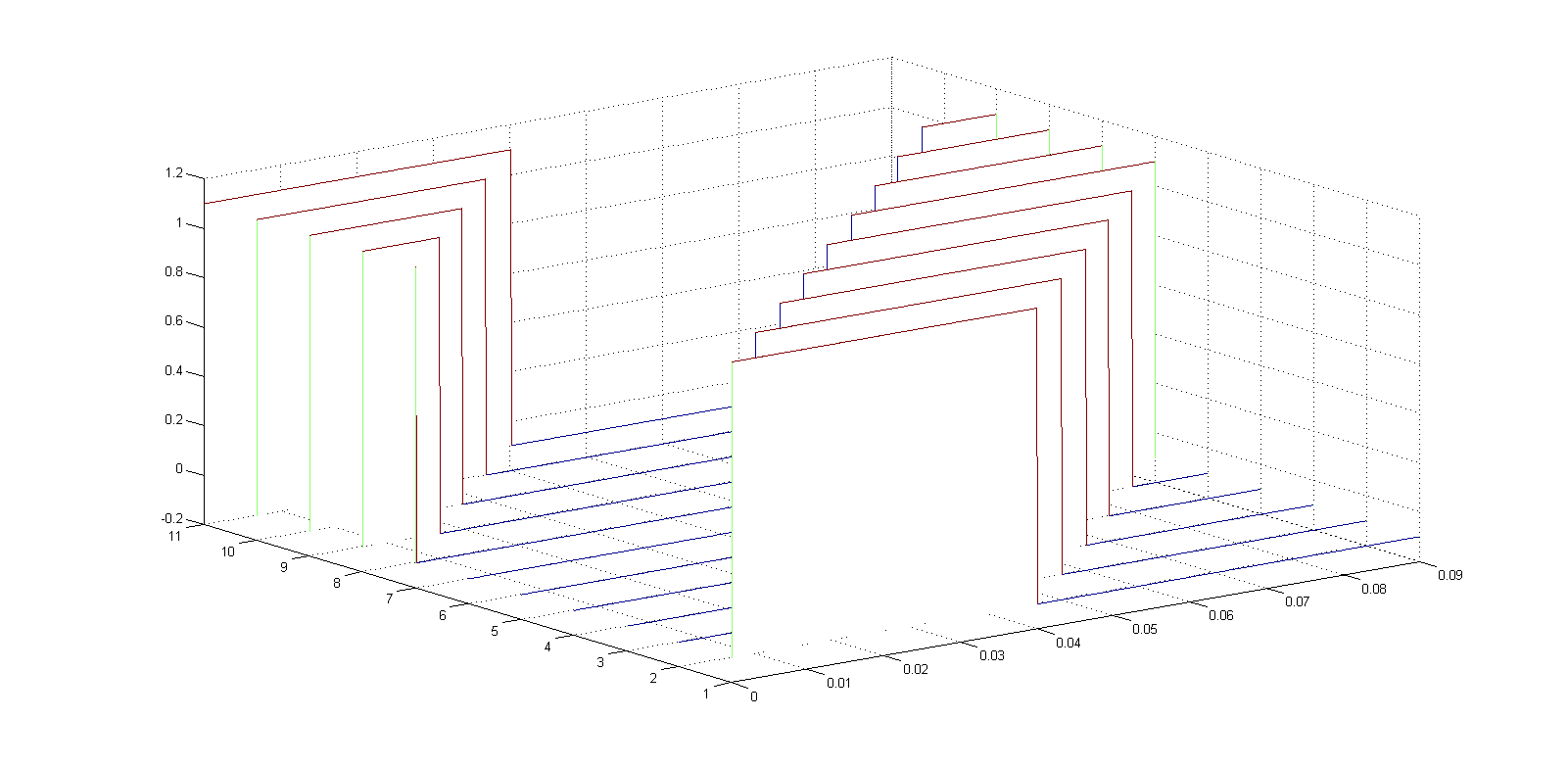}
\end{tabular}
\end{center}
\caption{Figure relative à la question~\ref{it:fig6}. Succession des trames représentées cette fois en fonction d'une base de temps propre à la trame. }
\label{fig6c}
\end{figure}

\begin{figure}
\begin{center}
\begin{tabular}{cc}
\includegraphics[width=0.5\linewidth]{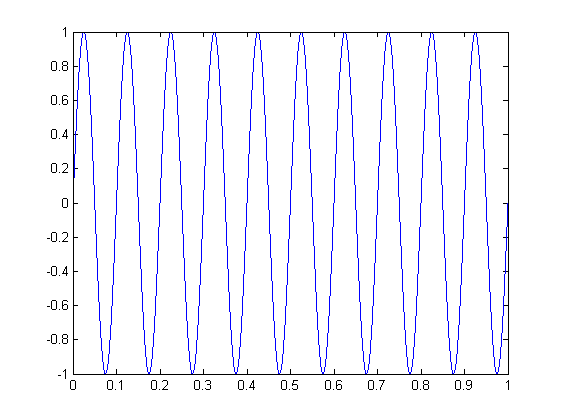} &\includegraphics[width=0.5\linewidth]{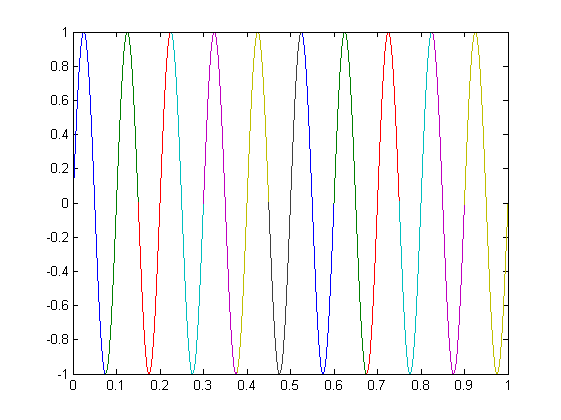}\\
\includegraphics[width=0.5\linewidth]{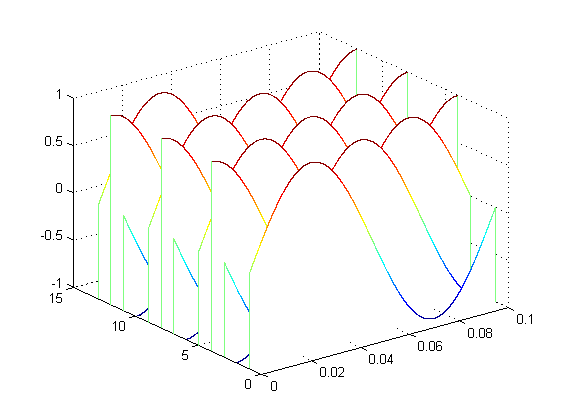} & \includegraphics[width=0.5\linewidth]{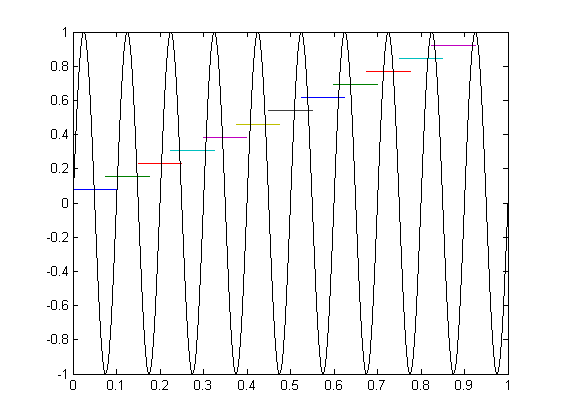}
\end{tabular}
\end{center}
\caption{En haut à gauche~: signal {\tt x} en fonction du temps. En haut à droite~: trames affichées avec des couleurs. En bas à gauche~: trames avec base de temps propre à chaque trame. En bas à droite~: 
 signal {\tt x} en noir et indication colorée associée à chaque trame.}
\label{fig2}
\end{figure}

\begin{enumerate}
\deb
\item Découpez le signal {\tt x} en trames de $100$ms sans chevauchement, et donnez les quatre visualisations en 
utilisant les programmes {\tt decoupe\_signal} et {\tt visualisation} qui sont disponibles.

Pour la suite, nous utiliserons  les notations suivantes, en référence aux notations utilisées en annexe~\ref{ssec:decoupe_signal}. 
\begin{itemize}
\item {\tt nb\_trames}~: nombre de trames, $K$
\item {\tt nb\_ech}~: nombre d'échantillons par trames, $N_K$
\item {\tt chevauchement}~: $\alpha$
\item {\tt duree\_trame}~: durée d'une trame, $T_{x}$
\item {\tt centres\_trames}~: instants associés au centres des différentes trames, $t_k^{(ce)}$ 
\item {\tt ech\_trames}~: échelle de temps associées au différentes trames, $t_{k,n}$. 
\end{itemize}

\begin{corr}
\relax \sol
\begin{verbatim}
clear
f0=10; duree=1; fe=44100; t=0:1/fe:(duree-1/fe); x=sin(2*pi*f0*t); 
chevauchement=0;  duree_trame=100e-3;
[xk,ech_trames,centres_trames]=decoupe_signal(x,fe,duree_trame,chevauchement);
visualisation(x,fe,duree_trame,xk,ech_trames,centres_trames);
\end{verbatim}
\end{corr}

\item Découpez le signal {\tt x} en trames de $100$ms avec $25\%$ de chevauchement, et donnez les quatre visualisations. 
Le résultat avec chevauchement est représenté sur la figure~\ref{fig2}. 
\finenum
\end{enumerate}

\begin{corr}
\relax \sol
\begin{verbatim}
clear
f0=10; duree=1; fe=44100; t=0:1/fe:(duree-1/fe); x=sin(2*pi*f0*t); 
chevauchement=0.25;  duree_trame=100e-3;
[xk,ech_trames,centres_trames]=decoupe_signal(x,fe,duree_trame,chevauchement);
visualisation(x,fe,duree_trame,xk,ech_trames,centres_trames);
\end{verbatim}
\end{corr}

\begin{enumerate}
\deb
\item\label{it:fig6} 
La figure~\ref{fig6c} 
représente différentes trames découpées sans chevauchement. La fréquence 
d'échantillonnage utilisée de $8$kHz. Essayez de retrouver les paramètres et le signal de départ
utilisées pour obtenir cette figure. 

\finenum
\end{enumerate}

\begin{corr}
\relax \sol

Le signal recherché est un signal périodique de période $T=100$ms dont le motif sur un 
période est \beqns 1.1\left (\1_{[0,\tau[}(t)-\1_{[\tau,T[}(t)\right )\eeqns avec $\tau=40$ms. Les trames considérées sont de durées 
$90$ms. Il est représenté figure~\ref{fig6d}.
\begin{verbatim}
clear
f0=10; duree=1; fe=8e3; t=0:1/fe:(duree-1/fe); x=1.2*((10*t-floor(10*t))*10<=4)-0.1; 
chevauchement=0;  duree_trame=90e-3;
[xk,ech_trames,centres_trames]=decoupe_signal(x,fe,duree_trame,chevauchement);
visualisation(x,fe,duree_trame,xk,ech_trames,centres_trames);
\end{verbatim}
\begin{figure}
\begin{center}
\begin{tabular}{c}
\includegraphics[width=0.5\linewidth]{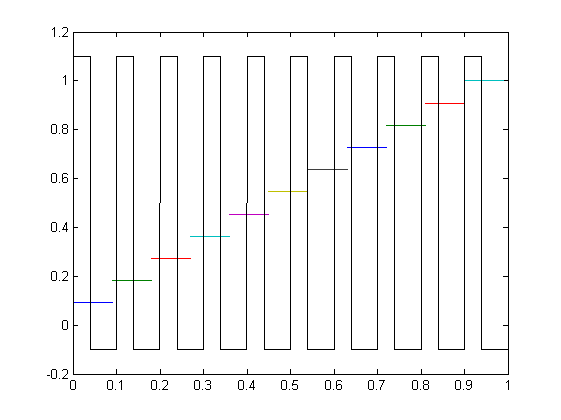}
\end{tabular}
\end{center}
\caption{Figure relative à la question~\ref{it:fig6}. Signal d'origine et indication des durées des trames en couleur. }
\label{fig6d}
\end{figure}
\end{corr}

\begin{enumerate}
\deb
\item Question facultative. La fonction {\tt v\_decoupe\_signal} est chargée de vérifier le bon fonctionnement 
de la fonction {\tt decoupe\_signal}. Essayez de justifier certaines des lignes de codes utilisées dans cette fonction. 
\finenum
\end{enumerate}

\subsection{Détection du début et de la fin d'un son}
\label{ssec:deb}

Ce qu'on appelle la puissance moyenne à court terme (PMCT)~\footnote{{\tt PMCT}
est défini dans \cite{Camastra07} p.~41-42.} et la magnitude moyenne à court terme (MMCT)
sont ainsi définies.
\beqn
P_\tau=\frac{{\int_\Rr x^2(t)w_{T_{x}}(t-\tau)\,dt}}{{\int_\Rr w_{T_{x}}(t-\tau)\,dt}}
\mbox{ et }
M_\tau=\frac{{\int_\Rr \left |x(t)\right|w_{T_{x}}(t-\tau)\,dt}}{{\int_\Rr w_{T_{x}}(t-\tau)\,dt}}
\eeqn
où $w_{Tx}(t)$ est une fenêtre à temps continu de support~\footnote{Le support est un terme mathématique équivalent 
ici à l'intervalle sur lequel le signal est non-nul.} $[-\frac{Tx}{2},\frac{Tx}{2}]$. 

On considère maintenant pour {\tt x} un des sons enregistrés. 

\`A partir du signal découpé temps continu, PMCT se calcule 
aux centres des différentes trames et 
non plus pour toutes les valeurs de $\tau$.
\beqns
\PMCT_k=P_{t_k^{(ce)}}\mbox{ et }\MMCT_k=M_{t_k^{(ce)}}
\eeqns


\`A partir d'un signal découpé temps discret, {\tt PMCT} et {\tt MMCT} sont alors ainsi définis.
\beqns
\PMCT_k=\frac{{\sum_{n=0}^{N_K-1} x_k^2[n]w_{N_K}[n-\frac{N_K}{2}]}}{{\sum_{n=0}^{N_K-1} w_{N_K}[n-\frac{N_K}{2}]}}
\mbox{ et }
\MMCT_k=\frac{{\sum_{n=0}^{N_K-1} \left |x_k[n]\right|w_{N_K}[n-\frac{N_K}{2}]}}{{\int_{n=0}^{N_K-1} w_{N_K}[n-\frac{N_K}{2}]}}
\eeqns
où $w_{N_K}$ est une fenêtre de taille $N_K$ centrée en zéro. 

En pratique, les différences entre {\tt PMCT} et {\tt MMCT} n'étant pas très significatives, on utilisera plutôt {\tt PMCT}.

\begin{enumerate}
\deb
\item Découpez {\tt x} en trames de $30$ms avec $0.25$ de chevauchement. Observez les quatre visualisations et en particulier 
l'importance du silence. 

\begin{corr}
\relax \sol
\begin{verbatim}
clear
horiz=@(u)u(:)';
[y,fe]=lireSon('..\donnees\un\',0); 
t_trame=30e-3; chevauchement=0.25;
[signal,ech_trames,c_trame]=decoupe_signal(y,fe,t_trame,chevauchement);
visualisation(y,fe,t_trame,signal,ech_trames,c_trame);
\end{verbatim}
\end{corr}
\item Calculez pour chaque trame {\tt PMCT}. 
Représentez sur un graphe le signal superposé à {\tt PMCT} en multipliant {\tt PMCT} par 
un facteur de sorte que le maximum de {\tt PMCT} coïncide avec le maximum de la valeur absolue du signal {\tt x}. 
Observez qu'on peut considérer {\tt PMCT} comme une approximation de l'enveloppe énergétique 
de {\tt x}~\footnote{voir page~3, section~2.1 dans~\cite{Peeters04}}. 
La figure~\ref{fig11} représente ces signaux de façon séparée. 

En pratique 
$w_n$ est une fenêtre égale au nombre d'indices dans chaque trame. 
Ici on utilise la fenêre de Hamming. 

\begin{figure}[htbp]
\begin{center}
\begin{tabular}{c}
\includegraphics[width=0.4\linewidth]{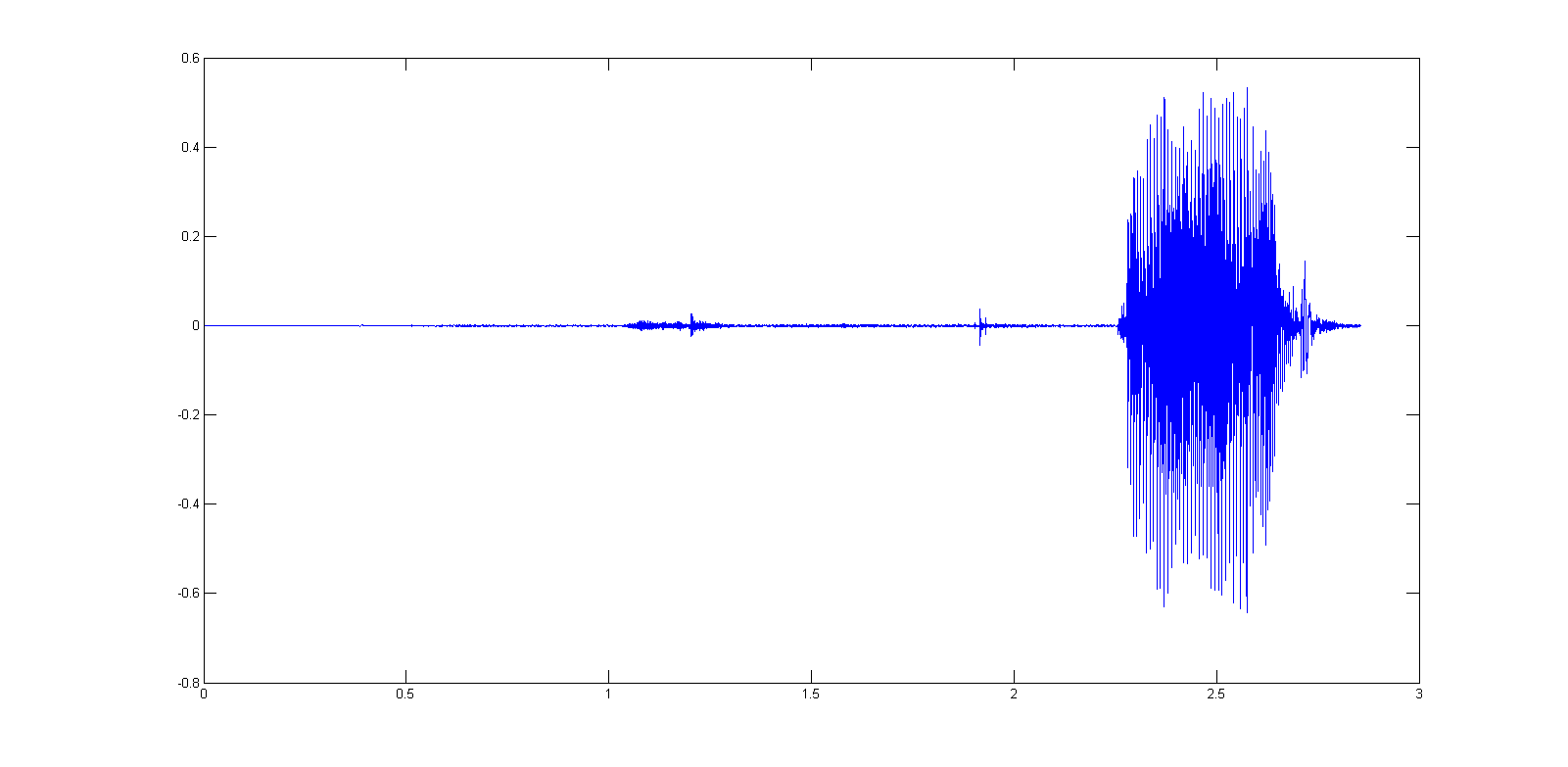}\\
\includegraphics[width=0.4\linewidth]{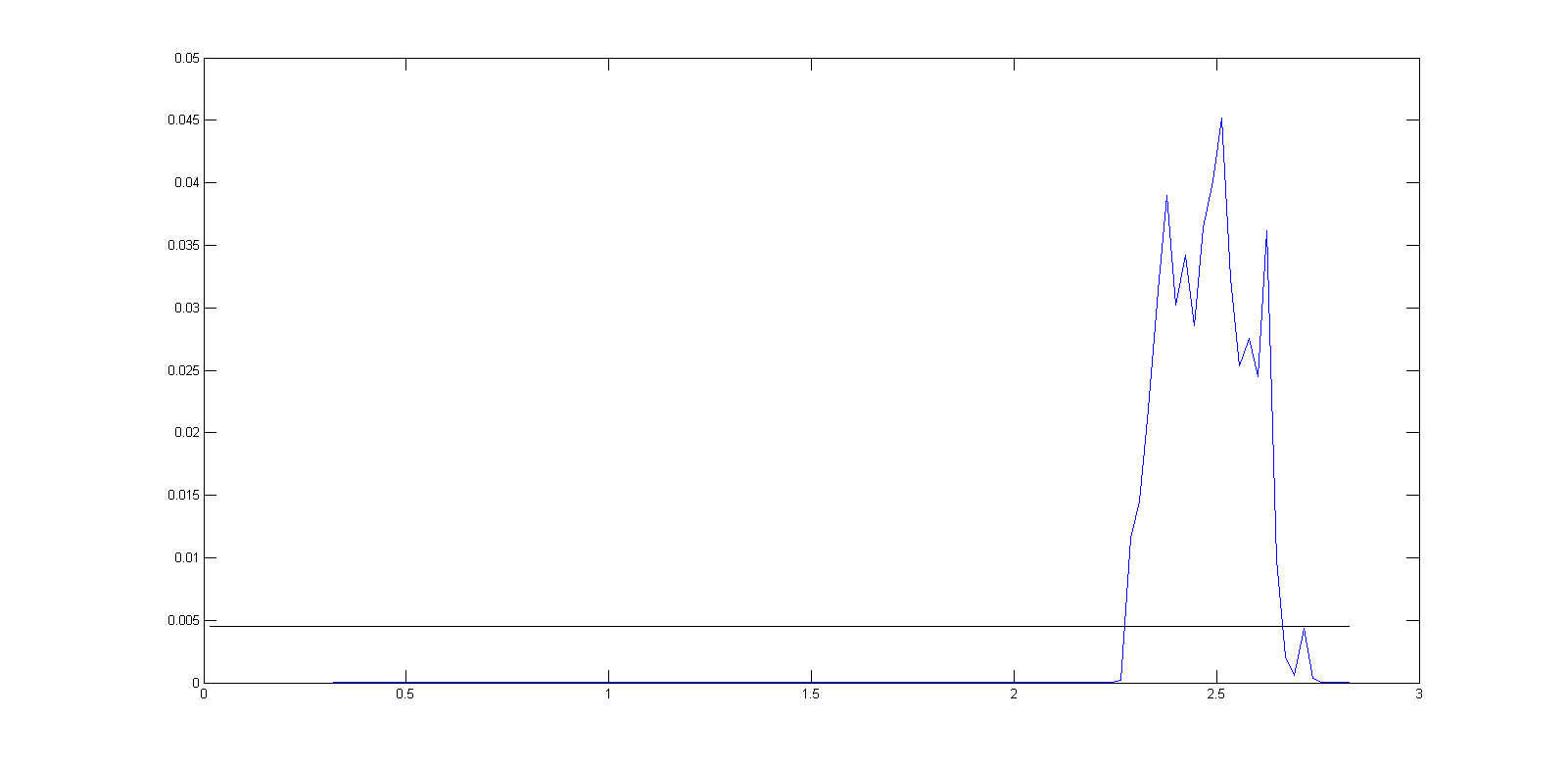}\\
\end{tabular}
\end{center}
\caption{En haut~: le signal sonore est représenté en fontion du temps et en bas la puissance moyenne 
en fonction du temps pour chaque trame. La ligne noire indique $10$\% du maximum.}
\label{fig11}
\end{figure}

\begin{corr}
\relax \sol 
\begin{verbatim}
clear
[x,fe]=lireSon('..\donnees\un\',0);
[xk,ech_trames,centres_trames]=decoupe_signal(x,fe,30e-3,0.25);
K=size(xk,1); NK=size(xk,2); 
fenetre=window(@hamming,NK)'; S_fen=sum(fenetre);
PMCT=zeros(K,1);
for k=1:K
  PMCT(k)=sum(xk(k,:).^2.*fenetre)/S_fen;
end
PMCT_mod=PMCT/max(PMCT)*max(abs(x));  
%la multiplication par max(abs(x)) est juste 
%là pour mettre à la même échelle sur le graphique
t=0:1/fe:(length(x)-1)/fe;
figure(1); plot(centres_trames,PMCT_mod,t,x)
\end{verbatim}
\end{corr}

\item\label{it:c_PMCT} Créez une fonction {\tt c\_PMCT} qui calcule {\tt PMCT} à partir du signal découpé
suivant la syntaxe suivante. 
\begin{verbatim}
function PMCT=c_PMCT(xk)
\end{verbatim}
\begin{corr}
\relax \sol
\begin{verbatim}
function PMCT=c_PMCT(xk)
  K=size(xk,1); NK=size(xk,2); 
  fenetre=window(@hamming,NK)'; S_fen=sum(fenetre);
  PMCT=zeros(K,1);
  for k=1:K
    PMCT(k)=sum(xk(k,:).^2.*fenetre)/S_fen;
  end
\end{verbatim}
\end{corr}

\finenum
\end{enumerate}
Pour déterminer le début et la fin, le seuil considéré est fixé à $10\%$ de la valeur maximale, et 
le temps de début $t_{de}, t_{dm}$ et de fin $t_{fe}, t_{fm}$ sont, respectivement, la première et dernière occurence où le seuil est dépassé. Ce seuil est indiqué par une ligne noire sur la figure~\ref{fig11} en bas. 

\beqns
s_e=\frac{1}{10}\max_k({\PMCT_k})
\eeqns

On cherche les trames associées à un son particulier différent du silence. 

\beqn
k_{de}=\underset{n}{\min}\,\{n|\,\PMCT_n>s_e\}
\mbox{ et }
k_{fi}=\underset{n}{\max}\,\{n|\,\PMCT_n>s_e\}
\eeqn

Ces indices correspondent à des instants particuliers que l'on peut retrouver sur la figure~\ref{fig11} comme 
étant les intersections entre la ligne noire et la courbe bleue. 

\beqn
t_{de}=t_{k_{de}}^{(ce)}\mbox{ et }t_{fi}=t_{k_{fi}}^{(ce)}
\eeqn
où $t_k^{(ce)}$ désigne l'instant associé au centre de la trame $k$.

\begin{enumerate}
\deb
\item\label{it:deb} Calculez les instants à partir duquel et en deça duquel le signal découpé 
a une puissance moyenne à court terme supérieure à $10\%$ de sa valeur maximale. 
Représentez le signal entre ces deux instants. 

\begin{corr}
\relax \sol
\begin{verbatim}
clear 
s_e=max(PMCT)/10;
k_de=find(PMCT>=s_e,1,'first');
k_fi=find(PMCT>=s_e,1,'last');
xk=xk(k_de:k_fi,:); 
ech_trames=ech_trames(k_de:k_fi,:); 
centres_trames=centres_trames(k_de:k_fi)-centres_trames(k_e);
PMCT=PMCT(k_de:k_fi);
MMCT=MMCT(k_de:k_fi);
figure(1); plot(centres_trames,PMCT)
figure(2); plot(centres_trames,MMCT)
\end{verbatim}
\end{corr}

\item\label{it:supprime_silence} 
Réalisez une fonction qui modifie le signal découpé en supprimant les trames antérieures $t_{de}$ et postérieures
à $t_{fi}$. La syntaxe de cette fonction est 
\begin{verbatim}
function [xk,ech_trames,centres_trames]=supprime_silence(xk,ech_trames,centres_trames)
\end{verbatim}
\begin{corr}
\relax \sol
\begin{verbatim}
function [xk,ech_trames,centres_trames]=supprime_silence(xk,ech_trames,centres_trames)
  K=size(xk,1); NK=size(xk,2); 
  fenetre=window(@hamming,NK)'; S_fen=sum(fenetre);
  PMCT=zeros(K,1);
  for k=1:K
    PMCT(k)=sum(xk(k,:).^2.*fenetre)/S_fen;
  end
  s_e=max(PMCT)/10;
  k_de=find(PMCT>=s_e,1,'first');
  k_fi=find(PMCT>=s_e,1,'last');
  xk=xk(k_de:k_fi,:); 
  ech_trames=ech_trames(k_de:k_fi,:); 
  centres_trames=centres_trames(k_de:k_fi)-centres_trames(k_e);
\end{verbatim}
\end{corr}
\finenum
\end{enumerate}

La normalisation du signal a pour but de rendre comparable 
un son qui serait prononcé de façon forte et un son prononcé à voix plus basse. 
Cette normalisation consiste à imposer que la puissance moyenne soit fixe et en l'occurrence égale à $1$, mais 
cette étape se fait après avoir retiré les silences du signal.~\footnote{Si on effectuait cette normalisation du 
signal avant de retirer les silences, un son avec beaucoup de silences 
que l'on aurait normalisé  avec ses silences, aurait une 
puissance moyenne plus forte, une fois retirés les silences.}

La puissance moyenne du signal complet~\footnote{En pratique elle peut être calculé 
en appliquant moyennant le résultat de la fonction {\tt c\_PMCT}.} est.
\beqn
P=\frac{1}{KN_K}\left ({\sum_k{\sum_n \left (x_k[n]\right)^2w_{N_K}[n]}}\right )
\eeqn
La normalisation consiste à diviser chaque trame du signal par $\sqrt{P}$
\beqn
x_k[n]\mapsto \frac{x_k[n]}{\sqrt{P}}
\eeqn

\begin{enumerate}
\deb
\item On considère ici deux sons associés respectivement à {\em un} et {\em trois}. 
Retirez les silences de ces deux sons et représentez sur un premier graphe les deux évolutions
des puissances moyennes à court termes lorsque les signaux sont normalisés~; et sur 
un deuxième graphe les deux évolutions des puissances moyennes à court terme lorsque 
les signaux ne sont pas normalisés. 
\begin{corr}
\relax \sol
\begin{verbatim}
clear
[x1,fe]=lireSon('..\donnees\un\',0); 
[x3,fe]=lireSon('..\donnees\trois\',0); 
[xk1,ech1_trames,centres1_trames]=decoupe_signal(x1,fe,30e-3,0.25);
[xk1,ech1_trames,centres1_trames]=supprime_silence(xk1,ech1_trames,centres1_trames);
[xk3,ech3_trames,centres3_trames]=decoupe_signal(x3,fe,30e-3,0.25);
[xk3,ech3_trames,centres3_trames]=supprime_silence(xk3,ech3_trames,centres3_trames);
P_xk1=mean(c_PMCT(xk1)); xk1a=xk1/sqrt(P_xk1); xk1b=xk1;
P_xk3=mean(c_PMCT(xk3)); xk3a=xk3/sqrt(P_xk3); xk3b=xk3;
PMCT1a=c_PMCT(xk1a); PMCT1b=c_PMCT(xk1b);
PMCT3a=c_PMCT(xk3a); PMCT3b=c_PMCT(xk3b);
figure(1); plot(centres1_trames,PMCT1a,'b-',centres3_trames,PMCT3a,'r-');
figure(2); plot(centres1_trames,PMCT1b,'b-',centres3_trames,PMCT3b,'r-');
\end{verbatim}
On peut alors vérifier que la puissance du signal normalisé est de $1$. 
\begin{verbatim}
mean(c_PMCT(xk1a)),
\end{verbatim}
\end{corr}

\finenum
\end{enumerate}

\fbox{\begin{minipage}{0.8\linewidth}
\imp
Une fonction {\tt preparation} est disponible~\footnote{{\tt preparation} est expliquée en annexe~\ref{sec:fct}.}
elle contient une partie des traitements proposés. Mais c'est à vous de la compléter pour prendre en compte d'autres 
traitements sans modifier les paramètres en entrée et en sortie.~\footnote{La première ligne de la fonction 
est {\tt function [xk,ech\_trames,centres\_trames]} {\tt =preparation(x,fe,duree\_trame,chevauchement)}, les 
variables entre crochets sont les paramètres de sortie et les variables entre parenthèses sont les paramètres 
d'entrée.}

\end{minipage}}

\section{Séance 7\\ Obtenir une classification grâce à des descripteurs}

L'objectif au cours de cette séance est de classer les sons à partir de descripteurs. Un descripteur peut 
soit être une valeur pour l'ensemble du signal, soit un ensemble de valeurs, une pour chaque trame.


\subsection{Obtenir un descripteur par son à partir d'un descripteur par trame}

Il est possible de définir des descripteurs pour l'ensemble du son à partir de descripteurs par trames. 
$\mu^{(j)}$ et $\sigma^{(j)}$ sont définis dans l'équation~(\ref{eq:mu}) et dans (\ref{eq:sigma}), pages~\pageref{eq:mu} 
et~\pageref{eq:sigma}. 
\beqns
\mu^{(j)}=\frac{1}{K}{\sum_{k=1}^K \bz_k^{(j)}}
\eeqns
\beqns
\sigma^{(j)}=\sqrt{\frac{1}{K}{\sum_{k=1}^K \left (\bz_k^{(j)} -\mu^{(j)}\right)^2}}
\eeqns

On pourrait aussi envisager estimer un coefficient d'asymétrie ou un coefficient d'aplatissement
\beqn
\gamma_1^{(j)}=\frac{1}{(\sigma^{(j)})^3}{\frac{1}{K}{\sum_{k=1}^K \left (\bz_k^{(j)} -\mu^{(j)}\right)^3}}\mbox{ et }
\gamma_2^{(j)}=\frac{1}{(\sigma^{(j)})^4}{\frac{1}{K}{\sum_{k=1}^K \left (\bz_k^{(j)} -\mu^{(j)}\right)^4}}
\eeqn

\`A partir de descripteurs de sons, on peut définir une distance entre sons. Par exemple 
\beqns
d(x_a,x_b)=\sqrt{{\sum_{j=1}^J \left( \mu_a^{(j)}-\mu_b^{(j)}\right )^2+\left( \sigma_a^{(j)}-\sigma_b^{(j)}\right )^2}}
\eeqns

\begin{enumerate}
\deb
\item\label{it:fig4} On considère ici le descripteur par trame {\tt PMCT}~\footnote{Ce descripteur est calculé avec {\tt s\_c\_PMCT}.}
\`A partir de ce descripteur, générez deux descripteurs par son en utilisant la moyenne et l'écart-type. 
Représentez sur un graphique 2D l'ensemble des données en plaçant en 
bleu les sons associés à un, vert les sons associés à deux et rouge les sons associés à trois. 
La position de chaque son apparaît sous la forme d'un point dont l'ordonnée est déterminée 
par l'estimation de l'écart-type de {\tt PMCT} et l'abscisse est déterminée par 
la moyenne de {\tt PMCT}. 

\ind {\em
Pour réaliser cette simulation, vous pouvez utiliser le programme tout fait 
{\tt s\_c\_PMCT}. 
Pour calculer la moyenne et l'écart-type, il existe les fonctions {\tt mean} et {\tt std}. 
}
\begin{figure}
\begin{center}
\begin{tabular}{c}
\includegraphics[width=0.5\linewidth]{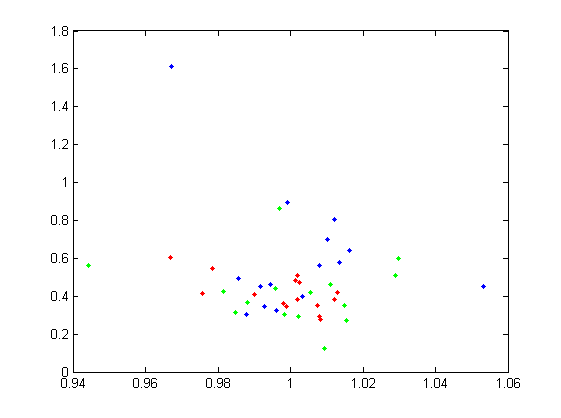}
\end{tabular}
\end{center}
\caption{Graphe relatif à la question~\ref{it:fig4}. Chaque point correspond à un
 son coloré respectivement en rouge, vert et bleu, en fonction de ce que le son correspond à un, deux ou trois. 
La position du son est déterminée en abscisse par la moyenne de {\tt PMCT} et en ordonnée à par l'écart-type 
de {\tt PMCT}.}
\label{fig4}
\end{figure}

\begin{corr}
\relax \sol
\begin{verbatim}
clear
rep_l={'..\donnees\un\','..\donnees\deux\','..\donnees\trois\'};
duree_trame=30e-3; chevauchement=0.25;
X1=[]; X2=[]; Y=[];
for rep_=1:length(rep_l)
  rep=rep_l{rep_};
  M=lireSon(rep,-1);
  for m=1:M
    [x,fe]=lireSon(rep,m);
    [xk,ech_trames,centres_trames]=preparation(x,fe,duree_trame,chevauchement);
    X1=[X1;mean(s_c_PMCT(xk))];
    X2=[X2; std(s_c_PMCT(xk))];
    Y=[Y;rep_];
  end
end
ind1=find(Y==1); ind2=find(Y==2); ind3=find(Y==3);
figure(1); plot(X1(ind1),X2(ind1),'r.',X1(ind2),X2(ind2),'g.',X1(ind3),X2(ind3),'b.');
\end{verbatim}
\end{corr}
\finenum
\end{enumerate}

La distance~\footnote{La distance euclidienne dans un graphe de deux points de coordonnée 
$(x_1,y_1)$ et $(x_2,y_2)$ est définie par $\sqrt{(x_2-x_1)^2+(y_2-y_1)^2}$. Dans certaines 
conditions ce calcul coïncide approximativement avec un coefficient de proportionnalité à la distance sur le papier 
mesurée par une règle.}
 entre deux points de la figure~\ref{fig4} pour cette application s'écrit ainsi. 
\beqn\label{eq:d1}
d_1(x_1,x_2)=\sqrt{(\mu_1-\mu_2)^2+(\sigma_1-\sigma_2)^2}
\eeqn
où $x_1$ et $x_2$ désignent les signaux associés à deux sons et $\mu_1,\mu_2,\sigma_1,\sigma_2$ désignent 
les quantités calculées avec (\ref{eq:mu}) et (\ref{eq:sigma}) sur chacun de ces signaux. 

\begin{enumerate}
\deb
\item\label{it:distance1} Réalisez une fonction notée {\tt distance1} qui à partir de deux sons détermine 
la distance entre ces sons définie par (\ref{eq:d1}).
La durée de chaque trame est de $30$ms et le chevauchement est de $0.25$. 
Vous utiliserez la syntaxe suivante
\begin{verbatim}
function d=distance1(rep1,m1,rep2,m2)
\end{verbatim}

\ind {\em
La fonction {\tt v\_distance} définie en annexe~\ref{ssec:verif} teste sur un exemple pris au hasard si 
cette fonction vérifie deux propriétés d'une distance la symétrie et l'inégalité triangulaire. 
}

\begin{corr}
\relax \sol
\begin{verbatim}
function d=distance1(rep1,m1,rep2,m2)
  [x1,fe]=lireSon(rep1,m1); [x2,fe]=lireSon(rep2,m2);
  duree_trame=30e-3; chevauchement=0.25; 
  xk1=preparation(x1,fe,duree_trame,chevauchement);
  xk2=preparation(x2,fe,duree_trame,chevauchement);
  PMCT1=s_c_PMCT(xk1); PMCT2=s_c_PMCT(xk2);
  mu1=mean(PMCT1);     mu2=mean(PMCT2);
  sigma1=std(PMCT1);   sigma2=std(PMCT2);
  d=sqrt((mu1-mu2)^2+(sigma1-sigma2)^2);
end
\end{verbatim}
\end{corr}

\finenum
\end{enumerate}

\subsection{Comment tester la performance d'un algorithme de classification à partir d'une distance entre sons}
\label{ssec:kNN}

La figure~\ref{fig4} permet d'illustrer le fonctionnement de l'algorithme de classification proposé~\footnote{C'est
l'algorithme du kNN avec $k=1$, (en français, recherche des plus proches voisins)}. 
Un son donné est représenté par un point $M$ dans ce graphe et l'objectif est de déterminer sa couleur (rouge, vert ou bleu). 
On cherche le point le plus proche de $M$ et l'algorithme estime que la couleur de $M$ est la couleur du point le plus 
proche de $M$. 
Cette tâche qui semble simple est effectuée en quatre étapes. 
\begin{itemize}
\item[1.] Les points dont la couleur est connue sont regroupés en fonction de leur couleur 
en trois groupes de points.
\item[2.] On définit la distance entre $M$ et un ensemble de points comme étant la plus petite de toutes 
les distance entre $M$ et chacun des points de cet ensemble. 
\item[3.] Muni de cette définition de la distance entre un point et un ensemble de points, on évalue trois distances,
 celle de $M$ avec le groupe des points de couleur rouge,
de $M$ avec le groupe vert, et de $M$ avec le groupe bleu.
\item[4.] L'algorithme affecte à $M$ la couleur correspondant à la couleur du groupe de point ayant la distance 
la plus faible avec $M$. 
\end{itemize}

Formellement, les quatre étapes sont ainsi décrites. 
\begin{itemize}
\item[1.] $\MX_1,\MX_2,\MX_3$ désignent les sons associés à un, deux et trois.  
\item[2.] $d(\bx,\MX)$  est une distance entre un son noté $\bx$ et un ensemble de sons $\MX$
\beqn
d(\bx,\MX)=\underset{\bx'\in \MX}{\min}\,d(\bx,\bx')
\eeqn

\item[3.] Les trois distances entre $\bx$ et $\MX_1,\MX_2,\MX_3$ sont
\beqns
d(\bx,\MX_{i})
\eeqns
avec $i\in\{1,2,3)$
\item[4.] Le fait que le son $\bx$ affecté à un, deux ou trois est respectivement noté par $\widehat{\bx}=1$, $\widehat{\bx}=2$ ou  
$\widehat{\bx}=3$.
Cette affectation est déterminée par 
\beqn
\widehat{\bx}=\underset{i\in\{1,2,3\}}{\mbox{argmin}}\,d(\bx,\MX_{i})
\eeqn
\end{itemize}

\begin{enumerate}
\deb
\item\label{it:genere_un_pb_classification} \'Etant donné un son particulier on cherche à savoir si phonétiquement il ressemble plus à  $1,2,3$. 
Pour cela choisissez un son au hasard et testez sa ressemblance avec les autres sons de la même catégorie 
et avec ceux d'une autre catégorie en utilisant la distance définie à la question~\ref{it:distance1}.

\ind {\em
La fonction {\tt genere\_un\_pb\_classification} définie en annexe~\ref{ssec:fct_code}
fabrique des répertoires avec les fichiers pour réaliser la classification. 
La fonction {\tt s\_distance1} associée à la question~\ref{it:distance1} est aussi disponible en annexe~\ref{ssec:fct_ss_code}. 
}

\begin{corr}
\relax \sol
\begin{verbatim}
clear
rep_l={'..\donnees\un\','..\donnees\deux\','..\donnees\trois\'}; 
rep_simulation='..\simulation\';
rep_requete='..\simulation\requete\';
rep_apprend_l={'..\simulation\un\','..\simulation\deux\','..\simulation\trois\'}; 
i_ref=genere_un_pb_classification(rep_l,rep_simulation,rep_requete,rep_apprend_l);
d_min=zeros(1,3);
for i=1:3
  M=lireSon(rep_apprend_l{i},-1); 
  d=-ones(1,M);
  for m=1:M
    d(m)=s_distance1(rep_requete,1,rep_apprend_l{i},m);
  end
  d_min(i)=min(d);
end
[~,ind]=min(d_min);
disp(['d_min=',num2str(d_min),' son predit = ',num2str(ind),' verite=',num2str(i_ref)]),
\end{verbatim}
\end{corr}
\item\label{it:predit} Réalisez une fonction appelée {\tt predit} qui étant donné un son identifié par {\tt rep\_requete,m}
et un ensemble d'apprentissage {\tt rep\_apprend\_l} et une fonction {\tt distance}, prédit l'appartenance 
de ce son à un, deux ou trois en fonction du son le plus proche de la requête. 
La syntaxe de cette fonction est 
\begin{verbatim}
function y=predit(rep_requete,rep_apprend_l,distance)
\end{verbatim}
\begin{corr}
\relax \sol
\begin{verbatim}
function y=predit(rep_requete,m,rep_apprend_l,distance)
%cette fonction predit le son identifie par rep_requete et m en 
%fonction du son le plus proche dans rep_apprend_l 
%pour la fonction distance dont la syntaxe est
%function d=distance(rep1,m1,rep2,m2)
%Un exemple d'utilisation de la fonction predit est 
% rep_l={'..\donnees\un\','..\donnees\deux\','..\donnees\trois\'}; 
% rep_simulation='..\simulation\';
% rep_requete='..\simulation\requete\';
% rep_apprend_l={'..\simulation\un\','..\simulation\deux\','..\simulation\trois\'}; 
% i_ref=genere_un_pb_classification(rep_l,rep_simulation,rep_requete,rep_apprend_l);
% y=predit(rep_requete,1,rep_apprend_l,@s_distance1);
% disp(['y=',num2str(y),' i_ref=',num2str(i_ref)])
  d_min=zeros(1,3);
  for i=1:3
    Ma=lireSon(rep_apprend_l{i},-1); 
    d=-ones(1,Ma);
    for ma=1:Ma
      d(ma)=distance(rep_requete,m,rep_apprend_l{i},ma);
    end
    d_min(i)=min(d);
    if ~(min(d)>=0) error('predit'), end
  end
  [~,y]=min(d_min);
end
\end{verbatim}
\end{corr}

\finenum
\end{enumerate}

Une appréciation de la performance d'un algorithme de classification requière 
d'une part une façon de mesurer la performance. On définit la sensibilité globale ({\em overall accuracy})
\beqn
\mbox{OA}=\frac{1}{{\sum_i M_i^{(r)}}}{\sum_{i=1}^3{\sum_{m=1}^{M_i^{(r)}} \1\left (\widehat{\bx}_{i,m}=i\right)}}
\eeqn
avec les notations 
suivantes.
\begin{itemize}
\item  $i$ désigne un type de son, un, deux ou trois
\item  $M_i^{(r)}$ désignent le nombre de sons testés de type $i$ (i.e. qui à la fois sont 
de type $i$ et 
font parti de l'ensemble des requêtes).
\beqn
M_i^{(r)}={\sum_{m=1}^{M^{(r)}}\1(y_m=i)}
\eeqn
\item  $\widehat{\bx}_{i,m}$ désigne la prédiction qui 
est faite par l'algorithme de classification testée pour un son qui est de type $i$. 
\item $\1\left (\widehat{\bx}_{i,m}=i\right)$ vaut $1$ si
$\widehat{\bx}_{i,m}=i$ et $0$ sinon. 
\item  $\widehat{\bx}_{m}$ désigne la prédiction qui 
est faite par l'algorithme de classification testée.
\end{itemize}
Ainsi on peut voir $\mbox{OA}$ comme la proportion de réponses correctes de l'algorithme de classification, tous sons 
confondus. 

La matrice de confusion est définie 
par 
\beqn
C_{ij}={\sum_{m=1}^{M^{(r)}}\1(y_m=i)\1(\widehat{\bx}_m=j)}
\eeqn
OA peut alors s'exprimer en fonction de la matrice de confusion 
\beqn
\mbox{OA}=\frac{{\sum_i\,C_{ii}}}{{\sum_{ij}\,C_{ij}}}
\eeqn
$M_i^{(r)}$ peut s'exprimer en fonction de la matrice de confusion 
\beqn
M_i^{(r)}={\sum_j C_{ij}}
\eeqn

D'autre part, une appréciation de la performance d'un algorithme requière de faire un plus grand nombre de 
tests que lors de la question~\ref{it:genere_un_pb_classification}.
Une façon de mettre en place un plus grand nombre de test consiste à faire une validation croisée. 

 L'ensemble des sons disponibles sont divisés  en $K$ sous-ensembles de tailles similaires notées 
${\sum_{i=1}^3 M_{k,i}}$. Ces sous-ensembles sont aussi appelés partitions. 
Une de ces partition est utilisée comme un ensemble de requêtes. Les $K-1$ autres partitions sont regroupées 
pour former un ensemble d'apprentissage. Une première valeur de sensibilité globale,
 $\OA_1$ est alors calculée en autorisant 
l'algorithme de classification à n'utiliser que  l'ensemble d'apprentissage pour réaliser ces prédictions vis-à-vis 
de toutes les requêtes. 
Puis on répète cette opération en sélectionnant un autre ensemble de requêtes parmi les $K-1$ partitions non-encore testées. 
Le première partition testée ainsi que les $K-2$ partitions non-testées sont utilisées comme ensemble d'apprentissage. Ceci 
permet de calculer une deuxième valeur de sensibilité globale $\OA_2$. Cette opération est répétée au totale $K$ fois de façon à 
ce que chacune des $K$ partitions servent une fois d'ensemble de requêtes et chacune de ces opérations permet 
le calcul d'une valeur de sensibilité globale notée $\OA_k$. L'estimation de la sensibilité globale consiste à moyenner ces valeurs $\OA_k$ pour obtenir 
une estimation notée $\widehat{\OA}_K$. 

L'estimateur ainsi obtenue avec la validation croisée s'écrit ainsi.~\footnote{L'expression du dénominateur 
est ici plus simple parce que on peut remarquer que 
${\sum_{k=1}^K{\sum_{i=1}^3{\sum_{m=1}^{M_{k,i}}1}}}=M$.} 
\beqn
\widehat{\OA}_K=\frac{1}{M}{\sum_{k=1}^K{\sum_{i=1}^3{\sum_{m=1}^{M_{k,i}} \1\left (\widehat{\bz}_{k,i,m}=i\right)}}}
\eeqn

\begin{enumerate}
\deb
\item\label{it:genere_validation_croisee} 
Utilisez une validation croisée avec $K=5$ en utilisant la distance définie à la question~\ref{it:distance1}
et l'algorithme défini préalablement à la question~\ref{it:genere_un_pb_classification}.
Estimez la  sensibilité globale de cet algorithme de classification avec cette distance.

\ind {\em
La fonction {\tt genere\_validation\_croisee},  définie en annexe~\ref{ssec:fct_code}, permet 
de générer des répertoires contenant les fichiers associés à la validation croisée $k$.
}

\fbox{\parbox{0.8\linewidth}{
\imp
La fonction {\tt lireSon} permet de simplifier la programmation, elle présente un inconvénient,
parce qu'elle utilise l'ordre des fichiers géré par Windows et susceptible d'être modifié. 
Il semble que cela pose un problème seulement si cet ordre est modifié entre l'appel à la fonction {\tt genere\_pb} 
et l'utilisation des valeurs de {\tt y} générées par cette fonction.
L'appel un grand nombre de fois de la fonction {\tt v\_genere\_pb(rep\_l,rep\_simulation,rep\_requete,rep\_apprend\_l,y)}
ou l'appel à la fonction assez lente {\tt v\_mesure\_sensibilite}
permet de vérifier s'il y a un problème de cette nature. 
}}

\begin{corr}
\relax \sol
\begin{verbatim}
clear
rep_l={'..\donnees\un\','..\donnees\deux\','..\donnees\trois\'}; 
rep_simulation='..\simulation\';
rep_requete='..\simulation\requete\';
rep_apprend_l={'..\simulation\un\','..\simulation\deux\','..\simulation\trois\'}; 
K=5;
genere_pb=genere_validation_croisee(rep_l,rep_simulation,rep_requete,rep_apprend_l,K);
num=0; den=0;
for k=1:K
  y=genere_pb(k);
  v_genere_pb(rep_l,rep_simulation,rep_requete,rep_apprend_l,y);
  Mr=length(y);
  den=den+Mr;
  yp=zeros(size(y));
  for mr=1:Mr
   yp(mr)=s_predit(rep_requete,mr,rep_apprend_l,@s_distance1);
  end
  num=num+sum(yp==y);
end
OA=num/den;
disp(['OA=',num2str(OA)])
\end{verbatim}
Avec $\OA\sim 1/3$, les résultats trouvés montrent que ce classifieur est aussi bon qu'une prédiction aléatoire.
\end{corr}

\item\label{it:mesure_sensibilite}
Réalisez une foncton notée {\tt mesure\_sensibilite} qui mesure la sensibilité globale de l'algorithme 
défini préalablement à la question~\ref{it:genere_un_pb_classification} pour une fonction distance donnée 
et pour une valeur de $K$ étant donné un ensemble de données. 
La syntaxe de cette fonction est 
\begin{verbatim}
function OA=mesure_sensibilite(distance,K)
\end{verbatim}
\begin{corr}
\relax \sol
\begin{verbatim}
function OA=mesure_sensibilite(distance,K)
%cette fonction mesure la sensibilité globale de l'algorithme de classification 
%consistant à chercher le son le plus proche du son requête et à lui attribuer un, deux ou trois 
% en fonction de ce son le plus proche. 
%distance est une fonction qui définit la distance. 
%Sa syntaxe est function d=distance(rep1,m1,rep2,m2)
%K indique le nombre de validations croisées utilisées. 
%Voici un exemple d'utilisation 
%P=mesure_sensibilite(@s_distance1,5); disp(['P=',num2str(P)])
  rep_l={'..\donnees\un\','..\donnees\deux\','..\donnees\trois\'}; 
  rep_simulation='..\simulation\';
  rep_requete='..\simulation\requete\';
  rep_apprend_l={'..\simulation\un\','..\simulation\deux\','..\simulation\trois\'}; 
  genere_pb=genere_validation_croisee(rep_l,rep_simulation,rep_requete,rep_apprend_l,K);
  num=0; den=0;
  for k=1:K
    disp(['K=',num2str(K),' k=',num2str(k)])
    y=genere_pb(k);
    if rand(1)<0.05  v_genere_pb(rep_l,rep_simulation,rep_requete,rep_apprend_l,y); end
    Mr=length(y);
    den=den+Mr;
    yp=zeros(size(y));
    for mr=1:Mr
     yp(mr)=s_predit(rep_requete,mr,rep_apprend_l,distance);
    end
    num=num+sum(yp==y);
  end
  OA=num/den;
end
\end{verbatim}
\end{corr}
\finenum
\end{enumerate}

Il est possible d'avoir une apprèciation plus fine de la performance de l'algorithme en 
analysant son impact sur la détection d'une classe à l'exclusion des deux autres. 
On appelle  précision ($P_i$) et rappel ($R_i$) par classe $i\in\{1,2,3\}$. 

\beqn
P_i=
\frac{\sum_{m=1}^{M_i} \1\left (\widehat{\bz}_{i,m}=i\right)}
{{\sum_{m=1}^{M_1} \1\left (\widehat{\bz}_{1,m}=i\right)}
+{\sum_{m=1}^{M_2} \1\left (\widehat{\bz}_{2,m}=i\right)}
+{\sum_{m=1}^{M_3} \1\left (\widehat{\bz}_{3,m}=i\right)}}
=\frac{C_{ii}}{{\sum_j C_{ji}}}
\eeqn

\beqn
R_i=\frac{1}{M_i}{\sum_{m=1}^{M_i} \1\left (\widehat{\bz}_{i,m}=i\right)}
=\frac{C_{ii}}{{\sum_j C_{ij}}}
\eeqn

$M_1,M_2,M_3$ désignent respectivement le nombre de sons de type $1,2,3$.

Ces notions peuvent être utilisées avec une validation croisée où pour chaque valeur de $k$, on obtient 
une valeur de précision et de rappel par classe notée 
$P_{k,i}$ et $R_{k,i}$, et ces valeurs sont moyennées par rappport à $k$.

\fbox{\parbox{0.8\linewidth}{
\imp
Dans la mesure où ces programmes prennent du temps pour s'exécuter, il est 
tentant d'ouvrir une deuxième session Matlab pour continuer d'autres simulations. 
En fait il est important d'éviter des accès disques simultané. Si une session 
utilise les fonctions {\tt lireSon}, 
{\tt genere\_un\_pb\_classification}, {\tt genere\_validation\_croisee}, 
{\tt mesure\_precision}, {\tt distance}, il ne faudrait pas que 
l'autre session utilise ce genre de programme tout particulièrement 
si c'est en écriture, mais même en lecture ce n'est pas conseillé.
}}

\subsection{Utilisation d'une distance entre trames}

On choisit maintenant $J$ descripteurs ayant une valeur par trames notées $\bz_k=[z_k^{(j)}]_j$. 
Avec la norme euclidienne, il est possible de mesurer la distance entre deux trames aux instants associés à  $k$ et $k'$~:
\beqn
d(\bx_k,\bx'_{k'})=\sqrt{\frac{1}{J}{\sum_{j=1}^J\left  (z_k^{(j)}-z_{k'}^{'(j)}\right)^2}}
\eeqn

\`A partir d'une distance entre trames, si deux sons ont le même nombre de trames $K$, on peut définir 
une distance entre sons 
\beqn
d(\bx,\bx')=\sqrt{{\sum_{k=0}^{K-1}d^2(\bx_k,\bx'_k)}}
\eeqn

\begin{enumerate}
\deb
\item
Choisissez deux sons correspondant au même chiffre et un troisième son associé à un chiffre différent.
Ici le seul descripteur à considérer est 'PMCT'.
Visualisez l'évolution de la distance entre les trames de deux sons associés au même chiffre
et entre les trames de deux sons associés à des chiffres différents. Ici, on ne compare que des trames aux mêmes 
instants comptés à partir de l'instant détecté en section~\ref{ssec:deb} et la rubrique~\ref{it:deb}. 
Affichez aussi la moyenne des distances entre les deux premiers sons et entre le premier et le troisième. 

\ind{\em
La simulation peut se faire en complétant les lignes suivantes.
\begin{verbatim}
duree_trame=30e-3; chevauchement=0.25;
M=lireSon('..\donnees\un\',-1);
M_vect=randperm(M); m1=M_vect(1); m2=M_vect(2); 
[x,fe,~]=lireSon('..\donnees\un\',m1);
[xk1,ech_trames,centres_trames1]=preparation(x,fe,duree_trame,chevauchement);
[x,fe2,~]=lireSon('..\donnees\un\',m2); 
[xk2,ech_trames,centres_trames2]=preparation(x,fe,duree_trame,chevauchement);
...
PMCT1=...
PMCT2=...
PMCT3=...
distancesA=abs(PMCT1-PMCT2);
distancesB=abs(PMCT1-PMCT3);
disp(['moy distA=',num2str(mean(distancesA)),' moy distB=',num2str(mean(distancesB))])
figure(1); plot(centres_trames,distancesA,'b-',centres_trames,distancesB,'r-');
\end{verbatim}

}

\begin{corr}
\relax \sol
\begin{verbatim}
duree_trame=30e-3; chevauchement=0.25;
M=lireSon('..\donnees\un\',-1);
M_vect=randperm(M); m1=M_vect(1); m2=M_vect(2); 
[x,fe,~]=lireSon('..\donnees\un\',m1);
[xk1,ech_trames,centres_trames1]=preparation(x,fe,duree_trame,chevauchement);
[x,fe2,~]=lireSon('..\donnees\un\',m2); 
if ~(fe==fe2)             error('le programme suppose que fe1=fe2'), end
[xk2,ech_trames,centres_trames2]=preparation(x,fe,duree_trame,chevauchement);
[x,fe,~]=lireSon('..\donnees\deux\',0);
[xk3,ech_trames,centres_trames3]=preparation(x,fe,duree_trame,chevauchement);
K1=size(xk1,1); K2=size(xk2,1); K3=size(xk3,1);
K=min([K1,K2,K3]);
centres_trames=centres_trames1(1:K);
PMCT1=s_c_PMCT(xk1(1:K,:)); 
PMCT2=s_c_PMCT(xk2(1:K,:)); 
PMCT3=s_c_PMCT(xk3(1:K,:)); 
distancesA=sqrt((PMCT1-PMCT2).^2);
distancesB=sqrt((PMCT1-PMCT3).^2);
disp(['moy distA=',num2str(mean(distancesA)),' moy distB=',num2str(mean(distancesB))])
figure(1); plot(centres_trames,distancesA,'b-',centres_trames,distancesB,'r-');
\end{verbatim}
Cette expérience montre que calculée de cette façon, on ne peut pas dire que 
la distance entre deux sons phonétiquement identiques soient inférieures 
à la distance entre deux sons phonétiquement différent. 
\end{corr}
\finenum
\end{enumerate}

Cette expérimentation montre que la variabilité entre des sons similaires est aussi importante 
que la variabilité entre des sons différents.

\subsection{Obtenir une distance entre sons à partir d'une distance entre trames}

Il existe une autre technique permettant de créer une distance entre sons à partir d'une 
distance entre trames. Cette technique est appelée {\em dynamic time warping} ou 
{\em déformation temporelle dynamique}.~\footnote{Cette notion est présentée en détail dans~\cite{Cassisi12}.}

\subsubsection{Déformer la base de temps d'un signal}
\label{sssec:def_tps}

\begin{enumerate}
\deb
\item 
On considère un signal $x_n=n$ échantillonné à $f_e=1$Hz d'une durée $20$s. Observez comme la forme du signal est modifiée 
lorsqu'on remplace $t_n=nT_e$ par $t'_n=t_n+B_n$ où $B_n$ est un bruit blanc gaussien d'écart-type $1$. 
\begin{verbatim}
xn=0:19;
tn=0:19;
tpn=tn+randn(size(tn));
figure(1); plot(tn,xn,tpn,xn); 
\end{verbatim}
Trouvez un moyen de vérifier si dans la simulation on a effectivement $t'_{n+1}\geq t'_n$. 
Qu'en pensez-vous~? 
\begin{corr}
\relax \sol
\begin{verbatim}
xn=0:19;
tn=0:19;
tpn=tn+randn(size(tn));
delta_tpn=tpn(2:end)-tpn(1:(end-1)); 
ind=find(delta_tpn<0); 
figure(1); plot(tn,xn,tpn,xn,tpn(ind),xn(ind),'k^',tpn(ind+1),xn(ind+1),'kv'); 
\end{verbatim}
Cette propriété n'est manifestement pas vérifiée dans la simulation. 
Pourtant le genre de déformation temporelle dans un vrai signal devrait vérifier cette propriété. 
\end{corr}

\item Complétez les commandes suivantes de manière à transformer le signal $x(t)=t\1_{[0,1]}(t)$ en le 
signal $y(t)=t^2\1_{[0,1]}(t)$ au moyen d'une déformation de la base de temps. 
\begin{verbatim}
fe=1e4; t=0:1/fe:(1-1/fe); 
x=t; 
y=t.^2;
...
figure(1); plot(t,x,'b',t,y,'r+',tp,x,'g');
\end{verbatim}
\begin{corr}
\relax \sol
\begin{verbatim}
fe=1e4; t=0:1/fe:(1-1/fe); 
x=t; 
y=t.^2;
tp=sqrt(t);
figure(1); plot(t,x,'b',t,y,'r+',tp,x,'g');
\end{verbatim}
\end{corr}
\finenum
\end{enumerate}

\subsubsection{Déformation au moyen d'une table d'indexation}

\`A la différence de la section~\ref{sssec:def_tps}, nous souhaitons ici pouvoir 
représenter le signal déformé du fait de la modification de la base de temps
 comme un signal temps discret échantillonné à la même fréquence d'échantillonnage. 
Ici nous modifions à la fois la base de temps et l'affectation des valeurs sur $x_n$ à chaque base de temps à travers 
deux tables d'indexations. 
\beqn
y_{\Phi_y[n]}\approx x_{\Phi_x[n]}
\eeqn
Il est donc nécessaire que  $\Phi_x[n]$
 et $\Phi_{y}[n]$ soient deux entiers respectivement contenus entre $0$ et $N_x-1$
et entre $0$ et $N_y-1$, $N_x$ et $N_y$ étant le nombre d'échantillons de $x_n$ et $y_n$ c'est-à-dire 
 la durée de chacun des  signaux multipliées par la fréquence d'échantillonnage. 
De plus, on cherche à ce que $\Phi_{x}$ et $\Phi_y$ vérifient les propriétés suivantes
\beqn
\label{eq:pro1}
\forall n\in\{0,\ldots,N_{\Phi}-1\}
\left \{
\begin{array}{l}
\Phi_x[n]\leq \Phi_x[n+1]\mbox{ et }\Phi_x[0]=0,\,\Phi_x[N_{\Phi}-1]=N_x-1\\[0.2cm]
\Phi_{y}[n]\leq \Phi_{y}[n+1]\mbox{ et }\Phi_{y}[0]=0,\,\Phi_{y}[N_{\Phi}-1]=N_y-1\\[0.2cm]
\left |\Phi_x[n]-\Phi_{y}[n]\right |\leq \delta
\end{array}
\right . 
\eeqn
où $N_{\Phi}$ est le nombre d'index dans $\Phi_x$ et dans $\Phi_{y}$. 


\begin{enumerate}
\deb
\item Les lignes de commandes donnent trois signaux, $(t_n,x_n)$, $(t'_n,y_n)$ et un signal transformé 
$(t'_{\Phi_{y}[n]},x_{\Phi_x[n]})$ visualisés 
sur la figure~\ref{fig7} (p.~\pageref{fig7}). Indiquez quelle couleur correspond à quelle courbe. 
Vérifiez si les lignes de commandes suivantes vérifient les propriétés indiquées en (\ref{eq:pro1})
\begin{verbatim}
clear
xn=0:20;
tn=0:20;
tpn=0:15; 
yn=sin(tpn/15*pi/2)*20;
delta=5;
e=@()ceil(rand(1)*2)-1;
ok=0;
while(~ok)
  Phi_x=[1];
  Phi_y=[1];
  while(1)
    Phi_x=[Phi_x(:)' Phi_x(end)+e()];
    Phi_y=[Phi_y(:)' Phi_y(end)+e()];
    if Phi_x(end)>=length(xn) break; end
    if Phi_y(end)>=length(tpn) break; end
  end
  ok=1;
  if ~(Phi_x(end)==length(xn)) ok=0; end
  if ~(Phi_y(end)==length(tpn)) ok=0; end
  if ~(max(abs(Phi_y-Phi_x))<=delta) ok=0; end
end
figure(1); plot(tn,xn-0.05,'b',tpn,yn-0.05,'g',tpn(Phi_y),xn(Phi_x),'r-'); 
axis([-0.1 max(max(tpn),max(tn))+0.1 -0.1 max(max(xn),max(yn))+0.1])
\end{verbatim}

\begin{corr}
\relax \sol 
Sur la figure~\ref{fig7}, $(t_n,x_n)$ correspond au signal bleu, $(t'_n,y_n)$ au signal vert et 
$(t'_{\Phi_{y}[n]},x_{\Phi_x[n]})$ au signal rouge.
\begin{verbatim}
if ~all(diff(Phi_y)>=0)           disp('Phi_y non croissant'), end
if ~all(diff(Phi_x)>=0)           disp('Phi_x non croissant'), end
if ~(Phi_y(1)==1)                 disp('Premiere valeur de Phi_y non conforme'), end
if ~(Phi_x(1)==1)                 disp('Premiere valeur de Phi_x non conforme'), end
if ~(Phi_y(end)==length(tpn))     disp('Derniere valeur de Phi_y non conforme'), end
if ~(Phi_x(end)==length(xn))      disp('Derniere valeur de Phi_x non conforme'), end
if ~all(abs(Phi_y-Phi_x)<=delta)  disp('Majoration de abs(Phi_y-Phi_x) non conforme'), end
\end{verbatim}
\end{corr}
\finenum
\end{enumerate}
Grâce à ces tables d'indexations, nous avons maintenant un moyen de comparer des signaux de tailles différentes. 
Ainsi s'il n'était pas possible d'écrire $\frac{1}{N}{\sum_n (x_n-y_n)^2}$ car $x_n$ et $y_n$ ne sont pas de mêmes
tailles il est maintenant possible d'écrire l'équation~(\ref{eq:tw_d1}) page~\pageref{eq:tw_d1}. 
\beqns
d_{\Phi_x,\Phi_{y}}(x,y)=\sqrt{\frac{1}{N_x+N_y}{\sum_{n=0}^{N_{\Phi}-1} \left(x_{\Phi_x[n]}-y_{\Phi_{y}[n]} \right )^2}}
\eeqns
Plus précisément, 
une façon de définir la distance avec prise en compte de la déformation temporelle dynamique 
est de chercher $\Phi_x$ et $\Phi_y$ vérifiant les conditions (\ref{eq:pro1}) et minimisant 
$d_{\Phi_x,\Phi_{y}}(x,y)$ et cette valeur minimale est la distance entre $x_n$ et $y_n$. 
\beqn
d_\delta(x,y)=\underset{\Phi_x,\Phi_y}{\min}\,d_{\Phi_x,\Phi_{y}}(x,y)
\eeqn
\begin{enumerate}
\deb
\item Ici nous réalisons une approximation très grossière de la minimisation en tirant aléatoirement 
$200$ tables d'indexations $\Phi_x$ et $\Phi_y$ et en considérant les tables qui réalisent cette minimisation. 
Modifiez les lignes de code plus haut de façon à trouver une approximation de cette distance pour $\delta=6$. Quelle valeur avez-vous trouvée~?
\begin{corr}
\relax \sol
\begin{verbatim}
clear
xn=0:20; I=length(xn);
tn=0:20;
tpn=0:15; 
yn=sin(tpn/15*pi/2)*20; J=length(yn);
delta=6;
d_min=Inf;
e=@()ceil(rand(1)*2)-1;
for exp=1:200
  ok=0;
  while(~ok)
    Phi_x=[1];
    Phi_y=[1];
    while(1)
      Phi_x=[Phi_x(:)' Phi_x(end)+e()];
      Phi_y=[Phi_y(:)' Phi_y(end)+e()];
      if Phi_x(end)>=length(xn) break; end
      if Phi_y(end)>=length(tpn) break; end
    end
    ok=1;
    if ~(Phi_x(end)==length(xn)) ok=0; end
    if ~(Phi_y(end)==length(tpn)) ok=0; end
    if ~(max(abs(Phi_y-Phi_x))<=delta) ok=0; end
  end
  d=sum((yn(Phi_y)-xn(Phi_x)).^2)/(I+J);
  if d<d_min 
    d_min=d;
    disp(['exp=',num2str(exp),' d_min=',num2str(d_min)]),
    figure(1); plot(tn,xn-0.05,'b',tpn,yn-0.05,'g',tpn(Phi_y),xn(Phi_x),'r-'); 
    axis([-0.1 max(max(tpn),max(tn))+0.1 -0.1 max(max(xn),max(yn))+0.1])
  end
end
\end{verbatim}
\end{corr}
\finenum
\end{enumerate}

\subsubsection{Heuristique permettant de trouver rapidement une distance tenant compte d'une déformation temporelle dynamique}

On peut voir les deux tables d'indexations comme un chemin particulier permettant de relier le point $(0,0)$ au 
point $(N_x-1,N_y-1)$. Ce chemin est composé d'arêtes horizontales, verticales et en diagonales et 
à chaque arête, il y a  une valeur correspondant à 
\beqn
d_{\Phi_x,\Phi_y}(x,y,n)=\left |x_{\Phi_x[n]}-y_{\Phi_y[n]}\right|^2
\eeqn
En sommant les carrés de ces différentes valeurs et avec une normalisation on obtient une distance associée à un chemin
\beqn
d_{\Phi_x,\Phi_{y}}(x,y)=\sqrt{\frac{1}{N_x+N_y}{\sum_{n=0}^{N_{\Phi}-1} d_{\Phi_x,\Phi_y}(x,y,n)}}
\eeqn

\begin{enumerate}
\deb
\item\label{it:quest_fig8} En observant attentivement les deux figures qui s'affichent dont un exemple 
se trouve à la figure~\ref{fig8} (p.~\pageref{fig8}), expliquez concrètement les valeurs affichées 
et comment la distance globale entre $x_n$ et $y_n$ peut se retrouver à partir des valeurs affichées. 
Les valeurs affichées sont associées aux ronds qui figurent des noeuds ({\em nodes} ou {\em vertices}). Un trait joignant 
deux sommets voisins est appelé une arête ({\em edge}).

Les deux signaux à comparer sont les suivants, ils sont représentés sur le graphe de gauche de la figure~\ref{fig8}.   
\beqns
\begin{array}{llllll}
x_0=0 & x_1=1 & x_2=2 & x_3=3 & x_4=4 & x_5=5\\[0.2cm]
y_0=0 & y_1=2 & y_2=4 & y_3=5\\[0.2cm]
\end{array}
\eeqns
Les tables d'indexations créées sont~: 
\beqns
\begin{array}{llllll}
\Phi_x[0]=0 & \Phi_x[1]=1 & \Phi_x[2]=2 & \Phi_x[3]=3 & \Phi_x[4]=4 & \Phi_x[5]=5 \\[0.2cm]
\Phi_y[0]=0 & \Phi_y[1]=0 & \Phi_y[2]=0 & \Phi_y[3]=1 & \Phi_y[4]=2 & \Phi_y[5]=3 \\[0.2cm]
\end{array}
\eeqns


\begin{verbatim}
clear
xn=0:5;
tn=0:5;
tpn=0:3; 
yn=[0 2 4 5];
delta=5;
Phi_x =[   1     2     3     4     5     6];
Phi_y =[     1     1     1     2     3     4];
ok=1; %ok=0 pour regénérer une nouvelle solution
e_=[1 0;1 1; 0 1]; 
e=@()e_(ceil(rand(1)*3),:);
while(~ok)
  Phi_x=[1];
  Phi_y=[1];
  while(1)
    e_v=e(); 
    Phi_x=[Phi_x(:)' Phi_x(end)+e_v(1)];
    Phi_y=[Phi_y(:)' Phi_y(end)+e_v(2)];
    if Phi_x(end)>=length(xn) break; end
    if Phi_y(end)>=length(tpn) break; end
  end
  ok=1;
  if ~(Phi_x(end)==length(xn)) ok=0; end
  if ~(Phi_y(end)==length(tpn)) ok=0; end
  if ~(max(abs(Phi_y-Phi_x))<=delta) ok=0; end
end
figure(1); plot(tn,xn-0.05,'bx-',tpn,yn-0.05,'gx-',tpn(Phi_y),xn(Phi_x),'rx-','Linewidth',2); 
axis([-0.1 max(max(tpn),max(tn))+0.1 -0.1 max(max(xn),max(yn))+0.1])
xlabel('t_n'), 
text(0.5*(tn(4)+tn(5)),0.5*(xn(4)+xn(5))-0.1,'\leftarrow (t_n,x_n)','FontSize',14);
text(0.5*(tn(2)+tn(3)),0.5*(yn(2)+yn(3))-0.1,'\leftarrow (t_n,y_n)','FontSize',14);
text(0.5*(tpn(Phi_y(4))+tpn(Phi_y(5))),0.5*(xn(Phi_x(4))+xn(Phi_x(5)))-0.1,'\leftarrow (t_{\Phi,y}[n],x_{\Phi,x}[n])','FontSize',14);
Axis=[-1.1 max(max(length(tpn)),max(length(tn)))+1.1...
 -1.1 max(max(length(xn)),max(length(yn)))+1.1];
figure(2); plot(Phi_x-1,Phi_y-1,'o-',...
  [Axis(1), Axis(2)],delta+[Axis(1), Axis(2)],'k-',...
  [Axis(1), Axis(2)],delta+[Axis(1), Axis(2)]+0.05,'k:',...
  [Axis(1), Axis(2)],delta+[Axis(1), Axis(2)]+0.1,'k:',...
  [Axis(1), Axis(2)],-delta+[Axis(1), Axis(2)],'k-',...
  [Axis(1), Axis(2)],-delta+[Axis(1), Axis(2)]-0.05,'k:',...);
  [Axis(1), Axis(2)],-delta+[Axis(1), Axis(2)]-0.1,'k:',...
  'LineWidth',2);
axis(Axis)
for k=1:length(Phi_x)
  dk=(yn(Phi_y(k))-xn(Phi_x(k))).^2;
  text(Phi_x(k)-0.4-1,Phi_y(k)+0.4-1,num2str(dk,'%2.1f'));
end
text(3,5,'Lieu permis'),
xlabel('\Phi_x[n]'), ylabel('\Phi_y[n]'),
\end{verbatim}

\begin{corr}
\relax \sol 
La figure~\ref{fig8} est  à sa gauche similaire à la figure~\ref{fig7} avec 
$(t_n,x_n)$, $(t'_n,y_n)$  et 
$t_{\Phi_y[n]},x_{\Phi_x[n]}$ en bleu, rouge et vert.
\`A droite, les tables d'indexations sont représentées avec  en ordonnée $\Phi_y[n]$ et 
 en abscisse $\Phi_x[n]$. Il y a sept "ronds" ce qui correspond à sept valeurs de $n$ de $0$ à $6$
identique au nombre de "croix" sur la courbe rouge à gauche. 
En partant de la gauche la première arête  est verticale signifiant que l'évolution de $y_n$ est 
au début bloqué et que seul $x_n$ est modifié ce qui est figuré par le fait que 
le premier segment de la courbe rouge est horizontal. L'arête suivante diagonale correspond aussi 
à un déplacement diagonal de la courbe rouge. Par la suite une arête du graphe de droite est horizontale ce qui 
correspond à un déplacement vertical  de la courbe rouge du graphe de gauche. 
Les nombres sont affichés  légèrement à gauche et légèrement en hauteur chaque noeud, de manière 
à ne pas recouvrir les traits. Le premier nombre $0.0$ indique que le premier plus rouge en bas à gauche coïncide 
avec le premier plus de la courbe verte. 
Le deuxième nombre $4.0$ traduit le fait que pour $n=1$, il y a une différence de $2$ entre le deuxième plus de la courbe rouge 
qui s'est éloigné du deuxième plus de la courbe verte dont la deuxième croix s'est aussi éloignée. C'est différence de $2$ une fois au carré 
reste égale à $4$.

\end{corr}
\item En exécutant à de multiples reprises les lignes de codes de la question précédente, 
combien y a-t-il de noeuds possibles et combien d'arêtes possibles. 
Observez qu'il n'y a qu'une seule valeur possible par arête. 

\underline{Indication}
Le premier noeud appelé racine ({\em root}) correspond au point de départ. Le dernier noeud appelé ({\em sink}) 
est le point d'arrivé. Ces deux noeuds sont communs à tous les graphes. 
Une méthode pour compter le nombre d'arêtes consiste à considérer chaque noeud et à comptabiliser le nombre d'arêtes 
partant de ce noeud. 
\begin{corr}
\relax \sol
En plus de la racine et du puit, il y a  11 noeuds et 23 arêtes possibles (il est possible que j'ai oublié des noeuds et 
des arêtes). 
\begin{itemize}
\item 
2 noeuds avec une valeur de $0$.
\item 4 noeuds avec une valeur de $1$
\item 4 noeuds avec une valeur de $4$
\item 1 noeud avec une valeur de $9$
\end{itemize}
\end{corr}
\item Les chemins acceptables sont uniquement ceux passant par les noeuds que vous avez pu visualiser 
en appliquant plusieurs fois les lignes de commandes. Ces chemins ne doivent être composés que d'arêtes 
allant vers la droite, en diagonale montante et vers la droite et en allant verticalement vers le haut et 
ce dans les trois cas en se déplaçant que d'un pas à la fois. Les trois types d'arêtes sont respectivement
notés $\Hh,\Dd,\Vv$, de manière à ce qu'un chemin puisse être codé sous la forme d'une succession de ces symboles. 
Observez que les chemins ne sont pas tous de la même longueur. 
Proposez un chemin le plus court, le plus long, celui dont la somme des valeurs associées est la plus faible et 
la plus élevée. 

\begin{corr}
\relax \sol
\begin{itemize}
\item Chemin le plus court~: $\Dd,\Dd,\Hh,\Hh,\Dd$. 
\item Chemin le plus long~: $\Hh,\Hh,\Vv,\Vv,\Hh,\Hh,\Dd$. 
\item Chemin dont la somme des valeurs associées à chaque noeud traversé est la plus faible~: 
$\Dd,\Hh,\Dd,\Hh,\Dd$. Cela vaut $2$. 
\item Chemin dont la somme des valeurs associées à chaque noeud traversé est la plus élevée~: 
$\Vv,\Dd,\Hh,\Hh,\Hh,\Dd$. Cela vaut $18$.
\end{itemize}
\end{corr}
\item\label{it:c_ex1} L'algorithme généralement utilisé apporte une petite modification visant à diminuer l'utilisation des arêtes $\Dd$. 
Pour cela la valeur de la distance est non pas exactement moyenne des valeurs associées à chaque noeud traversé, 
il s'agit d'une moyenne ou les valeurs associées aux noeuds traversées comptent double lorsqu'elles terminent une arête $\Dd$. 
Ainsi l'équation (\ref{eq:tw_d1}) est en fait remplacée par 
\beqn
d'_{\Phi_x,\Phi_{y}}(x,y)=\sqrt{\frac{1}{N_x+N_y}{\sum_{n=0}^{N_{\Phi}-1} \left(x_{\Phi_x[n]}-y_{\Phi_{y}[n]} \right )^2
\left (\Phi_x[n]-\Phi_x[n-1]+\Phi_y[n]-\Phi_y[n-1]\right )}}
\eeqn
où $\left (\Phi_x[n]-\Phi_x[n-1]+\Phi_y[n]-\Phi_y[n-1]\right )$ vaut $1$ lorsque le noeud $(\Phi_x[n],\Phi_y[n])$ 
est l'arrivée 
d'une arête $\Hh$ ou $\Vv$ et cette expression vaut $2$ lorsque ce noeud est l'arrivée d'une arête $\Dd$. 
Trouvez un chemin permettant de minimiser $d'_{\Phi_x,\Phi_{y}}(x,y)$.

\begin{corr}
\relax \sol 
Le chemin $\Hh,\Dd,\Hh,\Dd,\Dd$
 est associée à une distance de 
\beqns
\sqrt{\frac{1}{6+4}{\times}(1^2+0^2+1^2+0^2+0^2)}=\frac{\sqrt{5}}{5}
\eeqns
\end{corr}
\item (Facultatif)
{\tt dist\_DTD.m} est une implémentation  l'algorithme de Dijkstra~\footnote{Cet algorithme 
est par exemple présenté dans {\tt https://fr.wikipedia.org/wiki/Algorithme\_de\_Dijkstra}.}
Expliquez en gros comment ce programme fonctionne.

\begin{corr}
\relax \sol
Le programme parcourt l'ensemble de noeuds du graphe en partant de la racine et en progressant niveau de profondeur 
par niveau de profondeur avec une initialisation, la boucle while conditionnée à  la fonction {\tt est\_terminer} et la fonction 
{\tt suivant}. Le niveau de profondeur est déterminé par la valeur de $i+j$. 
La fonction {\tt suivant} cherche d'abord à progresser au sein d'un niveau de profondeur en incrémentant $j$ 
puis ensuite à passer au niveau 
suivant et cherche dans ce niveau d'après le noeud ayant la valeur de $j$ la plus faible compte tenu des conditions. 
Comme il est possible que dans le niveau suivant il n'y a aucun noeud possible, le programme réalise une 
boucle au sein des niveaux suivants jusqu'à trouver un niveau suivant où existe un noeud disponible. 
L'équation utilisée donnant cette valeur la plus faible est 
\beqns
j'=\max\, \left( 1, i+j+s-n_I,\left\lceil \frac{i+j+s-\delta}{2}\right \rceil\right )
\mbox{ et }i'=i+j+s-j'
\eeqns
où $i'$ et $j'$ sont la nouvelle valeur de $j$ et $i+j+s$ est la nouvelle profondeur considérée, $s$ étant un entier supérieur ou égale 
à $1$. 
Cette équation est en fait la résolution de ce problème d'optimisation 
\beqns
\mbox{Trouver $j'$ maximal tel que }\left \{\begin{array}{l}
1\leq i'\leq N_I\\[0.2cm]
1\leq j'\leq N_J\\[0.2cm]
\left |i'-j'\right |\leq \delta\\[0.2cm]
i'+j'=i+j+s
\end{array}
\right . 
\eeqns

L'algorithme de Dijkstra consiste d'une part à parcourir les noeuds de l'arbre et pour chaque noeud à déterminer 
la distance minimale. Le calcul de la distance minimale est possible si pour toutes les arêtes joignant ce noeud, 
on connait leur distance minimale. Il se trouve qu'ici il est simple de déterminer les arêtes arrivant à ce noeud 
ou partant depuis ce noeud. Ici l'implémentation proposée consiste à initialiser
la distance associée à la racine avec sa valeur et les distances associées aux autres noeuds 
à l'infini. Ensuite l'algorithme parcourt les noeuds de l'arbre par niveau 
de profondeur et détermine pour chaque noeud la distance des trois noeuds voisins comment leur valeur peut être modifié 
en fonction de leur distance avec ce noeud. 
\end{corr}
\item (Facultatif)
{\tt v\_dist\_DTD.m} est un programme qui vérifie le fonctionnement, expliquez les différents tests 
réalisés. 
\begin{corr}
\relax \sol
\begin{itemize}
\item[{\tt test1}] vérifie la valeur trouvée à la question~\ref{it:c_ex1}.
\item[{\tt test2}] vérifie 
\beqns
d_\delta(x_n,x_n)=0\mbox{ et }d_0(x_n,y_n)=\frac{1}{\sqrt{2N}}\sqrt{(x_0-y_0)^2+2{\sum_{n=1}^{N-1}(x_n-y_n)^2}}
\eeqns
où $x_n$ et $y_n$ sont tous de longueur $N$.
\item [{\tt test3}] est un cas particulier de {\tt test2} sur des valeurs très simples. 
\item [{\tt test4}] vérifie que 
\beqns
d_\delta(x_n,y_n)\leq d_{\delta-1}(x_n,y_n)
\eeqns
pour $\delta\geq 1$. 
\item [{\tt test5}] vérifie que le fait d'insérer un échantillon $y_k$ de $y_n$ dans $x_n$ à la $k$ème position n'augmente 
pas la distance si on utilise une valeur de $\delta$ incrémentée. Ici on ne tient pas compte de la normalisation qui est modifiée
du fait de l'insertion d'un  nouvel échantillon. 
\item [{\tt test6}] vérifie que 
\beqns
\sqrt{N_1+N_2}d_{\delta}([x^{(1)}_n\, x^{(2)}_n],[y^{(1)}_n\, y^{(2)}_n])\leq 
\sqrt{N_1}d_{\delta}(x^{(1)}_n,y^{(1)}_n)+\sqrt{N_2}d_{\delta}(x^{(2)}_n,y^{(2)}_n)
\eeqns
où $[x^{(1)}_n\, x^{(2)}_n]$ et $[y^{(1)}_n\, y^{(2)}_n]$ sont respectivement la concaténation 
de $x^{(1)}_n$, $x^{(2)}_n$ et de  $y^{(1)}_n$, $y^{(2)}_n$. $x^{(1)}_n$, $y^{(1)}_n$ et $x^{(2)}_n$, $y^{(2)}_n$ sont 
respectivement de longueurs $N_1$ et $N_2$. 
\item [{\tt test7}] vérifie que 
\beqns
d_{\delta}(\bx_n,\by_n)
\leq \sqrt{{\sum_{f} (d_{\delta}(x_n^f,y_n^f))^2}}
\mbox{ et }
d_{0}(\bx_n,\by_n)
= \sqrt{{\sum_{f} (d_{0}(x_n^f,y_n^f))^2}}
\eeqns
où $x_n^f$ sont les composantes de $\bx_n$. 
\item [{\tt test8}] vérifie dans le cadre multidimensionnel que 
\beqns
\forall \lambda, d_{\delta}(\lambda x_n,\lambda y_n)=|\lambda|d_{\delta}(x_n,y_n)
\mbox{ et }d_{\delta}(x_n, y_n)=d_{\delta}( y_n,x_n)
\eeqns
\end{itemize}
\end{corr}
\finenum
\end{enumerate}

\subsubsection{Utilisation de cette distance en simulation numérique}

\begin{enumerate}
\deb
\item\label{it:distance2}
 De façon un peu similaire 
à la question~\ref{it:distance1}, créez une fonction {\tt distance2} permettant de comparer deux sons avec prise en compte de la déformation 
temporelle dynamique en l'utilisant non pas sur les valeurs des deux signaux 
mais sur leur puissance moyenne à court terme sur chaque trame.
La durée de chaque trame est de $30$ms et le chevauchement est de $0.25$. 
Pour $\delta$, vous utiliserez un nombre de trame égale à un dizième du nombre de trames du signal le plus court. 
Vous utiliserez la syntaxe suivante
\begin{verbatim}
function d=distance2(rep1,m1,rep2,m2)
\end{verbatim}
Attention cette distance ne vérifie pas l'inégalité triangulaire et il est donc 
tout à fait normal que {\tt v\_distance} déclenche une erreur. 

\begin{corr}
\relax \sol
\begin{verbatim}
function d=distance2(rep1,m1,rep2,m2)
  [x1,fe]=lireSon(rep1,m1); [x2,fe]=lireSon(rep2,m2);
  duree_trame=30e-3; chevauchement=0.25; 
  xk1=preparation(x1,fe,duree_trame,chevauchement);
  xk2=preparation(x2,fe,duree_trame,chevauchement);
  PMCT1=s_c_PMCT(xk1); PMCT2=s_c_PMCT(xk2);
  delta=ceil(min(length(PMCT1),length(PMCT2))/10);
  d=dist_DTD(PMCT1,PMCT2,delta);
end
\end{verbatim}
\end{corr}
\item Utilisez la question~\ref{it:mesure_sensibilite} pour calculer la sensibilité globale (OA) par une validation croisée
et comparez les performances entre celles obtenues avec {\tt distance1} et {\tt distance2}.

\begin{corr}
\relax \sol
\begin{verbatim}
OA1=s_mesure_sensibilite(@s_distance1,5);
OA2=s_mesure_sensibilite(@s_distance2,5);
\end{verbatim}
Avec mes données, j'observe {\tt OA1}=0.26 tandis que {\tt OA2}=0.44.
\end{corr}
\finenum
\end{enumerate}

\section{Séance 8\\Descripteurs de trames}

\subsection{\'Estimation spectrale}
\subsubsection{\'Estimation de la fréquence fondamentale, {\tt F0\_ZCR}}

\begin{figure}
\begin{center}
\begin{tabular}{c}
\includegraphics[width=0.5\linewidth]{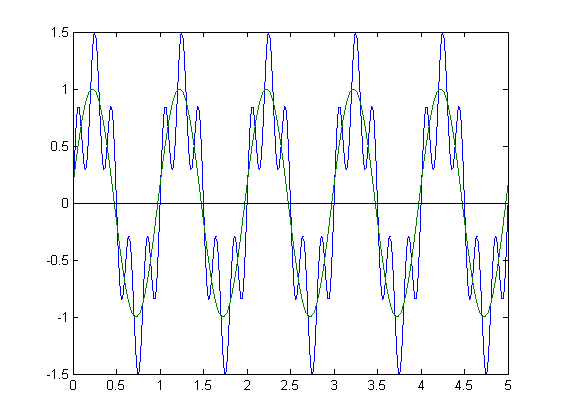}
\end{tabular}
\end{center}
\caption{Les courbes bleues et vertes ont le même nombre de passages à zéro. Mais la courbe bleue est une sinusoïde de fréquence $1$Hz, 
tandis que la courbe verte est une somme de sinusoïde dont la fréquence la plus basse est aussi $1$Hz.}
\label{fig3}
\end{figure}

On cherche maintenant la fréquence fondamentale du signal.~\footnote{Pour un son musical assez simple, cette 
fréquence correspond au {\em pitch} défini section~C.3, p.~215-217 de~\cite{Rocchesso03}.} 
L'utilité de s'intéresser à la fréquence fondamentale pour reconnaître une voyelle est 
expliqué en détail dans~\cite{Meunier07}. On modélise ici le signal comme 
une superposition de signaux sinusoïdaux et on suppose 
que le nombre de fois que le signal traverse l'axe des abscisses ne dépend que de la fréquence la plus 
basse.~\footnote{Cette technique appelée {\em zero crossing rate} est évoquée dans la section~2.4 p.~12 de~\cite{Peeters04}, 
et définie dans~\cite{Camastra07} p.~43-44.}
La simulation suivante visible sur la figure~\ref{fig3} permet d'illustrer l'idée. Elle est obtenue avec les 
commandes suivantes.~\footnote{Comme le suggère la simulation, il serait faux d'affirmer que le nombre de passage par zéros 
est déterminée par la fréquence la plus basse pour toutes combinaisons linéaires de sinusoïdes.}
\begin{verbatim}
t=0:1e-3:5;
x1=sin(2*pi*t)+1/2*sin(2*pi*5*t);
x2=sin(2*pi*t+pi/20);
figure(1); plot(t,x1,t,x2,t,0,'k-')
\end{verbatim}
Le nombre de passage par $0$ est noté $Z$
\beqn
Z_k=\frac{1}{\sum_n w_n}\frac{1}{2}{\sum_n w_n\left | \mbox{signe}\left (s_k[n+1]\right )-\mbox{signe}\left (s_k[n]\right)\right |}
\eeqn
où $\mbox{signe}(x)$ vaut $1$ pour $x>0$, $-1$ pour $x<0$ et $0$ pour $x=0$.

\`A partir de $Z_k$, on calcule une estimation de la fréquence fondamentale pour chaque trame, notée $f^{(ZCR)}_k$ ou {\tt F0\_ZCR}. 
\beqn
f^{(ZCR)}_k=\frac{Z_kf_e}{2}
\eeqn

Dans ce document, on utilise 4 indices muets, $n,k,l$ et $m$.
\begin{itemize}
\item $n$ est utilisé pour au sein d'un signal ou d'une trame pour numéroter les différents instants espacés de $T_e$. 
\item $k$ est utilisé pour numéroter les trames. 
\item $l$ est utilisé pour numéroter les fréquences dans le cadre du calcul de la transformée de Fourier d'une trame dans le 
cadre de la section~\ref{ssec:feq_moy_sp}.
$l$ est à valeurs entières entre $0$ et $\lfloor N_K/2\rfloor$. 
\item $m$ est utilisé pour numéroter le filtre utilisé au sein d'un banc de filtres dans le cadre de la section~\ref{ssec:banc_de_filtres}.
$m$ est à valeurs entre $1$ et $N_f$. 
\end{itemize}

\begin{enumerate}
\deb
\item
Pour un exemple de son dont le silence a été supprimé, calculez une estimation de ce descripteur pour chaque trame 
et représentez l'évolution de ce descripteur en fonction du temps. 
\begin{verbatim}
[x,fe]=lireSon('..\donnees\un\',0);
[xk,ech_trames,centres_trames]=preparation(x,fe,30e-3,0.25);
K=size(xk,1); 
F0_ZCR=zeros(K,1);
fenetre=window(@hamming,length(xk(1,:)))';
for trame=1:K
  Z=abs(sign(xk(trame,2:end))-sign(xk(trame,1:end-1)))/2;
  F0_ZCR(trame)=mean(Z.*fenetre(1:end-1))*fe/2/mean(fenetre(1:end-1));
end
figure(1); plot(centres_trames,F0_ZCR);
\end{verbatim}
\item\label{it:c_F0_ZCR} \'Ecrivez un programme qui calcule pour chaque trame la fréquence fondamentale en utilisant 
la syntaxe suivante.
\begin{verbatim}
function F0_ZCR=c_F0_ZCR(xk,fe)
\end{verbatim}
\begin{corr}
\relax \sol
\begin{verbatim}
function F0_ZCR=c_F0_ZCR(xk,fe)
  K=size(xk,1); NK=size(xk,2);
  F0_ZCR=zeros(K,1);
  fenetre=window(@hamming,NK)';
  for trame=1:K
    Z=abs(sign(xk(trame,2:end))-sign(xk(trame,1:end-1)))/2;
    F0_ZCR(trame)=mean(Z.*fenetre(1:end-1))*fe/2/mean(fenetre(1:end-1));
  end
end
\end{verbatim}
\end{corr}
\item\label{it:distance3} \'Ecrivez un programme évaluant la distance entre deux signaux en tenant compte de la déformation 
temporelle dynamique et en l'appliquant à l'estimation de {\tt F0\_ZCR}, de façon assez similaire 
à la question~\ref{it:distance2}. Cette distance est appelée {\tt distance3}. 

\begin{corr}
\relax \sol
\begin{verbatim}
function d=distance3(rep1,m1,rep2,m2)
  [x1,fe]=lireSon(rep1,m1); [x2,fe]=lireSon(rep2,m2);
  duree_trame=30e-3; chevauchement=0.25; 
  xk1=preparation(x1,fe,duree_trame,chevauchement);
  xk2=preparation(x2,fe,duree_trame,chevauchement);
  x=s_c_F0_ZCR(xk1,fe); y=s_c_F0_ZCR(xk2,fe);
  delta=ceil(min(length(x),length(y))/10);
  d=dist_DTD(x,y,delta);
end
\end{verbatim}
\end{corr}

\item Au moyen de {\tt distance3}, déduisez-en la sensibilité globale en utilisant {\tt F0\_ZCR}. 

\begin{corr}
\relax \sol 
\begin{verbatim}
OA3=s_mesure_sensibilite(@s_distance3,5);
\end{verbatim}
Avec mes données, je trouve {\tt OA3}=0.76.
\end{corr}
\finenum
\end{enumerate}

Une autre technique existe pour mesurer la fréquence fondamentale en s'appuyant sur l'autocorrélation~\footnote{Cette 
technique est brièvement présentée p.~45-46 dans~\cite{Camastra07}.}

\subsubsection{Estimation de la fréquence moyenne du spectre}
\label{ssec:feq_moy_sp}

\begin{corr}
\relax Réalisation du graphe illustrant la notion de moyenne pondérée.
\begin{verbatim}
f=1e-2:10:10^4;
%f=-10:1e-3:10;
P=abs(sin(pi*f/3000)./(pi*f/3000));
%P=exp(-1/2*f.^2/4);
f0=sum(f.*P)/sum(P);
figure(1); plot(f,P,'k-',[f0,f0],[0, max(P)],'k-.','Linewidth',1.5);
sigma=sqrt(sum((f-f0).^2.*P)/sum(P));
P_app=sum(P)/sum(P_app)*exp(-0.5*(f-f0).^2/sigma^2);
figure(2); plot(f,P,'k-',[f0,f0],[0, max(P)],'k-.',...
  [f0-sigma f0-sigma],[0 max(P)],'k:',...
  [f0+sigma f0+sigma],[0 max(P)],'k:',...
  f,P_app,'k--','Linewidth',1.5); 
axis([f0-sigma*(1.05) max(f) 0 max(P)]);

\end{verbatim}
\end{corr}

Un estimateur de la fréquence moyenne du spectre $\mu_k$~\footnote{Cet estimateur 
est défini dans la section~6.1.1 p.~13 de~\cite{Peeters04}.}
 est estimée avec  l'équation~(\ref{eq:mu_sp}) page~\pageref{eq:mu_sp}. 
\beqns
\mu_k=\frac{{\sum_{l=0}^{\lfloor \frac{N_K}{2}\rfloor}\,\,\frac{lf_e}{N_K}\,\,\left |\widehat{X}_k[l]\right |^2}}
{{\sum_{l'=0}^{\lfloor \frac{N_K}{2}\rfloor}\left |\widehat{X}_k[l']\right |^2}}
\eeqns 
où $\widehat{X}_k[l]$ est la transformée de Fourier discrète~\footnote{Cette application 
de la transformée de Fourier discrète sur le signal découpé peut être vu comme 
l'utilisation d'une transformée de Fourier à court terme, p.~99 de~\cite{Rocchesso03}.} à $k$ fixé de $x_k[n]$ définie par 
$\widehat{X}_k[l]=\frac{1}{N_K}{\sum_{n=0}^{N_K-1}x_k[n]e^{-j2\pi\frac{ln}{N_K}}}$~;
et ${\lfloor \frac{N_K}{2}\rfloor}$ signifie le plus grand entier inférieur à $N_K/2$. 
Un indice $l'$ est utilisé parce qu'il n'y a en fait aucun lien entre le compteur utilisé dans la sommation au niveau du numérateur et
le compteur utilisé dans la sommation au niveau du dénominateur.

L'expression~(\ref{eq:mu_sp}) est en fait constituée de trois éléments. 
\beqns
\frac{lf_e}{N_K}\mbox{ pour }0\leq l\leq {\lfloor \frac{N_K}{2}\rfloor}\quad \mbox{et}\quad \left |\widehat{X}_k[l]\right |^2\quad \quad 
\eeqns
\begin{itemize}
\item 
$f_l=\frac{lf_e}{N_K}$ désigne la fréquence en Hz associée à chaque valeur de $l$ dans le calcul de la transformée de Fourier discrète.
\item $0\leq l\leq {\lfloor \frac{N_K}{2}\rfloor}$ indiquent les valeurs de $l$ pour lesquelles cette transformée de Fourier 
discrète est calculée, ici elle est calculée entre $0$ et $f_e/2$. 
\item $p_l=\left |\widehat{X}_k[l]\right |^2$ est certes le module au carré de la transformée de Fourier discrète, mais
ici cette expression est utilisée comme une pondération dans une formule classique de moyenne. 
\end{itemize}
Ainsi on pourrait réécrire (\ref{eq:mu_sp}) de la façon suivante
\beqn\label{eq:mu_pl}
\mu=\frac{{\sum_{l=0}^{l_{\max{}}}f_l p_l}}{{\sum_{l'=0}^{l_{\max{}}} p_{l'}}}
\eeqn
Le dénominateur est la somme des pondérations et $l_{\max}$ désigne ici ${\lfloor \frac{N_K}{2}\rfloor}$.

\begin{enumerate}
\deb
\item Utilisez~(\ref{eq:mu_sp}) pour estimer pour chaque trame la fréquence moyenne du spectre et 
représentez l'évolution de cette fréquence moyenne en fonction du temps pour un son pris au hasard. 
\begin{corr}
\relax \sol
\begin{verbatim}
[x,fe]=lireSon('..\donnees\un\',0);
[xk,ech_trames,centres_trames]=preparation(x,fe,30e-3,0.25);
K=size(xk,1); NK=size(xk,2); 
SPC=zeros(K,1);
fenetre=window(@hamming,NK)';
for trame=1:K
  Xk=fft(fenetre.*signal(trame,:)); k_l=1:floor(NK/2); Xk=Xk(k_l); 
  SPC(trame)=fe/NK*sum(abs(Xk).^2.*k_l)/sum(abs(Xk.^2));
end
figure(1); plot(centres_trames,SPC);
\end{verbatim}
\end{corr}
\item\label{it:c_SPC} Réalisez une fonction qui calcule pour chaque trame la fréquence moyenne du spectre avec la 
syntaxe suivante. 
\begin{verbatim}
function SPC=c_SPC(xk,fe)
\end{verbatim}
\begin{corr}
\relax \sol
\begin{verbatim}
function SPC=c_SPC(xk,fe)
  K=size(xk,1); NK=size(xk,2); 
  SPC=zeros(K,1);
  fenetre=window(@hamming,NK)';
  k_l=1:floor(NK/2); 
  for trame=1:K
    Xk=fft(fenetre.*xk(trame,:)); Xk=Xk(k_l); 
    SPC(trame)=fe/NK*sum(abs(Xk).^2.*k_l)/sum(abs(Xk.^2));
  end
end\end{verbatim}
\end{corr}
\finenum
\end{enumerate}

\subsubsection{Estimation de la largeur du spectre}

En utilisant les mêmes notations que pour (\ref{eq:mu_sp}), une estimation de la largeur du spectre~\footnote{Cet estimateur est défini dans la section~6.1.2 p.~13 de~\cite{Peeters04}.} est
\beqns
\sigma_k=\sqrt{\frac{{\sum_{l=0}^{\lfloor \frac{N_K}{2}\rfloor}\,\,\left (\frac{lf_e}{N_K}-\mu_k\right )^2\,\,\left |\widehat{X}_k[l]\right |^2}}{{\sum_{l'=0}^{\lfloor \frac{N_K}{2}\rfloor}\left |\widehat{X}_k[l']\right |^2}}}
\eeqns
Cette équation se trouve aussi définie page~\pageref{eq:mu_sp}.

En reprenant les notations utilisée pour (\ref{eq:mu_pl}), cet estimateur de la largeur du spectre peut s'écrire ainsi. 
\beqn
\sigma=\sqrt{
{\sum_{l=0}^{l_{\max{}}}(f_l-\mu)^2 \frac{p_l}{{\sum_{l'=0}^{l_{\max{}}} p_{l'}}}}
}
\eeqn
Ainsi $\sigma$ peut s'interpréter comme la racine carré de la variance d'une variable aléatoire discrète $F$ dont les probabilités sont 
données par $\frac{p_l}{{\sum_{l'=0}^{l_{\max{}}} p_{l'}}}$. 

\begin{enumerate}
\deb
\item Utilisez (\ref{eq:sigma_sp}) pour chaque trame la largeur de bande du spectre et représentez l'évolution 
de cette largeur de bande en fonction du temps pour un son pris au hasard. 
\begin{corr}
\relax \sol
\begin{verbatim}
[x,fe]=lireSon('..\donnees\un\',0);
[xk,ech_trames,centres_trames]=preparation(x,fe,30e-3,0.25);
K=size(xk,1); NK=size(xk,2); 
SPW=zeros(K,1);
fenetre=window(@hamming,NK)';
for trame=1:K
  Xk=fft(fenetre.*xk(trame,:)); k_l=1:floor(NK/2); Xk=Xk(k_l); 
  spc=fe/NK*sum(abs(Xk).^2.*k_l)/sum(abs(Xk.^2));
  SPW(trame)=sum(abs(Xk).^2.*abs(k_l*fe/NK-spc))/sum(abs(Xk).^2);
end
figure(1); plot(centres_trames,SPW);
\end{verbatim}
\end{corr}

\item\label{it:c_SPW}
Réalisez une fonction qui calcule pour chaque trame la fréquence moyenne du spectre avec la 
syntaxe suivante. 
\begin{verbatim}
function SPW=c_SPW(xk,fe)
\end{verbatim}
\begin{corr}
\relax \sol
\begin{verbatim}
function SPW=c_SPW(xk,fe)
  K=size(xk,1); NK=size(xk,2); 
  SPW=zeros(K,1);
  fenetre=window(@hamming,NK)';
  k_l=1:floor(NK/2); 
  SPC=c_SPC(xk,fe);
  for trame=1:K
    Xk=fft(fenetre.*xk(trame,:)); Xk=Xk(k_l); 
    SPW(trame)=sum(abs(k_l*fe/NK-SPC(trame)).*abs(Xk).^2)/sum(abs(Xk.^2));
  end
end
\end{verbatim}
\end{corr}
\finenum
\end{enumerate}

\subsubsection{\'Estimation d'autres caractéristiques spectrales}

Un estimateur du coefficient d'asymétrie ({\em skewness})~\footnote{Cet estimateur est défini dans la section~6.1.3 p.~13 de~\cite{Peeters04}.} est donné 
par 
\beqn
\gamma_1[k]=\frac{1}{\sigma_k^3}{\frac{{\sum_{l=0}^{\lfloor \frac{N_K}{2}\rfloor}\,\,\left (\frac{lf_e}{N_K}-\mu_k\right )^3\,\,\left |\widehat{X}_k[l]\right |^2}}{{\sum_{l'=0}^{\lfloor \frac{N_K}{2}\rfloor}\left |\widehat{X}_k[l']\right |^2}}}
\eeqn

Un estimateur du coefficient d'aplatissement ({\em kurtosis})~\footnote{Cet estimateur est défini dans la section~6.1.4 p.~14 de~\cite{Peeters04}.} est donné par 
\beqn
\gamma_2[k]=\frac{1}{\sigma_k^4}{\frac{{\sum_{l=0}^{\lfloor \frac{N_K}{2}\rfloor}\,\,\left (\frac{lf_e}{N_K}-\mu_k\right )^4\,\,\left |\widehat{X}_k[l]\right |^2}}{{\sum_{l'=0}^{\lfloor \frac{N_K}{2}\rfloor}\left |\widehat{X}_k[l']\right |^2}}}
\eeqn

\subsection{Utilisation de bancs de filtres}
\label{ssec:banc_de_filtres}


\begin{corr}
\relax 
Simulation pour illustrer un banc de filtres. 
\begin{verbatim}
fe=1e5; 
t=0:1/fe:0.03-1/fe; 
ft=1000+500*sin(2*pi*t/t(end)); 
figure(1); plot(t,ft); 
x=cos(2*pi*ft.*t); 
figure(2); plot(t,x)
X=fftshift(fft(x)); 
ech_f=-fe/2:fe/length(t):(fe/2-fe/length(t)); 
f_lim=500+1000*(0:5)/5;
TFI=@(X)real(ifft(fftshift(X)));
H1=(abs(ech_f)<=f_lim(2)); 
H2=(abs(ech_f)>f_lim(2)).*(abs(ech_f)<=f_lim(3)); 
H3=(abs(ech_f)>f_lim(3)).*(abs(ech_f)<=f_lim(4)); 
H4=(abs(ech_f)>f_lim(4)); 
y1=TFI(H1.*X); y2=TFI(H2.*X);y3=TFI(H3.*X); y4=TFI(H4.*X);
y=y1+y2+y3+y4; 
figure(3); plot(t,x,'k-',t,y1-2,t,y2-4,t,y3-6,t,y4-8,t,y-10); 
legend('x','y1','y2','y3','y4','y1+y2+y3+y4'); 
ind=find(abs(ech_f)<=1500);
figure(4); plot(ech_f(ind),H1(ind),ech_f(ind),H2(ind),ech_f(ind),H3(ind),ech_f(ind),H4(ind))
legend('H1','H2','H3','H4'); axis([-1500 1500 -0.1 1.1]); 
\end{verbatim}
\end{corr}

On considère une fréquence d'échantillonnage $f_e$ et un nombre de filtres $N_f$ 
On appelle support d'un filtre $\MH$ noté $\supp({\MH})$ l'intervalle des fréquences pour lesquelles la réponse fréquentielle est non-nulle. 
Ainsi $\1_{[2,3]}(f)$ est la réponse fréquentielle d'un filtre de support $[2,3]$. 
On appelle un banc de filtre un ensemble de filtre $(\MH_m)_m$ qui valent $1$ sur leur support et 
dont les supports forment une partition de l'intervalle $[f_{\min},\frac{f_e}{2}]$. 
\beqn
\underset{1\leq m\leq N_f}{\bigcup}\,\supp(\MH_m)=\left[f_{\min},\frac{f_e}{2}\right ]\mbox{ et }
\forall m\neq m',\quad \supp({\MH_{m}})\,{\bigcap}\,\supp({\MH_{m'}})=\emptyset
\eeqn

Pour simplifier, nous considérons ici des filtres dont le spectre est rectangulaire.
\beqn
\widehat{H}_m(f)=\1_{\supp(\MH_m)}(f)
\eeqn

Le banc de filtre qu'on utilise ici suit une échelle logarithmique, c'est-à-dire que $\supp(\MH_m)$ sont des intervalles 
qui après application de la fontion $f\mapsto \ln(f)$ deviennent des intervalles de mêmes longeur. Cette longueur commune 
est 
\beqns
\mbox{lg}=\frac{1}{N_f}\left (\ln\frac{f_e}{2}-\ln f_{\min}\right )
\eeqns
Les supports des filtres sont alors donnés par 
\beqn
\supp(\MH_m)=
\left [   e^{\ln f_{\min}+(m-1)\mbox{lg}}, e^{\ln f_{\min}+m\mbox{lg}} \right ]
=\left [   f_{\min}r^{m-1}, f_{\min}r^m \right ]
\eeqn
où $r=e^{\mbox{lg}}$

Ici s'agissant de signaux de durée finie, la fréquence minimale se déduit de la durée du signal
\beqn
f_{\min}=\frac{1}{\mbox{duree}}
\eeqn

En pratique on élargit un petit peu le filtre basse fréquence de façon à y inclure la fréquence nulle. 
\beqn
\supp(\MH_1)
=\left [   0, f_{\min}r \right ]
\eeqn

\begin{enumerate}
\deb
\item 
On considère ici $f_e= 44100$Hz et 8 filtres dont les caractéristiques sont adaptées à des signaux de durée $30ms$. 
Donnez les instructions permettant d'obtenir la gauche de la figure~\ref{fig9} (p.~\pageref{fig9}) qui représente un banc de filtre.

\begin{corr}
\relax \sol
\begin{verbatim}
fe=44100; Nf=8; duree=30e-3; fmin=1/duree;
lg=(log(fe/2)-log(fmin))/Nf; 
r=exp(lg); 
F=zeros(Nf,6); H=zeros(Nf,6); 
for m=1:Nf
  F(m,:)=[0 fmin*r^(m-1) fmin*r^(m-1) fmin*r^m fmin*r^m  fe/2];
  H(m,:)=[0 0            1            1        0         0]; 
end
H(1,1:2)=1;
figure(1); plot(F',H');
axis([0 fe/2 -0.01 1.01])
tmp=F'; F_hor=tmp(:)'; [F_ord,ordre]=sort(F_hor); 
[F_mat,K_mat]=meshgrid(F_ord,1:Nf);
H_mat=zeros(size(F_mat));
for m=1:Nf
  Hm=[zeros(size(H(1:(m-1),:))); H(m,:); zeros(size(H((m+1):end,:)))];
  tmp=Hm'; Hm_hor=tmp(:)'; 
  H_mat(m,:)=Hm_hor(ordre);
end
figure(2); waterfall(F_mat,K_mat,H_mat);
\end{verbatim}
\end{corr}
\finenum
\end{enumerate}

Ce banc de filtre permet de définir de multiples descripteurs. 
Nous considérons ici {\tt LOG\_MAG\_FB\_LOG}~\footnote{La signification du nom se retrouve en lisant 
l'expression de droite à gauche. {\tt LOG} en dernier signifie qu'on considère une échelle logarithme, puis 
que cette échelle permet de définir un banc de filtre ({\em filter bank}), qu'on calcule ensuite le module ({\em magnitude}) 
et qu'avant de considérer la moyenne on considère le logarithme des modules.}

\beqn
\mbox{LOG\_MAG\_FB\_LOG}(k,m)=
\frac{1}{\left |\supp(\MH_m)\right |}{\int_{f\in \supp(\MH_m)} 10\log_{10}\left |\widehat{H}_m(f)\widehat{X}_k(f)\right|\,df}
\eeqn
où $|\supp(\MH_m)|$  désigne la longueur de l'intervalle sur lequel $\mathcal H$ est non-nul. 

En pratique, le spectre est discrétisé et 
\beqn
\mbox{LOG\_MAG\_FB\_LOG}(k,m)=
\frac{1}{\left |\{l|f_l\in \supp(\MH_m)\}\right |}{\sum_{\{l|f_l\in \supp(\MH_m)\}} 10\log_{10}\left |\widehat{H}_{m,l}\widehat{X}_{k,l}\right|}
\eeqn
où $|\{l|f_l\in \supp(\MH_m)\}|$ désigne le nombre d'indices $l$ pour lesquels $\widehat{H}_{m,l}$ est non-nul. 
\begin{enumerate}
\deb
\item Choisissez un signal, un découpage de ce signal en trames et un nombre de filtre. Représentez en fonction 
du temps et de la fréquence ce descripteur pour ce son. 
La figure~\ref{fig10} (p.~\pageref{fig10}) représente un exemple de représentation de ce descripteur temps-fréquence.

\begin{corr}
\relax \sol
\begin{verbatim}
fe=44100; Nf=5; duree_trame=30e-3; 
r=exp(lg); 
F=zeros(Nf,6); H=zeros(Nf,6); 
[x,fe]=lireSon('..\donnees\un\',0);
[xk,ech_trames,centres_trames]=preparation(x,fe,duree_trame,0.25);
N=size(xk,2); K=size(xk,1); 
lg=(log(fe/2)-log(fe/N))/Nf; 
r=exp(lg); 
fenetre=window(@hamming,N)'; S_fen=sum(fenetre);
%Xk=zeros(length(centres_trames),ceil(length(xk,1)/2));
LOG_MAG_FB_LOG=zeros(K,Nf);
for trame=1:K
  Xk=fft(xk(trame,:).*fenetre);  
  F=0:fe/N:(fe-fe/N); 
  for l=1:Nf
    if 1==l ind=find(F<=fe/N*r);
    else ind=find((F>=fe/N*r^(l-1))&(F<=fe/N*r^(l)));
    end
    LOG_MAG_FB_LOG(trame,l)=mean(10*log10(abs(Xk(ind))));
  end
end
ech_freq=fe/N*r.^(0.5:(Nf-0.5));
[centres_trames_mat,ech_freq_mat]=meshgrid(centres_trames,ech_freq); 
figure(9); waterfall(centres_trames_mat,ech_freq_mat,LOG_MAG_FB_LOG');
figure(10); plot(centres_trames,LOG_MAG_FB_LOG)
legend(num2str(ech_freq','%4.0f'));
\end{verbatim}
\end{corr}
\item\label{it:distance4} Construisez une distance appelée {\tt distance4} qui permette d'évaluer la différence entre 
deux sons en considérant toutes les valeurs de ce descripteurs ($N_f$=5) et en tenant d'une possible distorsion temporelle.

\begin{corr}
\relax \sol
\begin{verbatim}
function d=distance4(rep1,m1,rep2,m2)
  [x1,fe]=lireSon(rep1,m1); [x2,fe]=lireSon(rep2,m2);
  duree_trame=30e-3; chevauchement=0.25; 
  xk1=preparation(x1,fe,duree_trame,chevauchement);
  xk2=preparation(x2,fe,duree_trame,chevauchement);
  x=c_LOG_MAG_FB_LOG(xk1,fe); 
  y=c_LOG_MAG_FB_LOG(xk2,fe); 
  delta=ceil(min(length(x),length(y))/10);
  d=dist_DTD(x,y,delta);
end

function LOG_MAG_FB_LOG=c_LOG_MAG_FB_LOG(xk,fe)
  N=size(xk,2); K=size(xk,1); Nf=5;
  lg=(log(fe/2)-log(fe/N))/Nf; 
  r=exp(lg); 
  fenetre=window(@hamming,N)'; S_fen=sum(fenetre);
  %Xk=zeros(length(centres_trames),ceil(length(xk,1)/2));
  LOG_MAG_FB_LOG=zeros(K,Nf);
  for trame=1:K
    Xk=fft(xk(trame,:).*fenetre);  
    F=0:fe/N:(fe-fe/N); 
    for l=1:Nf
      if 1==l ind=find(F<=fe/N*r);
      else ind=find((F>=fe/N*r^(l-1))&(F<=fe/N*r^(l)));
      end
      LOG_MAG_FB_LOG(trame,l)=mean(10*log10(abs(Xk(ind))));
    end
  end
end
\end{verbatim}
\end{corr}

\item Calculez la sensibilité globale obtenue avec {\tt distance4}. 
\begin{corr}
\begin{verbatim}
OA4=s_mesure_sensibilite(@s_distance4,5);
\end{verbatim}
Avec mes données, je trouve  $0.93$.
\end{corr}
\finenum 
\end{enumerate}

\appendix
\chapter{Annexes}

\section{Fonctions disponibles}
\label{sec:fct}

\subsection{Fonctions disponibles avec le code}
\label{ssec:fct_code}
\begin{itemize}
\itb
La fonction {\tt m4a2mat} transforme les fichiers {\tt .m4a} en des fichiers {\tt .mat} contenant la variable {\tt y} et {\tt fe}.
Voici une utilisation possible. 
\begin{verbatim}
m4a2mat('..\donnees\un\');m4a2mat('..\donnees\deux\');m4a2mat('..\donnees\trois\')
\end{verbatim}
\itb La fonction {\tt lireSon} permet de dénombrer le nombre de fichiers {\tt .mat} disponibles dans un répertoire, ou 
de lire un fichier {\tt .mat} particulier, ou d'en sélectionner un au hasard. Voici une utilisation possible. 
\begin{verbatim}
M=lireSon('..\donnees\un',-1);
for m=1:M
  [y,fe]=lireSon('..\donnees\un',m);
  sound(y,fe);
end
\end{verbatim}
\itb La fonction {\tt decoupe\_signal} transforme un signal {\tt x} échantillonné à la fréquence {\tt fe}
en une matrice {\tt xk}. Chaque ligne de cette matrice représente une trame dont la durée est déterminée par
{\tt duree\_trame}. La proportion commune entre une trame et la suivante est déterminée par un coefficient entre 
$0$ et $1$ noté {\tt chevauchement}. L'instant associé au milieu de chaque trame est notée dans un vecteur colonne
appelé {\tt centres\_trames}. {\tt ech\_trames} est une matrice de même taille que {\tt xk}, 
et chaque composante indique l'instant associé à la composante correspondante de {\tt xk}. 
Les équations utilisée dans ce programme sont définies en annexe~\ref{ssec:decoupe_signal}. Leur 
adaptation doit tenir compte du fait que sous Matlab, les indices commencent par $1$ aussi bien pour $k$ que pour $n$. 
Voici un exemple d'utilisation.
\begin{verbatim}
[x,fe]=lireSon('..\donnees\un\',0);
duree_trame=30e-3; chevauchement=0.25;
[xk,ech_trames,centres_trames]=decoupe_signal(x,fe,duree_trame,chevauchement);
\end{verbatim}

\itb La fonction {\tt visualisation} donne une visualisation d'un signal découpé. Cette fonction requière 
le signal {\tt x}, la fréquence d'échantillonnage {\tt fe}, la durée de chaque trame {\tt duree\_trame}, 
le signal découpé {\tt xk} et ses instants associés {\tt ech\_trames} ainsi que les instants associés 
aux centres de chacune des trames {\tt centres\_trames}. 
Voici un exemple d'utilisation.
\begin{verbatim}
[x,fe]=lireSon('..\donnees\un\',0);
duree_trame=30e-3; chevauchement=0.25;
[xk,ech_trames,centres_trames]=decoupe_signal(x,fe,duree_trame,chevauchement);
visualisation(x,fe,duree_trame,xk,ech_trames,centres_trames);
\end{verbatim}
\itb La fonction {\tt preparation} prépare les données en transformant les fichiers disponibles dans {\tt .mat} 
\itb La fonction {\tt genere\_un\_pb\_classification} utile pour la question~\ref{it:genere_un_pb_classification}
génère un problème de classification en prélevant dans la base de données un son et en mettant tous les autres 
dans des répertoires d'apprentissages.
Voici un exemple d'utilisation. 
\begin{verbatim}
rep_l={'..\donnees\un\','..\donnees\deux\','..\donnees\trois\'}; 
rep_simulation='..\simulation\';
rep_requete='..\simulation\requete\';
rep_apprend_l={'..\simulation\un\','..\simulation\deux\','..\simulation\trois\'}; 
y=genere_un_pb_classification(rep_l,rep_simulation,rep_requete,rep_apprend_l);
\end{verbatim}
où y est un entier parmi 1, 2 ou 3 indiquant la nature du son requête. 
\itb La fonction {\tt genere\_validation\_croisee} utile pour la question~\ref{it:genere_validation_croisee} 
génère une fonction qui au moyen d'un argument $k$ génère 
$K$ validations croisées. Voici un exemple d'utilisation.
\begin{verbatim}
 rep_l={'..\donnees\un\','..\donnees\deux\','..\donnees\trois\'}; 
 rep_simulation='..\simulation\';
 rep_requete='..\simulation\requete\';
 rep_apprend_l={'..\simulation\un\','..\simulation\deux\','..\simulation\trois\'}; 
 K=5;
 genere_pb=genere_validation_croisee(rep_l,rep_simulation,rep_requete,rep_apprend_l,K);
 for k=1:K
   y=genere_pb(k);
   disp(['les repertoires pour la validation ',num2str(k),' sont presents'])
   pause(1),
 end
\end{verbatim}
\itb
{\tt [d,delta]=dist\_DTD(x,y,delta)} cette fonction évalue la distance avec l'algorithme du time warping 
(deformation temporelle dynamique)
en supposant qu'on ne peut décaler les indices de plus ou moins {\tt delta} ($\delta$)
{\tt x} et {\tt y} sont des matrices dont les lignes sont des listes d'attributs et chaque ligne est associée à un 
instant particulier. 
Voici un exemple d'utilisation 
\begin{verbatim}
x=[0 1 1 3 2 2]';
y=[0 1 3 2]'; 
delta=3;
[d,delta]=dist_DTD(x,y,delta);
\end{verbatim}

\end{itemize}

\subsection{Fonctions disponibles sans le code et utilisables seulement avec le préfixe {\tt s\_}}
\label{ssec:fct_ss_code}

\begin{itemize}
\itb La fonction {\tt synthetiseur\_vocal} convertit un vecteur composé de chiffres à valeurs dans {1,2,3} en un séquence de
mots, qui dictent oralement la séquence de chiffres, (question~\ref{it:seq_mots}).
Voici un exemple d'utilisation 
\begin{verbatim}
s_synthetiseur_vocal([1,2,3]); 
\end{verbatim}
\itb La fonction {\tt sous\_ech} modifie 
la fréquence d'échantillonnage d'un signal et l'amène à $8$kHz. 
(question~\ref{it:sous_ech}).
Voici un exemple d'utilisation 
\begin{verbatim}
[y,fe]=lireSon('..\donnees\un\',0); 
[y2,fe2]=s_sous_ech(y,fe);
\end{verbatim}
\itb La fonction {\tt rehausser} permet d'atténuer les fréquences plus basses sans modifier 
la perception humaine du son, conformément à la question~\ref{it:rehausser}. 
Voici un exemple d'utilisation
\begin{verbatim}
[y,fe]=lireSon('..\donnees\un\',0); 
y1=s_rehausser(y,fe); 
sound(y1,fe);
\end{verbatim}
\itb Conformément à la question~\ref{it:c_PMCT}, la fonction {\tt c\_PMCT} permet de calculer la puissance moyenne de chaque trame à partir du signal 
découpé. En moyennant l'ensemble des valeurs calculées, elle permet aussi de trouver une estimation de la puissance 
moyenne du signal. 
Voici un exemple d'utilisation
\begin{verbatim}
[y,fe]=lireSon('..\donnees\un\',0); 
[xk,ech_trames,centres_trames]=decoupe_signal(x,fe,30e-3,0.25);
PMCT=s_c_PMCT(xk);
\end{verbatim}
\itb Conformément à la question~\ref{it:supprime_silence}, la fonction {\tt supprime\_silence} permet de retirer les trames 
correspondant à un silence. 
Voici un exemple d'utilisation.
\begin{verbatim}
[x,fe]=lireSon('..\donnees\un\',0); 
[xk,ech_trames,centres_trames]=decoupe_signal(x,fe,30e-3,0.25);
xk=s_supprime_silence(xk,ech_trames,centres_trames); 
\end{verbatim}
\itb Les fonctions {\tt distance1, distance2, distance3, distance4} sont 
des exemples de distances. Elles sont respectivement définies 
dans les questions~\ref{it:distance1},~\ref{it:distance2},~\ref{it:distance3} et~\ref{it:distance4}. 
Voici un exemple d'utilisation. 
\begin{verbatim}
M1=lireSon('..\donnees\un\',-1); M2=lireSon('..\donnees\deux\',-1); 
m1=ceil(rand(1)*M1); m2=ceil(rand(1)*M2); 
disp(['d1(',num2str(m1),',',num2str(m2),')=',...
  num2str(s_distance1('..\donnees\un\',m1,'..\donnees\deux\',m2))]),
disp(['d2(',num2str(m1),',',num2str(m2),')=',...
  num2str(s_distance2('..\donnees\un\',m1,'..\donnees\deux\',m2))]),
disp(['d3(',num2str(m1),',',num2str(m2),')=',...
  num2str(s_distance3('..\donnees\un\',m1,'..\donnees\deux\',m2))]),
disp(['d4(',num2str(m1),',',num2str(m2),')=',...
  num2str(s_distance4('..\donnees\un\',m1,'..\donnees\deux\',m2))]),
\end{verbatim}
\itb Conformément à la question~\ref{it:predit}, la fonction {\tt predit} classifie un son identifié par {\tt rep\_requete,m}
vis-à-vis d'un ensemble d'apprentissage {\tt rep\_apprend\_l} et d'une fonction de distance {\tt distance}. 
Voici un exemple d'utilisation
\begin{verbatim}
rep_l={'..\donnees\un\','..\donnees\deux\','..\donnees\trois\'}; 
rep_simulation='..\simulation\';
rep_requete='..\simulation\requete\';
rep_apprend_l={'..\simulation\un\','..\simulation\deux\','..\simulation\trois\'}; 
i_ref=genere_un_pb_classification(rep_l,rep_simulation,rep_requete,rep_apprend_l);
y=s_predit(rep_requete,1,rep_apprend_l,@s_distance1);
disp(['y=',num2str(y),' i_ref=',num2str(i_ref)])
\end{verbatim}
\itb Conformément à la question~\ref{it:mesure_sensibilite}, la fonction {\tt mesure\_sensibilite} met en oeuvre une validation 
croisée avec $K$ et applique l'algorithme de recherche du son le plus proche au sens d'une distance à indiquer. 
Voici un exemple d'utilisation 
\begin{verbatim}
P=mesure_sensibilite(@s_distance1,5); disp(['P=',num2str(P)]),
\end{verbatim}
\itb Conformément à la question~\ref{it:c_F0_ZCR}, la fonction {\tt c\_F0\_ZCR} calcule pour chaque trame une estimation de la fréquence
fondamentale. 
Voici un exemple d'utilisation.
\begin{verbatim}
[x,fe]=lireSon('..\donnees\un\',0); 
[xk,ech_trames,centres_trames]=decoupe_signal(x,fe,30e-3,0.25);
xk=s_supprime_silence(xk,ech_trames,centres_trames); 
F0_ZCR=s_c_F0_ZCR(xk,fe); 
\end{verbatim}
\itb Conformément à la question~\ref{it:c_SPC}, la fonction {\tt c\_SPC} calcule pour chaque trame une estimation de la 
fréquence moyenne.
Voici un exemple d'utilisation. 
\begin{verbatim}
[x,fe]=lireSon('..\donnees\un\',0); 
[xk,ech_trames,centres_trames]=decoupe_signal(x,fe,30e-3,0.25);
xk=s_supprime_silence(xk,ech_trames,centres_trames); 
SPC=s_c_SPC(xk,fe); 
\end{verbatim}
\itb Conformément à la question~\ref{it:c_SPW}, la fonction {\tt c\_SPW} calcule pour chaque trame une estimation de la 
la larbeur du spectre.
Voici un exemple d'utilisation. 
\begin{verbatim}
[x,fe]=lireSon('..\donnees\un\',0); 
[xk,ech_trames,centres_trames]=decoupe_signal(x,fe,30e-3,0.25);
xk=s_supprime_silence(xk,ech_trames,centres_trames); 
SPW=s_c_SPW(xk,fe); 
\end{verbatim}

\end{itemize}

\subsection{Quelques fonctions de vérifications}
\label{ssec:verif}

L'ensemble des fonctions de vérifications peuvent être  déclenchées avec la fonction {\tt verification}, ce qui 
permet de s'assurer que l'installation fonctionne. 

\begin{itemize}
\itb La fonction {\tt v\_distance} vérifie si une fonction de distance teste sur des
exemples pris au hasard la propriété de symétrie et 
l'inégalité triangulaire.~\footnote{Une distance tenant compte de la déformation 
temporelle dynamique ne vérifie pas l'inégalité triangulaire.} 
Voici un exemple d'utilisation.
\begin{verbatim}
v_distance(@s_distance1);
\end{verbatim}

\itb La fonction {\tt v\_distance\_symetrie} vérifie si une fonction de distance teste sur des
exemples pris au hasard la propriété de symétrie.
Voici un exemple d'utilisation.
\begin{verbatim}
v_distance(@distance2);
\end{verbatim}
\itb La fonction {\tt v\_genere\_un\_pb} vérifie à la fois {\tt genere\_un\_pb\_classification} 
et la fonction générée par {\tt genere\_validation\_croisee}.
Voici un exemple d'utilisation.
\begin{verbatim}
rep_l={'..\donnees\un\','..\donnees\deux\','..\donnees\trois\'}; 
rep_simulation='..\simulation\';
rep_requete='..\simulation\requete\';
rep_apprend_l={'..\simulation\un\','..\simulation\deux\','..\simulation\trois\'}; 
y=genere_un_pb_classification(rep_l,rep_simulation,rep_requete,rep_apprend_l);
v_genere_pb(rep_l,rep_simulation,rep_requete,rep_apprend_l,y);
K=ceil(rand(1)*10);
genere_pb=genere_validation_croisee(rep_l,rep_simulation,rep_requete,rep_apprend_l,K);
for k=1:K
 y=genere_pb(k);
 v_genere_pb(rep_l,rep_simulation,rep_requete,rep_apprend_l,y);
end
\end{verbatim}
\itb La fonction {\tt v\_predit} vérifie {\tt s\_predit} et s'utilise ainsi. 
\begin{verbatim}
v_predit();
\end{verbatim}
\itb La fonction {\tt v\_mesure\_sensibilite} vérifie {\tt mesure\_sensibilite} et s'utilise ainsi. 
\begin{verbatim}
v_mesure_sensibilite();
\end{verbatim}

\end{itemize}

\section{Explications sur certaines questions}

\begin{figure}
\begin{center}
\begin{tabular}{cc}
\includegraphics[width=0.5\linewidth]{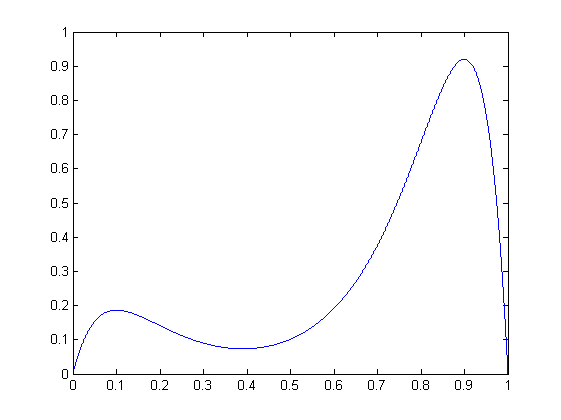} &\includegraphics[width=0.5\linewidth]{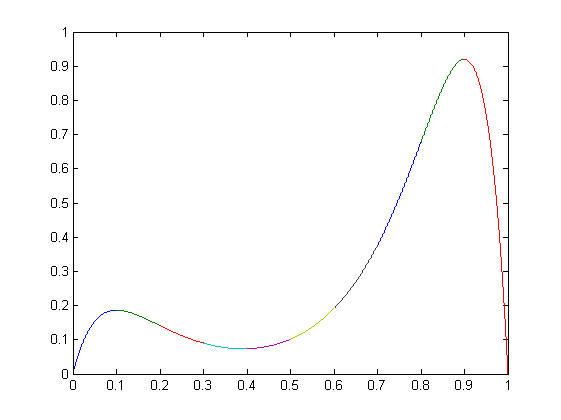}\\
\includegraphics[width=0.5\linewidth]{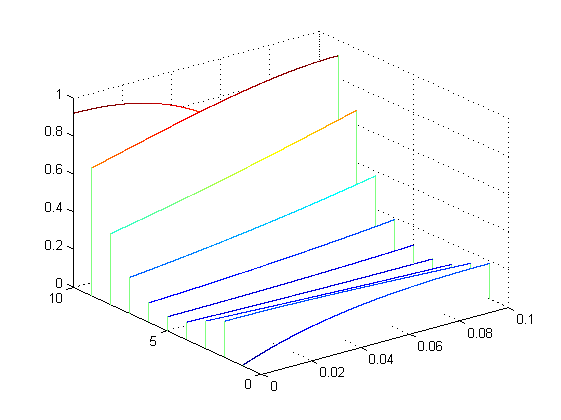} & \includegraphics[width=0.5\linewidth]{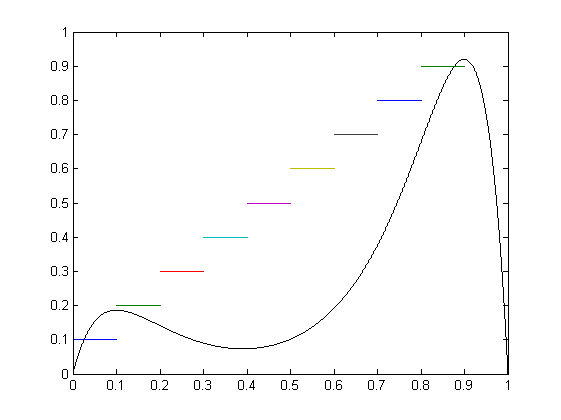}
\end{tabular}
\end{center}
\caption{En haut à gauche~: signal {\tt x} en fonction du temps. En haut à droite~: succession des trames affichées avec des couleurs différentes en fonction du temps. En bas à gauche~: succession des trames représentées cette fois en fonction d'une base de temps propre à la trame. En bas à droite~: signal {\tt x} en noir en fonction du temps et indication colorée des instants associés à chaque trame.}
\label{fig5}
\end{figure}

La figure~\ref{fig5} donne une autre illustration de la section~\ref{ssec:decoupe}
pour le signal $x(t)=25*(1-t)e^{-10*(1-t)}+5te^{-10*t}$ avec $f_e=44.1$kHz, une 
durée de $1$s. Chaque trame dure $100$ms et il n'y a pas de chevauchement.

\section{Explications}
\subsection{Découpage des signaux}
\label{ssec:decoupe_signal}

Pour simplifier, on suppose dans un premier temps qu'il s'agit de signaux à temps continu. 

Nous considérons les notations suivantes~:
\begin{itemize}
\item $x(t)$ signal à découper
\item $T_x$ durée du signal $x$
\item $x_k(t)$ succession de trames du signal $x$
\item $T_{xk}$ durée de chaque portion du signal
\item $t_k^{(de)},t_k^{(ce)},t_k^{(fi)}$~: instants associés au début, milieu et fin de chaque trame
\item $\alpha$ chevauchement entre une trame et la suivante. Il s'agit d'une valeur entre $0$ et $1$ qui indique la portion 
 commune entre une trame et la suivante. 
\end{itemize}

Les équations du découpage en trames sont les suivante~\footnote{Ces équations prennent en compte le fait que 
la partie du signal $x$ qui ne ferait pas une trame complète n'est pas prise en compte}~: 
\beqn
\left \{
\begin{array}{l}
t_0^{(de)}=0\\[0.1cm]
t_k^{(fi)}=t_k^{(de)}+T_s\\[0.2cm]
t_{k+1}^{(de)}+\alpha T_s=t_k^{(fi)}\\[0.2cm]
t_k^{(ce)}=\frac{1}{2}\left (t_k^{(de)}+t_k^{(fi)}\right )\\[0.2cm]
t_{K-1}^{(fi)}\leq T_x\leq t_{K}^{(fi)}\\[0.2cm]
x_k(t)=x\left (t-t_k^{(de)}\right )\1_{[t_k^{(de)},t_k^{(fi)}]}\left (t-t_k^{(de)}\right)\\[0.2cm]
\end{array}
\right . 
\eeqn

Ces équations se résolvent en cherchant par exemple d'abord à déterminer $t_k^{(de)}$.
\beqn
\left \{
\begin{array}{l}
t_k^{(de)}=kT_{xk}(1-\alpha)\\[0.2cm]
t_k^{(ce)}=kT_{xk}(1-\alpha)+\frac{T_{xk}}{2}\\[0.2cm]
t_k^{(fi)}=kT_{xk}(1-\alpha)+T_{xk}\\[0.2cm]
K=\left \lceil  \frac{T_x-T_{xk}}{T_{xk}(1-\alpha)} \right \rceil
\end{array}
\right . 
\eeqn
où $\lceil x\rceil$ désigne  le plus grand entier inférieur à $x$.

En réalité, il s'agit de signaux à temps discret. Il n'y a aucune raison en général que la durée du signal coïncide avec 
la fin de la dernière trame, il n'y a aucune raison que les débuts et fin des 
trames coïncident avec des échantillons. 

Considérant les nouvelles notations 
\begin{itemize}
\item $\beta=\frac{T_x}{T_{xk}}>1$
\item $\gamma=\frac{T_{xk}}{T_e}>1$
\end{itemize}
L'indice dans $x_n$ du premier élément de la trame $k$ est le plus petit  entier au dessus de $\frac{t_k^{(de)}}{T_e}$, 
tandis que l'indice dans $x_n$ du dernier élément de la trame $k$ 
doit d'une part correspondre à un instant proche de $t_k^{(fi)}$ mais doit être ajusté de telles sortes que toutes 
les trames soient toutes de même longueur notée $N_K$. 
est le plus grand  entier en dessous de $\frac{t_k^{(fi)}}{T_e}$. Chaque échantillon $x_{k}[n]$ est associé à un 
instant noté $t_{k,n}$.
Ainsi 
\beqn
\left \{
\begin{array}{l}
K=\left \lceil  \frac{\beta-1}{1-\alpha} \right \rceil\\[0.2cm]
n_k^{(de)}=\left \lceil \gamma k(1-\alpha) \right \rceil\\[0.2cm]
N_K=\left \lfloor \gamma\right \rfloor\\[0.2cm]
n_k^{(fi)}=n_k^{(de)}+N_k-1\\[0.2cm]
x_k[n]=x\left [n-n_k^{(de)}\right ]\1_{\left [n_k^{(de)},n_k^{(fi)}\right]}\left [n\right ]\\[0.2cm]
t_{k,n}=\left (n_k^{(de)}+n\right )T_e\mbox{ pour }0\leq n\leq N_K-1\\[0.2cm]
t_k^{(ce)}=\gamma(k(1-\alpha)+\frac{1}{2})\\
\end{array}
\right . 
\eeqn
où $\lfloor x\rfloor$ désigne  le plus grand entier inférieur à $x$.

\subsection{Calculs}
\label{sec:calcul}

Le filtre a la fréquence de coupure $f_c$ quand 
\beqns
c=\cos\left (2\pi \frac{f_c}{f_e}\right ) \mbox{ et }
\eta=2c+1-2\sqrt{c(c+1)}
\eeqns
\begin{itemize}
\item
Le filtre considéré a une fonction de transfert $H(z)=1-\eta z^{-1}$ et 
une réponse fréquentielle $\widehat{H}(f)=1-\eta e^{-j2\pi \frac{f}{f_e}}$.
\item 
Le calcul s'appuie sur une évaluation du module de la réponse fréquentielle en $\frac{f_e}{2}$
\beqns
|\widehat{H}(f_e)|^2=|1-\eta(-1)|^2=(1+\eta)^2
\eeqns
et sur une évaluation en $f=f_c$
\beqns
|\widehat{H}(f)|^2=|1-\eta e^{-j2\pi \frac{f}{f_e}}|^2=(1-\eta \cos(2\pi\frac{f}{f_e})+\eta^2 \cos^2(2\pi\frac{f}{f_e}))+
\eta^2\sin^2(2\pi\frac{f}{f_e})=1+\eta^2-2\eta c
\eeqns
\item 
La fréquence $f$ est la fréquence de coupure si $|\widehat{H}(f_e)|^2=2|\widehat{H}(f)|^2$
\item Ainsi $\eta$ est la solution de cette équation du second degré
\beqns
(1+\eta)^2=2(1+\eta^2-2\eta c)
\eeqns
où $c=\cos\left (2\pi \frac{f_c}{f_e}\right )$
\item Cette équation s'écrit 
\beqns
\eta^2-2\eta (2c+1)+1=0
\eeqns
et comporte deux solutions seulement si $c<0$, c'est-à-dire si $f_e>4f_c$. 
Une des deux solutions est $\eta_1=2c+1-2\sqrt{c(c+1)}$. L'autre solution est $\eta_2=2c+1+2\sqrt{c(c+1)}$.
\item Un calcul simple montre que $\eta_1=2c+1-2\sqrt{c(c+1)}<1<2c+1+2\sqrt{c(c+1)}=\eta_2$ prouvant 
que la première solution conduit à un filtre à minimum de phase mais pas la seconde solution.
\item Si on note respectivement $\widehat{H}_1(f)$ et $\widehat{H}_2(f)$ les filtres associés à la première et la deuxième solution, 
on a $\frac{|\widehat{H}_1(f)|^2}{|\widehat{H}_2(f)|^2}=\frac{1+\eta_1^2-2\eta_1\cos(\frac{2\pi f}{f_e})}
{1+\eta_2^2-2\eta_2\cos(\frac{2\pi f}{f_e})}$. Utilisant le fait que $\eta_1$ et $\eta_2$ sont solutions de la même équation du second 
degré, on peut affirmer que les deux filtres sont  en module proportionnels. 
$\frac{|\widehat{H}_1(f)|^2}{|\widehat{H}_2(f)|^2}=\frac{\eta_1}{\eta_2}$
\end{itemize}

\section{Simulation concernant de possible déformations d'un signal}

\begin{figure}[htbp]
\begin{center}
\begin{tabular}{c}
\includegraphics[width=0.4\linewidth]{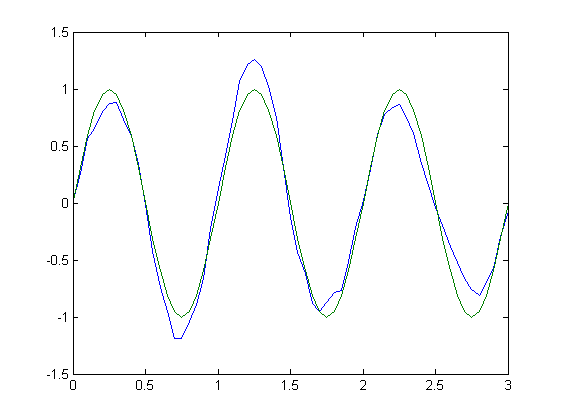}
\includegraphics[width=0.4\linewidth]{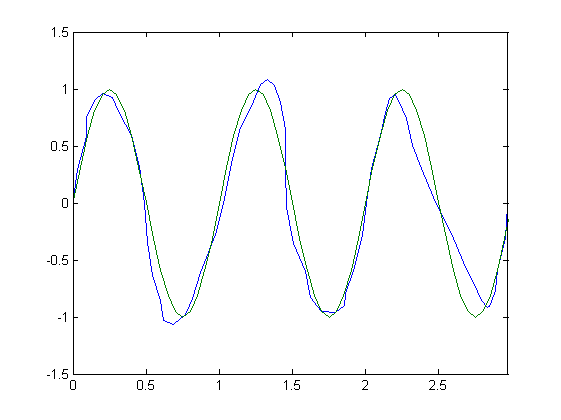}\\
\includegraphics[width=0.4\linewidth]{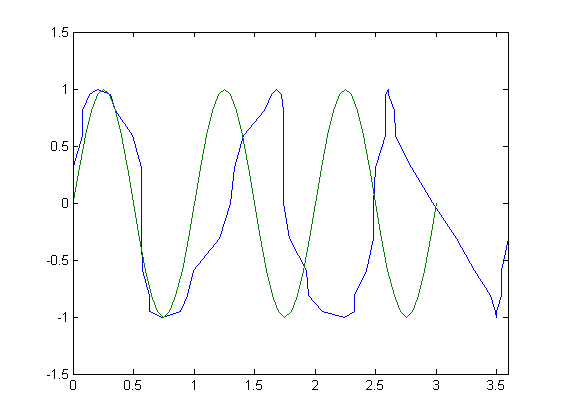}
\includegraphics[width=0.4\linewidth]{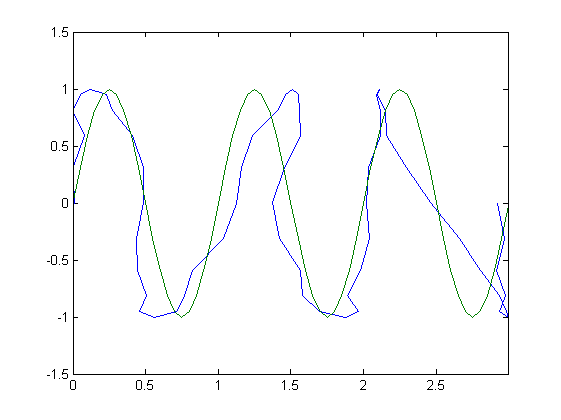}
\end{tabular}
\end{center}
\caption{En haut à gauche~: sinusoïde déformée par l'ajout d'un bruit blanc gaussien filtré par un moyenneur.
En haut à droite~: sinusoïde déformée par l'ajout d'un bruit blanc gaussien filtré par un moyenneur à la fois 
sur l'échelle des abscisses et des ordonnées.
En bas à gauche et à droite~: sinusoïde déformée par l'ajout d'un bruit blanc gaussien filtré par un moyenneur sur l'échelle des abscisses.
}
\label{figO2}
\end{figure}

La figure~\ref{figO2} montre différentes façons de déformer une sinusoïde. 
On considère dans cette expérimentation une sinusoïde de fréquence $1$Hz échantillonnée à la fréquence $f_e=20$Hz.
\beqns
\left \{
\begin{array}{lcl}
t_n&=&nT_e\\
x_n&=&\sin(2*\pi nT_e)
\end{array}
\right . 
\eeqns
On considère un bruit blanc gaussien d'écart-type $0.1$ et filtré par un moyenneur sur une durée de $250$ms noté $B_n$. 
La modélisation classique illustré en haut à gauche de la figure~\ref{figO2} est 
\beqns
\left \{
\begin{array}{lcl}
t_n&=&nT_e\\
y_n^{(1)}&=&x_n+3B_n
\end{array}
\right . 
\eeqns
L'opérateur $[\,]_{\oplus}$ consiste à ne considérer que des incréments positifs ou nuls, de sorte que $[t_n+B_n]_{\oplus}$
est une suite croissante. 
une modélisation parfois utilisée en traitement du signal consiste à déformer à la fois l'axe des abscisses et des ordonnées. 
Ici il s'agit d'une version où l'échelle des temps est une suite croissante, illustrée sur  en haut à droite de la figure~\ref{figO2}. 
\beqns
\left \{
\begin{array}{lcl}
t_n&=&\left [nT_e+B_n\right ]_{\oplus}\\
y_n^{(2)}&=&x_n+B_n
\end{array}
\right . 
\eeqns

En bas à gauche de la figure~\ref{figO2} est illustrée une déformation ne concernant que l'échelle des abscisses, c'est vis-à-vis de cette 
déformation que l'on cherche à être robuste. 
La modélisation associée est la suivante. 
\beqns
\left \{
\begin{array}{lcl}
t_n&=&\left [nT_e+3B_n\right ]_{\oplus}\\
y_n^{(3)}&=&x_n
\end{array}
\right . 
\eeqns

En bas à droite de la figure~\ref{figO2} est illustrée une déformation de l'échelle des abscisses où la contrainte que cette échelle
reste croissante n'est pas appliquée. 
\beqns
\left \{
\begin{array}{lcl}
t_n&=&nT_e+3B_n\\
y_n^{(4)}&=&x_n
\end{array}
\right . 
\eeqns

\begin{corr}
\relax 
Code permettant de réaliser la figure~\ref{figO2}
\begin{verbatim}
t=0:5e-2:3;
x=sin(2*pi*t);
b=filter(ones(1,5)/5,1,randn(size(t)));
figure(1); plot(t,x+0.3*b,t,x); axis([0 3 -1.5 1.5])
t1=diff(t+0.1*b); t2=[0 cumsum(t1/2+abs(t1)/2)];
figure(2); plot(t2,x+0.1*b,t,x); axis([0 max(t2) -1.5 1.5])
t3=diff(t+0.3*b); t4=[0 cumsum(t3/2+abs(t3)/2)]; 
figure(3); plot(t4,x,t,x); axis([0 max(t4) -1.5 1.5])
t5=t+0.3*b; 
figure(4); plot(t5,x,t,x); axis([0 max(t5) -1.5 1.5])
\end{verbatim}
\end{corr}

\section{Onde sonore stationnaire à une dimension}

L'application~\footnote{Cette annexe est inspirée de~\cite{DIsrael10ch5}} 
du principe fondamental de la dynamique et de la thermodynamique des fluides 
permettent d'écrire 
\beqn
\frac{1}{c^2}\frac{\partial^2 p(x,t)}{\partial t^2}-\frac{\partial^2 p(x,t)}{\partial x^2}=0
\eeqn
où $c$ est la vitesse du son. 

Les solutions de cette équation différentielle sont de la forme 
\beqn
p(x,t)=p_0+f_1(t-\frac{x}{c})+f_2(t+\frac{x}{c})
\eeqn

On considère une onde avec une seule fréquence $f_0$.
\beqn\label{eq:p_onde_freq1}
p(x,t)=P_0+\Delta P_1\cos\left [2\pi f_0\left(t-\frac{x}{c}\right )+\varphi-\psi\right]
+\Delta P_2\cos\left[2\pi f_0\left (t+\frac{x}{c}\right )+\varphi+\psi\right]
\eeqn 

On considère ici un tuyau fermé de longueur L. 
On suppose $x=0$ à son extrêmité gauche et $x=L$ à son extrêmité droite. 
Il n'y a pas de déplacement de l'air là où le tube est fermé et cela se traduit par 
\beqn
\left .\frac{\partial p}{\partial x}\right |_{x=0}=0\quad \mbox{et}\quad 
\left .\frac{\partial p}{\partial x}\right |_{x=L}=0
\eeqn
Le calcul de ces expressions sur (\ref{eq:p_onde_freq1})
aboutit à 
\beqn
\left .\frac{\partial p}{\partial x}\right |_{x=0}=
\frac{2\pi  f_0}{c}\Delta P_1\sin\left [2\pi f_0\left(t-\frac{0}{c}\right )+\varphi-\psi\right]
-\frac{2\pi f_0}{c}\Delta P_2\sin\left [2\pi f_0\left(t+\frac{0}{c}\right )+\varphi+\psi\right]
\eeqn
Le fait que cette expression est nulle montre que $\Delta P_1=\Delta P_2$ qui est noté $\Delta P$ et 
que $\psi=0$. 
\beqn
\left .\frac{\partial p}{\partial x}\right |_{x=L}=
\frac{2\pi f_0}{c}\Delta P\sin\left [2\pi f_0\left(t-\frac{L}{c}\right )+\varphi\right]
-\frac{2\pi f_0}{c}\Delta P\sin\left [2\pi f_0\left(t+\frac{L}{c}\right )+\varphi\right]
\eeqn
Une relation trigonométrique permet d'écrire 
\beqn
\cos(a+b)-\cos(a-b)=-2\sin(a)\sin(b)
\eeqn

Une  relation trigonométrique similaire montre que 
\beqn
\sin(a+b)-\sin(a-b)=2\sin(b)\cos(a)
\eeqn
Aussi 
\beqn
\left .\frac{\partial p}{\partial x}\right |_{x=L}=
-\frac{4\pi f_0}{c}\Delta P
\cos\left [2\pi f_0 t+\varphi\right]
\sin\left (\frac{2\pi f_0L}{c}\right )
\eeqn
Cette expression est bien nulle si et seulement si 
\beqn
\sin\left (\frac{2\pi f_0L}{c}\right )=0
\eeqn
Ce qui est vrai lorsque 
\beqn
f_0=k\frac{c}{L}
\eeqn
\underline{Plus la longueur augmente plus la fréquence de base diminue. }


\section{Moyenne pondérée et largeur}

\'Etant donnée une courbe de puissance spectrale, $P(f)$, on cherche la fréquence moyenne pondérée par cette puissance. 
On peut définir cette fréquence comme la fréquence qui minimise une erreur quadratique fréquentielle pondérée par $P(f)$. 
\beqn\label{eq:mu_min}
\mu=\underset{\nu}{\arg\min}{\int (f-\nu)^2P(f)\,df}
\eeqn
En posant une fonction de coût de $\nu$ définie par 
\beqns
J(\nu)={\int (f-\nu)^2P(f)\,df}
\eeqns
La définition de $\mu$ avec (\ref{eq:mu_min}) signifie qu'en $\nu=\mu$, $\frac{\partial J}{\partial \nu}=0$. 
Le calcul de cette dérivée partielle conduit à 
\beqn\label{eq:mu_partial}
\left .\frac{1}{2}\frac{\partial J}{\partial \nu}\right |_{\nu=\mu}={\int (f-\mu)P(f)\,df}=0
\eeqn
En coupant cette intégrale en deux parties, les valeurs en dessous de $\mu$ et les valeurs au dessus de $\mu$, on peut montrer $\mu$
comme un centre de gravité de la courbe. 
\beqn
{\int_{f\leq \mu} (\mu-f)P(f)\,df}={\int_{f\geq \mu} (f-\mu)P(f)\,df}
\eeqn
En séparant $f$ et $\mu$ dans (\ref{eq:mu_partial}), on trouve une formule permettant de calculer $\mu$
\beqn
{\int f\,P(f)\,df}={\int \mu\, P(f)\,df}
\eeqn
Par suite 
\beqn
\mu=\frac{{\int f\,P(f)\,df}}{{\int P(f)\,df}}
\eeqn

La largeur du spectre est donnée par $2\sigma$ 
où 
\beqn
\sigma=\sqrt{\frac{\int (f-\mu)^2\,P(f)\,df}{{\int P(f)\,df}}}
\eeqn

\begin{corr}
\relax 
Supports de banc de filtres avec une échelle linéaire. 
\beqn
\mbox{lg}=\frac{1}{N_f}\left (\frac{f_e}{2}-f_{\min}\right )
\eeqn
\beqn
\supp \widehat{H}_m=\left [f_{\min} +(m-1)\mbox{lg},f_{\min} +m\mbox{lg}\right ]
\eeqn
\end{corr}




\appendix 

\chapter{Onde sonore stationnaire à une dimension}
\label{app:onde_stationnaire_1D}

L'application~\footnote{Cette annexe est inspirée de~\cite{DIsrael10ch5}} 
du principe fondamental de la dynamique et de la thermodynamique des fluides 
permettent d'écrire 
\beqn
\frac{1}{c^2}\frac{\partial^2 p(x,t)}{\partial t^2}-\frac{\partial^2 p(x,t)}{\partial x^2}=0
\eeqn
où $c$ est la vitesse du son. 

Les solutions de cette équation différentielle sont de la forme 
\beqn
p(x,t)=p_0+f_1(t-\frac{x}{c})+f_2(t+\frac{x}{c})
\eeqn

On considère une onde avec une seule fréquence $f_0$.
L'équation~(\ref{eq:p_onde_freq1}) définie page~\pageref{eq:p_onde_freq1} est recopiée ici. 
\beqns
p(x,t)=P_0+\Delta P_1\cos\left [2\pi f_0\left(t-\frac{x}{c}\right )+\varphi-\psi\right]
+\Delta P_2\cos\left[2\pi f_0\left (t+\frac{x}{c}\right )+\varphi+\psi\right]
\eeqns 

On considère ici un tuyau fermé de longueur L. 
On suppose $x=0$ à son extrêmité gauche et $x=L$ à son extrêmité droite. 
Il n'y a pas de déplacement de l'air là où le tube est fermé et cela se traduit par 
\beqn
\left .\frac{\partial p}{\partial x}\right |_{x=0}=0\quad \mbox{et}\quad 
\left .\frac{\partial p}{\partial x}\right |_{x=L}=0
\eeqn
Le calcul de ces expressions à partir de $p(x,t)$ 
aboutit à 
\beqn
\left .\frac{\partial p}{\partial x}\right |_{x=0}=
\frac{2\pi  f_0}{c}\Delta P_1\sin\left [2\pi f_0\left(t-\frac{0}{c}\right )+\varphi-\psi\right]
-\frac{2\pi f_0}{c}\Delta P_2\sin\left [2\pi f_0\left(t+\frac{0}{c}\right )+\varphi+\psi\right]
\eeqn
Le fait que cette expression est nulle montre que $\Delta P_1=\Delta P_2$ qui est noté $\Delta P$ et 
que $\psi=0$. 
\beqn
\left .\frac{\partial p}{\partial x}\right |_{x=L}=
\frac{2\pi f_0}{c}\Delta P\sin\left [2\pi f_0\left(t-\frac{L}{c}\right )+\varphi\right]
-\frac{2\pi f_0}{c}\Delta P\sin\left [2\pi f_0\left(t+\frac{L}{c}\right )+\varphi\right]
\eeqn
Une relation trigonométrique permet d'écrire 
\beqn
\cos(a+b)-\cos(a-b)=-2\sin(a)\sin(b)
\eeqn

Une  relation trigonométrique similaire montre que 
\beqn
\sin(a+b)-\sin(a-b)=2\sin(b)\cos(a)
\eeqn
Aussi 
\beqn
\left .\frac{\partial p}{\partial x}\right |_{x=L}=
-\frac{4\pi f_0}{c}\Delta P
\cos\left [2\pi f_0 t+\varphi\right]
\sin\left (\frac{2\pi f_0L}{c}\right )
\eeqn
Cette expression est bien nulle si et seulement si 
\beqn
\sin\left (\frac{2\pi f_0L}{c}\right )=0
\eeqn
Ce qui est vrai lorsque 
\beqn
f_0=k\frac{c}{L}
\eeqn
\underline{Plus la longueur augmente plus la fréquence de base diminue. }

\bibliographystyle{plain}

\bibliography{biblio/bibtex_son}
  
\end{document}